\documentclass[12pt]{article}

\RequirePackage{amssymb}
\RequirePackage{amsmath}

\DeclareMathOperator*{\cov}{cov}
\DeclareMathOperator*{\var}{var}
\DeclareMathOperator*{\supp}{supp}
\newcommand{\R}{\ensuremath{\mathbb{R}}}
\newcommand{\Exp}{\ensuremath{\mathbb{E}}}
\newcommand{\Prob}{\ensuremath{\mathbb{P}}}

\usepackage{bbm}

\makeatletter
\newcommand*{\indep}{
  \mathbin{
    \mathpalette{\@indep}{}
  }
}
\newcommand*{\nindep}{
  \mathbin{
    \mathpalette{\@indep}{\not}
  }
}
\newcommand*{\@indep}[2]{
  \sbox0{$#1\perp\m@th$}
  \sbox2{$#1=$}
  \sbox4{$#1\vcenter{}$}
  \rlap{\copy0}
  \dimen@=\dimexpr\ht2-\ht4-.2pt\relax
  \kern\dimen@
  {#2}
  \kern\dimen@
  \copy0 
} 
\makeatother

\usepackage{xr-hyper}
\pdfoutput=1
\usepackage{hyperref}
\hypersetup{
    colorlinks=true,
    linkcolor=black,
    citecolor=black,
    filecolor=black,
    urlcolor=black,
}

\usepackage[longnamesfirst]{natbib}
\usepackage{ulem}
\usepackage{color}

\usepackage{wrapfig}
\usepackage{tikz-cd}
\usepackage{tikz}
\usepackage{pgfplots}

\usepackage{rotating}
\usepackage{lscape}

\usepackage{array}
\usepackage{chngpage}
\usepackage{multirow}
\usepackage{tabulary}

\usepackage{tabularx}
\usepackage{booktabs}

\usepackage{tabularray-2021}
\UseTblrLibrary{booktabs}

\newcommand{\vspan}{\text{span}}
\def\rx{\bar{r}_X}
\def\ry{\bar{r}_Y}
\def\uc{\underline{c}}

\def\Wob{W_1}

\def\Wu{W_2}
\def\dob{d_1}

\RequirePackage{amsthm}

\theoremstyle{definition}

\newtheorem{proposition}{Proposition}

\newtheorem{theorem}{Theorem}
\newtheorem{corollary}{Corollary}

\newtheorem{Aassumption}{Assumption}

\newcounter{homogSection}
\newcommand{\aAssump}{A\arabic{homogSection}}
\newcounter{het1section}
\newcommand{\bAssump}{B\arabic{het1section}}
\newcounter{het2section}
\newcommand{\cAssump}{C\arabic{het2section}}

\newtheoremstyle{theoremSuppressedNumber}{}{}{}{}{\bfseries}{.}{ }{\thmname{#1}\thmnote{ (\mdseries #3)}}
\theoremstyle{theoremSuppressedNumber}

\usepackage{setspace}
\usepackage{subfig}
\usepackage{wrapfig}
\usepackage{float}

\usepackage{epstopdf}

\usepackage{mathtools}

\newcommand{\Lin}{\ensuremath{\mathbb{L}}}

\usepackage{relsize}
\newcommand*{\sbullet}{\mathrel{\mathsmaller{\bullet}}}

\title{\textbf{Assessing Omitted Variable Bias \\ when the Controls are Endogenous}\footnote{This paper is a revised, shorter version of our now-superseded previous working paper titled ``Assessing Omitted Variable Bias when the Controls are Endogenous'' (\citealt{DMP2023}, arXiv:2206.02303v4), without the design-based framework of Section 3. The design-based framework and associated results can now be found in our companion paper ``An Axiomatic Approach to Comparing Sensitivity Parameters
'' (\citealt{DiegertMastenPoirier2025}). We thank audiences at Northwestern, Duke, the SEA 2021 conference, Oxford, Brown, Texas A\&M, the joint Bonn-Mannheim seminar, Jinan University, the 2022 Interactions Conference at The University of Wisconsin-Madison, University of Virginia, UC Irvine, UC San Diego, UCLA, Ohio State, Yale, Penn, the 2023 AEA winter meeting, Johns Hopkins, George Washington University, Stanford, UC Santa Cruz, Western Ontario, UIUC, Michigan State, Rochester, the Federal Reserve Bank of Cleveland, the 2023 American Causal Inference Conference, and the joint CREST-PSE seminar. We thank audiences at those seminars and conferences, as well as Joe Altonji, Isaiah Andrews, Peter Hull, Evan Rose, and Jon Roth for helpful conversations and comments. We thank Hongchang Guo and Muyang Ren for excellent research assistance. Masten thanks the National Science Foundation for research support under Grant 1943138.}}

\author{Paul Diegert\footnote{Toulouse School of Economics,        \texttt{paul.diegert@tse-fr.eu}} \qquad
Matthew A. Masten\footnote{Department of Economics, Duke University,
        \texttt{matt.masten@duke.edu}} \qquad Alexandre Poirier\thanks{
    Department of Economics, Georgetown University,
 \texttt{alexandre.poirier@georgetown.edu}}
}

\date{February 4, 2026}

\begin{document}
\maketitle
\begin{abstract}
Omitted variables are one of the most important threats to the identification of causal effects. Several widely used methods assess the impact of omitted variables on empirical conclusions by comparing measures of selection on observables with measures of selection on unobservables. The recent literature has discussed various limitations of these existing methods, however. This includes challenges that arise when the omitted variables are endogenous, meaning that they are correlated with the included controls. We develop a new approach to regression sensitivity analysis that avoids those limitations, while still allowing researchers to calibrate sensitivity parameters by comparing the magnitude of selection on observables with the magnitude of selection on unobservables as in previous methods. We illustrate our results in an empirical study of the effect of historical American frontier life on modern cultural beliefs. Finally, we implement these methods in the companion Stata module \texttt{regsensitivity} for easy use in practice.
\end{abstract}



\onehalfspacing
\normalsize

\newpage

\section{Introduction}\label{sec:intro}

For many years, researchers' best approach to dealing with the threat of omitted variables was to either sign the bias, or to use a different identification strategy altogether. These include, for example, measurement error models using proxies, instrumental variable models, or methods for combining datasets where some of the omitted variables are observed. But oftentimes no viable alternative identification strategies are available, and the direction of the bias works against them---meaning that plausible omitted variables could explain away their results. In response to this challenge, the literature developed a variety of results to quantify the impact of omitted variables. A particularly important breakthrough was the paper by \cite{AltonjiElderTaber2005}, and later extended by \cite{Oster2019}, who showed how to measure the impact of omitted variables by comparing the magnitude of selection on observed variables with the magnitude of selection on unobserved variables. The tools developed in these papers are now widely used in empirical economics. 

However, the recent statistics and econometrics literature has discussed several limitations with those early, first generation methods. This includes \cite{DeLucaMagnusPeracchi2019}, \citet[section 6.3]{CinelliHazlett2020}, \cite{Basu2022}, \cite{MastenPoirier2025}, and a companion paper of ours, \cite{DiegertMastenPoirier2025}.\footnote{See section 5 of \cite{MastenPoirier2025} for a detailed review and explanation of all of these limitations.} In particular, in our companion paper \cite{DiegertMastenPoirier2025}, we show that previous methods can lead to incorrect conclusions about robustness when the controls are endogenous, meaning that they are correlated with the omitted variables of concern. For example, consider a classic regression of wages on education and controls like parental education. Typically we are worried that the coefficient on education in this regression is a biased measure of the returns to schooling because unobserved ability is omitted. In this example, the controls would be exogenous if unobserved ability is uncorrelated with parents' education, along with all other included controls. Such exogeneity assumptions on the control variables are usually thought to be very strong and implausible. For example, \cite{AngristPischke2017} discuss how
\begin{quote}
``The modern distinction between causal and control variables on the right-hand side of a regression equation requires more nuanced assumptions than the blanket statement of regressor-error orthogonality that’s emblematic of the traditional econometric presentation of regression.'' (page 129)
\end{quote}
Put differently: We usually do not expect the included control variables to be uncorrelated with the omitted variables; instead we merely hope that the treatment variable is uncorrelated with the omitted variables after adjusting for the included controls. These control variables are therefore usually thought to be endogenous. 

In this paper, we develop a method for assessing sensitivity to omitted variables that allows the controls to be endogenous. Our approach uses a new set of sensitivity parameters that have several important features: 
\begin{enumerate}

    \item They are based a new and simple measure of selection applied to both the observables and unobservables. Similar to \cite{AltonjiElderTaber2005} and \cite{Oster2019}, we take ratios of these measures to compare selection on unobservables with selection on observables in a way that treats those variables symmetrically. Thus researchers can still use our results to measure how strong selection on unobservables must be relative to selection on observables in order to overturn their baseline findings, even if the controls are endogenous. This is the main contribution of our paper.
    
\item They have a natural causal interpretation as ratios of causal effects of covariates on either the treatment or the outcome.
    
\item They allow researchers to impose assumptions on the correlation between the observed controls and the omitted variables. Upper bounds on this correlation can be used to impose exogenous or partially exogenous controls. This allows us to assess the extent to which endogenous controls affect the robustness of one's results. Lower bounds on this correlation can also be imposed to reflect situations where the observed controls are believed to be sufficiently endogenous.

\item They do not suffer from one of the main critiques of previous methods, as proved in our companion paper \cite{DiegertMastenPoirier2025}.
\end{enumerate}
No other approaches in the literature have all of these features.  

\subsubsection*{Overview of Our Results}

In section \ref{sec:model} we describe the baseline model. The parameter of interest is $\beta_\text{long}$, the coefficient on a treatment variable $X$ in a long OLS regression of an outcome variable $Y$ on a constant, treatment $X$, the observed covariates $W_1$, and some unobserved covariates $W_2$. In section \ref{sec:causalModels} we discuss causal models which allow us to interpret this parameter causally, based on three different identification strategies: unconfoundedness, difference-in-differences, and instrumental variables. Since $W_2$ is unobserved, we cannot compute the long regression of $Y$ on $(1,X,W_1,W_2)$ in the data. Instead, we can only compute the medium regression of $Y$ on $(1,X,W_1)$. We begin by considering a baseline model with a ``no selection on unobservables'' assumption, which implies that the coefficients on $X$ in the long and medium regressions are the same. Importantly, this baseline model does \emph{not} assume the controls $W_1$ are exogenous.

In section \ref{sec:MainNewAnalysis} we develop new methods for assessing sensitivity to omitted variables that (a) do not rely on the exogenous controls assumption, (b) satisfy the consistency and monotonicity-in-selection criteria that we develop in our companion paper \cite{DiegertMastenPoirier2025}, and (c) still allow researchers to compare the magnitude of selection on observables with the magnitude of selection on unobservables. Our identification results can be used with either one or two sensitivity parameters. The first sensitivity parameter compares the relative magnitude of the coefficients on the observed and unobserved covariates in a treatment selection equation. This parameter thus measures the magnitude of selection on unobservables by comparing it with the magnitude of selection on observables. The second sensitivity parameter compares the relative magnitude of the coefficients on the observed and unobserved covariates in the outcome equation. In section \ref{sec:endogeneity_restrictions}, we extend our results to allow researchers to also restrict the magnitude of control endogeneity.

We provide two main identification results. Our first result (Theorem \ref{cor:IdsetRyANDcFree}) characterizes the identified set for $\beta_\text{long}$, the coefficient on treatment in the long regression of the outcome on the treatment and the observed and unobserved covariates. This theorem only requires that researchers make an assumption about a single sensitivity parameter---the relative magnitudes of selection on observables and unobservables. We provide a closed form, analytical expression for the identified set, which makes this result easy to use in practice. Using this result, we show how to do breakdown analysis: To find the largest magnitude of selection on unobservables relative to observables needed to overturn a specific baseline finding. This value, called a breakdown point, can be used to measure the robustness of one's baseline results. We provide a simple expression for the breakdown point and recommend that researchers report estimates of it along with their baseline estimates. This estimated breakdown point provides a scalar summary of a study's robustness to selection on unobservables while allowing for arbitrarily endogenous controls.

If researchers are willing to restrict the impact of unobservables on outcomes, then they can obtain results that are more robust to selection on unobservables. In this case, the identified set is more difficult to characterize analytically (see Theorem \ref{thm:mainMultivariateResult} in the appendix). However, our second main identification result (Theorem \ref{cor:BFCalculation_rx_and_ry}) shows that we can nonetheless easily numerically compute breakdown points. We generalize this result in section \ref{sec:endogeneity_restrictions} to also allow restrictions on the endogeneity of the controls.

In section \ref{sec:empirical} we show how to use our results in empirical practice. We use data from \citet[\emph{Econometrica}]{BFG2020} who studied the effect of historical American frontier life on modern cultural beliefs. Specifically, they test a well known conjecture that living on the American frontier cultivated individualism and antipathy to government intervention. Among other analyses, they use Oster's (2019) method to argue that omitted variables do not affect their conclusions. Using our results, we obtain richer conclusions about robustness. In particular, when allowing for endogenous controls, we find that effects obtained from questionnaire based outcomes are often sensitive to omitted variables while the effects from election and property tax outcomes remain robust.

\subsubsection*{Related Literature}

We conclude this section with a brief review of the literature. We focus on two literatures: The literature on endogenous controls and the literature on sensitivity analysis in linear or parametric models.

The idea that the treatment variable and the control variables should be treated asymmetrically in the assumptions goes back to at least \cite{BarnowCainGoldberger1980}. They developed the ``linear control function estimator'', which is based on an early parametric version of the now standard unconfoundedness assumption. \citet[page 190]{HeckmanRobb1985}, \cite{HeckmanHotz1989}, and \citet[page 5035]{HeckmanVytlacil2007part2} all provide detailed discussions of this estimator. It was also one of the main estimators used in \cite{LaLonde1986}. \cite{StockWatson2011} provide a textbook description of it on pages 230--233 and pages 250--251. \cite{AngristPischke2009} also discuss it at the end of their section 3.2.1. Also see \cite{Frolich2008}. Note that this earlier analysis was based on mean independence assumptions, while the analysis in our paper only uses linear projections. More recently, \cite{HuenermundLouw2020} remind researchers that most control variables are likely endogenous and hence their coefficients should not be interpreted as causal. 

Although control variables are often thought to be endogenous, the literature on sensitivity analysis generally either assumes the controls are exogenous or uses a method called ``residualization.'' This includes \cite*{AltonjiElderTaber2005, AltonjiElderTaber2008}, \cite{Krauth2016}, and \cite{Oster2019}. \cite{Imbens2003} starts from the standard unconfoundedness assumption which allows endogenous controls, but in his parametric assumptions (see his likelihood equation on page 128) he assumes that the unobserved omitted variable is independent of the observed covariates. \cite{AET2019} propose an approach to allow for endogenous controls based on imposing a factor model on all covariates, observable and unobservable. Their approach and ours impose different assumptions and therefore are not nested. \cite{CinelliHazlett2020} develop a regression sensitivity analysis based on partial R-squareds. Their results do not rely on endogenous controls, but they do not define any measures of selection and do not work with symmetric ratio parameters like \cite{AltonjiElderTaber2005}, \cite{Oster2019}, or us, and hence their results do not allow researchers to compare selection on observables with unobservables as in \cite{AltonjiElderTaber2005}, \cite{Oster2019}, and our results.

Our companion paper \cite{DiegertMastenPoirier2025} uses an axiomatic framework to formally prove that residualization, the literature's most popular approach to allowing for endogenous controls, can lead to incorrect conclusions about robustness.  \cite{DeLucaMagnusPeracchi2019}, \cite{CinelliHazlett2020}, \cite{Basu2022}, and \cite{MastenPoirier2025} discuss a variety of additional limitations with the existing approaches to assessing sensitivity to omitted variables in linear models. The contribution of the present paper is not to survey or analyze properties of previously proposed methods, but rather to derive new identification results based on a new set of sensitivity parameters. These identification results can then be used to perform a sensitivity analysis to assess the impact of omitted variables on one's regression results. In contrast, our companion paper \cite{DiegertMastenPoirier2025} does not derive any identification results or propose any new sensitivity analyses. That paper instead focuses on comparing different definitions of sensitivity parameters, including the ones we propose in the present paper as well as a variety of parameters proposed elsewhere in the literature.

Thus far we have discussed the linear regression based literature. Nonparametric methods, like those based on the standard unconfoundedness assumption (for example, chapter 12 in \citealt{ImbensRubin2015}), often allow for endogenous controls. For this reason, several papers that assess sensitivity to unconfoundedness also allow for endogenous controls. This includes \cite{Rosenbaum1995, Rosenbaum2002}, \cite{Tan2006}, and \cite{MastenPoirier2018}, among others. These methods, however, do not provide formal results for calibrating their sensitivity parameters based on comparing selection on observables with selection on unobservables. These methods also focus on binary or discrete treatments, whereas the analysis in our paper can be used for continuous treatments as well.

\subsubsection*{Notation Remarks}

For random vectors $A$ and $B$, let $\cov(A,B)$ be the $\text{dim}(A) \times \text{dim}(B)$ matrix whose $(i,j)$th element is $\cov(A_i, B_j)$. Define $A^{\perp B} \coloneqq A - \cov(A,B)\var(B)^{-1}B$, the intercept plus the residual from a regression of $A$ on $(1,B)$. For a scalar random variable $A$, let $R_{A \sim B \sbullet C}^2$ denote the R-squared from a regression of $A^{\perp C}$ on $(1,B^{\perp C})$; this is called the partial R-squared (of $A$ on $B$, partialling out $C$).

\section{The Baseline Model}\label{sec:model}

Let $Y$ and $X$ be observed scalar variables. Let $\Wob$ be a vector of observed covariates of dimension $\dob$ and $\Wu$ be an unobserved vector of dimension $d_2$. Let $W \coloneqq (\Wob,\Wu)$. Consider the population linear regression of $Y$ on $X$, $W_1$, $W_2$, and a constant. Let $\beta_\text{long}$ denote the coefficient on $X$, while $(\gamma_1,\gamma_2)$ denote the coefficients on $(W_1,W_2)$. The setup here is similar to our companion paper \cite{DiegertMastenPoirier2025}. We make the following assumption to ensure these linear regression coefficients are well-defined.

\begin{Aassumption}\label{assump:posdefVar}
The variance matrix of $(Y,X,\Wob,\Wu)$ is finite and positive definite.
\end{Aassumption}

We can write
\begin{equation}\label{eq:outcome}
	Y = \beta_\text{long} X + \gamma_1' \Wob + \gamma_2' \Wu + Y^{\perp X,W}.
\end{equation}
The term $Y^{\perp X,W}$ denotes the linear projection residual plus the intercept. By construction, it is uncorrelated with $(X,W)$. We take as given that $\beta_\text{long}$ is the parameter of interest. This interest may come from an underlying causal structure, where $\beta_\text{long}$ has an interpretation as an average of treatment effects. For example, section \ref{sec:causalModels} below discusses how unconfoundedness, difference-in-differences, or instrumental variables as identification strategies may yield a causal interpretation for $\beta_\text{long}$. These interpretations may or may not require associated potential outcomes to be linear in $X$.  Ultimately, the results we present for $\beta_\text{long}$ do not depend on the reason for the interest in this parameter, and we present our analysis without assuming any particular causal structure.

Next consider the ``first-stage'' linear regression of $X$ on $W_1$, $W_2$, and a constant. Let $(\pi_1,\pi_2)$ denote the coefficients on $(\Wob,\Wu)$. We can write
\begin{equation}\label{eq:XprojectionW1W2}
	X = \pi_1' \Wob + \pi_2' \Wu + X^{\perp W}.
\end{equation}
The term $X^{\perp W}$ denotes the linear projection residual plus the intercept. By construction it is uncorrelated with each covariate in $(W_1,W_2)$. Although equation \eqref{eq:XprojectionW1W2} is not necessarily causal, we can think of $\pi_1$ as representing ``selection on observables'' while $\pi_2$ represents ``selection on unobservables.'' The following is thus a natural baseline assumption.

\begin{Aassumption}[No selection on unobservables]\label{assump:baselineEndogControl1}
$\pi_2 = 0$.
\end{Aassumption}

Let $\beta_\text{med}$ denote the coefficient on $X$ in the linear regression of $Y$ on $(1,X,\Wob)$. With no selection on unobservables, we have the following result.

\begin{proposition}\label{thm:baselineEndogControl1}
Suppose the joint distribution of $(Y,X,\Wob)$ is known. Suppose \ref{assump:posdefVar} and \ref{assump:baselineEndogControl1} hold. Then the following hold.
\begin{enumerate}
\item $\beta_\text{long} = \beta_\text{med}$. Consequently, $\beta_\text{long}$ is point identified.

\item The identified set for $\gamma_1$ is $\R^{\dob}$.
\end{enumerate}
\end{proposition}

This result shows why Assumption \ref{assump:baselineEndogControl1} can also be interpreted as a ``selection on observables'' assumption: adjusting linearly for the observable $W_1$ is sufficient to recover $\beta_\text{long}$. Note that this result allows for endogenous controls, in the sense that the observed covariates $\Wob$ can be arbitrarily correlated with the unobserved covariates $\Wu$. But it restricts the relationship between $(X,\Wob,\Wu)$ in such a way that we can still point identify $\beta_\text{long}$ even though $\Wob$ and $\Wu$ are arbitrarily correlated. The coefficient $\gamma_1$ on the observed controls, however, is completely unidentified. Note that Proposition \ref{thm:baselineEndogControl1} is unsurprising and is not the main contribution of this paper; we include it for completeness because we have not seen it formally stated in any previous references, especially the second part that $\gamma_1$ is completely unidentified.

In practice, we are often worried that the no selection on unobservables assumption does not hold. For example, consider the regression of wages on education discussed in section \ref{sec:intro}. Even with a rich set of observed covariates $W_1$, like family background variables, researchers typically suspect that there may still be omitted variables that affect both wages and education, and which could be correlated with observed variables like family background. We therefore develop a new approach to assess the importance of this no selection on unobservables assumption in the next section.

\section{Sensitivity Analysis}\label{sec:MainNewAnalysis}

There are many ways to relax Assumption \ref{assump:baselineEndogControl1} ($\pi_2 = 0$) and allow for some selection on unobservables. A key challenge is to construct measures of selection on unobservables and observables that are interpretable and comparable, and therefore allow researchers to reason about their relative magnitudes. In this section we propose a new measure of selection, argue for its interpretability, and then show how to use it to perform a sensitivity analysis to the selection on observables assumption.

\subsection{Measuring Selection}\label{subsec:selectionratio}

We begin by focusing on the relationship between \emph{treatment} and the covariates, observed and unobserved. Later, in section \ref{sec:identrYtoo}, we will extend our analysis to also consider the relationship between \emph{outcomes} and the covariates. Several previous approaches in the literature use a single sensitivity parameter to measure the combined impact of omitted variables on outcomes and treatment; this leads to a variety of issues and limitations, however, as discussed in the papers cited in section \ref{sec:intro}, and hence we instead consider these two relationships separately. To that end, recall the regression of treatment on a constant and all the covariates:
\[
	X = \pi_1' \Wob + \pi_2' \Wu + X^{\perp W}.
	\tag{\ref{eq:XprojectionW1W2}}
\]
We measure selection on unobservables by $\sqrt{\var(\pi_2'\Wu)}$, the standard deviation of the index of unobservables in this regression. We measure selection on observables by $\sqrt{\var(\pi_1'\Wob)}$, the standard deviation of the index of observables in this regression. We compare them by taking the ratio
\begin{align*}
	r_X \coloneqq \frac{\sqrt{\var(\pi_2'\Wu)}}{\sqrt{\var(\pi_1'\Wob)}}.
\end{align*}
This is a \emph{relative} measure that measures the impact of unobservables as a multiple of the impact of observables. Consequently, in models where the included covariates $W_1$ are very important for treatment selection even a small value of $r_X$ might be considered to be a large impact of the unobservables; we discuss this further in section \ref{sec:interpretation}.

Both the numerator and denominator of $r_X$ are invariant to invertible linear transformations of $\Wob$ or $W_2$, including rescalings, since the indices $\pi_1'W_1$ and $\pi_2'W_2$ are both invariant with respect to invertible linear transformations. This invariance ensures that $r_X$ is a unit-free measure of the relative magnitude of selection on unobservables to selection on observables. $r_X = 0$ corresponds to baseline assumption \ref{assump:baselineEndogControl1} ($\pi_2 = 0$) while $r_X > 0$ allows for some selection on unobservables.

This measure has two additional interpretations. First, $r_X$ has an R-squared interpretation. Specifically, for scalar $W_1$ and $W_2$ we have
\begin{equation}\label{eq:rx_interpretation_scalarW_1W_2}
	r_X^2 
	= \frac{R^2_{X \sim W_1,W_2} - R^2_{X \sim W_1}}{R^2_{X \sim W_1,W_2} - R^2_{X \sim W_2}}.
\end{equation}
The numerator is the difference in the R-squared from the regression in \eqref{eq:XprojectionW1W2} and the short R-squared that only includes $W_1$. The denominator is the same but switching the roles of $W_1$ and $W_2$. For vector covariates,
\begin{equation}\label{eq:rx_interpretation_vectorW_1W_2}
	r_X^2 
	= \frac{R^2_{X \sim W_1,W_2} - R^2_{X \sim \pi_1'W_1}}{R^2_{X \sim W_1,W_2} - R^2_{X \sim \pi_2'W_2}}.
\end{equation}
Second, $r_X$ can have a causal interpretation. Suppose $W_1$ and $W_2$ are scalars. Let $X(w_1,w_2)$ denote potential treatments. Assume this function is linear in $w_1$ and $w_2$:
\[
	X(w_1,w_2) = \pi_1 w_1 + \pi_2 w_2 + V
\]
for some structural unobservable $V$. Assume $\cov(W,V) = \cov(W, X(w_1,w_2)) = 0$. Then $ \sqrt{\var(\pi_2 W_2)}$ is the causal effect of a one standard deviation increase in the unobserved variable on $X$. Likewise, $\sqrt{\var(\pi_1 W_1)}$ is the causal effect of a one standard deviation increase in the observed variable on $X$. And the measure $r_X$ is the magnitude of the ratio of these two causal effects.

Having defined and interpreted our measure of selection on unobservables, we next move on to study identification. In section \ref{subsec:identrXonly} we state our main identification results based on $r_X$ only. In section \ref{sec:identrYtoo} we extend these results to allow researchers to also restrict the impact of omitted variables on the outcome variable. In section \ref{sec:interpretation} we make several remarks regarding interpretation and calibration of the sensitivity parameters. In section \ref{sec:endogeneity_restrictions}, we extend these results to allow researchers to also restrict the magnitude of control endogeneity. Finally, in section \ref{sec:estimationInference} we briefly discuss estimation and inference. 

\subsection{Identification Using Selection Equation Assumptions Only}\label{subsec:identrXonly}

Recall from section \ref{sec:model} that our parameter of interest is $\beta_\text{long}$, the OLS coefficient on $X$ in the long regression of $Y$ on $(1,X,\Wob,\Wu)$. Since $W_2$ is not observed, we cannot compute this regression from the data. However, we can compute $\beta_\text{med}$, the OLS coefficient on $X$ in the medium regression of $(1,X,\Wob)$. The difference between these two regression coefficients is given by the classic omitted variable bias formula. To bound this difference, we restrict the selection ratio $r_X$ described in section \ref{subsec:selectionratio}. Specifically, consider the following assumption.

\begin{Aassumption}\label{assump:rx}
$r_X \leq \rx$ for a known value of $\rx \geq 0$.
\end{Aassumption}

Following our discussion in section \ref{subsec:selectionratio}, this assumption is invariant to any invertible linear transformations of $W_1$ or $W_2$, and thus $\rx$ is a unit-free sensitivity parameter. Assumption \ref{assump:rx} can also be written as
\[
	\sqrt{ \var(\pi_2' W_2) } \leq \rx \cdot \sqrt{ \var(\pi_1' W_1) },
\]
which means that the association between treatment $X$ and a one standard deviation increase in the index of unobservables is at most $\rx$ times the association between treatment and a one standard deviation increase in the index of observables. The baseline model of section \ref{sec:model} corresponds to the case $\rx = 0$, since it implies $\pi_2 = 0$. We relax the baseline model by considering values $\rx > 0$.

Let $\mathcal{B}_I(\rx)$ denote the identified set for $\beta_\text{long}$ under the positive definite variance assumption \ref{assump:posdefVar} and the restriction \ref{assump:rx}. In particular, this identified set does not impose any restrictions on the impact of omitted variables on the outcome variable. Let
\[
	\underline{B}(\rx) \coloneqq \inf \mathcal{B}_I(\rx)
	\qquad \text{and} \qquad
	\overline{B}(\rx) \coloneqq \sup \mathcal{B}_I(\rx)
\]
denote its greatest lower bound and least upper bound. Our first main result, Theorem \ref{cor:IdsetRyANDcFree} below, provides simple, closed form expressions for these sharp bounds. For this theorem, we also use the following assumption.

\begin{Aassumption}\label{assump:noKnifeEdge}
$\cov(W_1,Y) \neq \cov(W_1,X) \cov(X,Y) / \var(X)$ and $\cov(W_1,X) \neq 0$.
\end{Aassumption}

This assumption is not necessary, but it simplifies the proofs. Moreover, this assumption usually holds in practice since it holds whenever $\beta_\text{short} \neq \beta_{\text{med}}$, where $\beta_\text{short}$ is the coefficient on $X$ in the short OLS regression of $Y$ on $(1,X)$.

\begin{theorem}\label{cor:IdsetRyANDcFree}
Suppose the joint distribution of $(Y,X,\Wob)$ is known. Suppose \ref{assump:posdefVar}, \ref{assump:rx}, and \ref{assump:noKnifeEdge} hold. If $\rx <  \sqrt{1 - R^2_{X \sim W_1}}$, then
\begin{align*}
	&[\underline{B}(\rx),\overline{B}(\rx)] \\[0.5em]
	&= \left[\beta_{\text{med}} - \sqrt{\frac{\var(Y^{\perp X,W_1})}{\var(X^{\perp W_1})} \cdot \frac{\rx^2 R^2_{X \sim W_1} }{1 - R^2_{X \sim W_1} - \rx^2 }}, 
	\; \beta_{\text{med}} + \sqrt{\frac{\var(Y^{\perp X,W_1})}{\var(X^{\perp W_1})} \cdot \frac{\rx^2 R^2_{X \sim W_1} }{1 - R^2_{X \sim W_1} - \rx^2 }} \; \right].
\end{align*}
Otherwise, $\underline{B}(\rx) = -\infty$ and $\overline{B}(\rx) = +\infty$.
\end{theorem}

Appendix \ref{sec:mainResultProofIntuition} gives a sketch derivation of and intuition for this result, while the complete formal derivation and sharpness proof is in Appendix \ref{sec:proofsForMainNewAnalysis}. Theorem \ref{cor:IdsetRyANDcFree} characterizes the largest and smallest possible values of $\beta_\text{long}$ when some selection on unobservables is allowed, the observed covariates are allowed to be arbitrarily correlated with the unobserved covariate, and we make no restrictions on the coefficients in the outcome equation. In fact, we prove that, with the exception of at most three singletons, the interval $[\underline{B}(\rx), \overline{B}(\rx)]$ is the identified set for $\beta_\text{long}$ under these assumptions. Here we focus on the smallest and largest elements to avoid technical digressions that are unimportant for applications.

There are two important features of Theorem \ref{cor:IdsetRyANDcFree}: First, it only requires researchers to reason about \emph{one} sensitivity parameter, unlike some existing approaches (our generalizations in sections \ref{sec:identrYtoo} and \ref{sec:endogeneity_restrictions} use additional sensitivity parameters to allow researchers to impose further restrictions). Second, it allows for arbitrarily endogenous controls since it imposes no restrictions on the association between $W_1$ and $W_2$. So this result allows researchers to examine the impact of selection on unobservables on their baseline results without also having to reason about the magnitude of endogenous controls.

Since Theorem \ref{cor:IdsetRyANDcFree} provides explicit expressions for the bounds, we can immediately derive a few of their properties. First, when $\rx = 0$, the bounds collapse to $\beta_\text{med}$, the point estimand from the baseline model with no selection on unobservables. Thus, we recover the baseline model as a special case. For small values of $\rx > 0$, the bounds are no longer a singleton, but their width increases continuously as $\rx$ increases away from zero. The rate at which the bounds increase depends on just a few features of the data: All can be derived from the variance matrix of $(Y,X,W_1)$. We also see that the bounds are symmetric around $\beta_\text{med}$. Finally, the bounds are finite only if $\rx < \sqrt{1 - R^2_{X \sim W_1}}$. 

\subsubsection*{Breakdown Analysis}

In practice, researchers often ask:
\begin{quote}
How strong does selection on unobservables have to be relative to selection on observables in order to overturn our baseline findings?
\end{quote}
We can use Theorem \ref{cor:IdsetRyANDcFree} to answer this question. Suppose in the baseline model we find $\beta_\text{med} \geq 0$. We are concerned, however, that $\beta_\text{long} \leq 0$, in which case our positive finding is driven solely by selection on unobservables. Define
\[
	\bar{r}_X^\text{bp} 
	\coloneqq \sup \{ \bar{r}_X \geq 0 : b \geq 0 \text{ for all } b \in \mathcal{B}_I(\bar{r}_X) \}.
\]
This value is called a \emph{breakdown point}. It is the largest amount of selection on unobservables we can allow for while still concluding that $\beta_\text{long}$ is nonnegative. The breakdown point when $\beta_\text{med} \leq 0$ can be defined analogously.

\begin{corollary}\label{corr:breakdownPointRXonly}
Suppose the assumptions of Theorem \ref{cor:IdsetRyANDcFree} hold. Then
\[
	\bar{r}_X^\text{bp} = \left(
	\frac{
	R_{Y \sim X \sbullet W_1}^2
	}{
	\frac{R_{X \sim W_1}^2}{1 - R_{X \sim W_1}^2}  + R_{Y \sim X \sbullet W_1}^2
	}
	\right)^{1/2}.
\]
\end{corollary}

The breakdown point described in Corollary \ref{corr:breakdownPointRXonly} characterizes the magnitude of selection on unobservables relative to selection on observables needed to overturn one's baseline findings. Corollary \ref{corr:breakdownPointRXonly} explicitly shows that this breakdown point depends on just two features of the observed data: The relationship between treatment and the outcome, after adjusting for the observed covariates, and the first stage relationship between treatment and the observed covariates. In particular, as the covariate adjusted relationship between outcomes and treatment strengthens, the breakdown point increases too. In contrast, as the relationship between treatment and covariates strengthens, the breakdown point decreases. This follows since we are using effects of the observed covariates to calibrate the magnitude of selection on unobservables. So when these observed covariates are strongly related to treatment, we need relatively less selection on unobservables to overturn our baseline findings. We further discuss this point in section \ref{sec:interpretation}.

\subsubsection*{Practical Recommendations for Applied Researchers}

First, we recommend that researchers present estimates of the breakdown point $\bar{r}_X^\text{bp}$ as a scalar measure of the robustness of their results. This can be displayed alongside the main result point estimates, as in Table \ref{table:mainTable1} in our empirical application. Second, we recommend that researchers present estimates of the bounds described in Theorem \ref{cor:IdsetRyANDcFree} as a function of $\rx$. Researchers may also find it useful to present estimates of the calibration parameters $\rho_k$ that we describe in section \ref{sec:interpretation}. We illustrate all of these recommendations  in our empirical application in section \ref{sec:empirical}.

\subsection{Adding Assumptions on the Outcome Equation}\label{sec:identrYtoo}

In some empirical settings researchers may be willing to restrict the impact of unobservables on outcomes. To do this, by analogy to $r_X$, define
\[
	r_Y \coloneqq \frac{\sqrt{\var(\gamma_2'W_2)}}{\sqrt{\var(\gamma_1'W_1)}}.
\]
This is the ratio of the standard deviation of the index of unobservables in the \emph{outcome} equation \eqref{eq:outcome} to the standard deviation of the index of observables in the same equation. Thus $r_Y$ measures the relative impact of unobservables and observables on outcomes, whereas $r_X$ measures their relative impact on treatment. Note that, like the discussion of $r_X$ in section \ref{subsec:selectionratio}, $r_Y$ can also be interpreted as a ratio of causal effects under a linearity assumption on potential outcomes. We now consider a bound on this ratio $r_Y$.

\begin{Aassumption}\label{assump:ry}
$r_Y \leq \ry$ for a known value of $\ry \geq 0$.
\end{Aassumption}

Assumption \ref{assump:ry} has a similar interpretation as \ref{assump:rx}: It says that the association between the outcome and a one standard deviation increase in the index of unobservables is at most $\bar{r}_Y$ times the association between the outcome and a one standard deviation increase in the index of observables. Like $r_X$, the parameter $r_Y$ is invariant to invertible linear transformations of $W_1$ and $W_2$, making $\bar{r}_Y$ a unit-free sensitivity parameter.

In our next result, which uses restrictions on both $r_X$ and $r_Y$, we assume that the omitted variable of interest $W_2$ is a scalar, and then normalize its variance to 1. We leave the extension to vector $W_2$ to future work. Let $\mathcal{B}_I(\rx,\ry)$ denote the identified set for $\beta_\text{long}$ under this scalar $W_2$ assumption, \ref{assump:posdefVar}, \ref{assump:rx}, and \ref{assump:ry}. Unlike the identified set of Theorem \ref{cor:IdsetRyANDcFree}, this set is less analytically tractable. We characterize it in Appendix \ref{sec:generalIdentSet}. In this section we show that it is nonetheless very easy to numerically implement a breakdown analysis, which is often researchers' main focus.
 
Suppose we are interested in the robustness of the conclusion that $\beta_\text{long} \geq \underline{b}$ for some known scalar $\underline{b}$. For example, $\underline{b} = 0$. Define the function
\[
	\bar{r}_Y^\text{bf}(\rx, \underline{b}) 
	\coloneqq \sup \{ \bar{r}_Y \geq 0 : b \geq \underline{b} \text{ for all } b \in \mathcal{B}_I(\rx, \ry) \}.
\]
This function is a \emph{breakdown frontier} (\citealt{MastenPoirier2019BF}), which captures trade-offs in the relative impacts of $\rx$ and $\ry$ on the identified set. In particular, we can use it to define the set
\[
	\text{RR} \coloneqq \{ (\rx, \ry) \in \R_{\geq 0}^2 : \ry \leq \ry^\text{bf}(\rx,\underline{b}) \}.
\]
This set is called the \emph{robust region} because the conclusion of interest, $\beta_\text{long} \geq \underline{b}$, holds for any combination of sensitivity parameters in this region. The size of this region is therefore a measure of the robustness of our baseline conclusion.

Although we do not have a closed form expression for the smallest and largest elements of $\mathcal{B}_I(\rx, \ry)$, our next main result shows that we can still easily compute the breakdown frontier numerically. To state the result, we first define some additional notation. Let $\| \cdot \|$ denote the Euclidean norm. Define the set
\begin{align*}
	\mathcal{D}^0 &\coloneqq \left\{(z,c,b) \in \mathbb{R} \times \mathbb{R}^{d_1} \times \mathbb{R}: z\sqrt{1 - \|c\|^2}\frac{\cov(W_1,Y-bX)}{\var(X^{\perp W_1})} - (\beta_\text{med} - b)c \ne 0, \|c\|<1 \right\}
\end{align*} 
and the functions 
\begin{align*}
  \underline{r}_Y(z,c,b) &\coloneqq \begin{cases}
    0 & \text{ if } b = \beta_{\text{med}} \\    
    \dfrac{|\beta_\text{med} - b|}{\|z\sqrt{1 - \|c\|^2}\frac{\cov(W_1,Y-bX)}{\var(X^{\perp W_1})} - (\beta_\text{med} - b)c\|} & \text{ if } (z,c,b) \in \mathcal{D}^0 \text{ and } b \neq \beta_{\text{med}} \\ 
    +\infty & \text{ otherwise} 
    \end{cases} \\
    p(z,c;\rx) &\coloneqq \rx^2\|\cov(W_1,X)\sqrt{1 - \|c\|^2} - cz\|^2 - z^2.
\end{align*}
For simplicity, we also normalize the treatment variable so that $\var(X) = 1$ and the covariates so that $\var(W_1) = I$. All of the results below can be rewritten without these normalizations at the cost of additional notation. We can now state our second main result. 

\begin{theorem}\label{cor:BFCalculation_rx_and_ry}
Suppose the joint distribution of $(Y,X,\Wob)$ is known. Suppose \ref{assump:posdefVar} and \ref{assump:noKnifeEdge} hold. Normalize $\var(X) = 1$ and $\var(W_1) = I$. Assume $W_2$ is a scalar and normalize $\var(W_2) = 1$. Suppose $\cov(W_1,Y)$ and $\cov(W_1,X)$ are linearly independent. Suppose $d_1 \geq 2$. Let $\underline{b} \in \R$ and $\rx, \ry \geq 0$.
\begin{enumerate}
\item If $\underline{b} \geq \beta_{\text{med}}$ then $\ry^\text{bf}(\bar{r}_{X}, \underline{b}) = 0$.

\item If $\underline{B}(\rx) > \underline{b}$, then $\ry^\text{bf}(\bar{r}_{X}, \underline{b}) = +\infty$.
    
\item If $\underline{B}(\rx) \leq \underline{b} < \beta_\text{med}$, then
\begin{align*}
      \bar{r}^\text{bf}_Y(\rx, \underline{b})
      = \min_{(z,c_1,c_2,b) \in (-\sqrt{\var(X^{\perp W_1})},\sqrt{\var(X^{\perp W_1})}) \times \R \times \R \times (-\infty, \underline{b}]} \ \ \underline{r}_Y(z,\cov(W_1,c_1 Y + c_2 X),b)& \\
	\text{subject to } \ p(z, \cov(W_1,c_1 Y + c_2 X); \rx) \geq 0& \\
	(b - \beta_{\text{med}})^2 < \frac{\var(Y^{\perp X, W_1})}{\var(X^{\perp W_1})} \frac{z^2}{\var(X^{\perp W_1}) - z^2}& \\
	\|\cov(W_1,c_1 Y + c_2 X)\| < 1.&
\end{align*}
\end{enumerate}
\end{theorem}

Theorem \ref{cor:BFCalculation_rx_and_ry} shows that the breakdown frontier can be computed as the solution to a smooth optimization problem. Importantly, this problem only requires searching over a 4-dimensional space. In particular, this dimension does not depend on the dimension of the covariates $W_1$. Consequently, it remains computationally feasible even with a large number of observed covariates, as is often the case in empirical practice. For example, the results for our empirical application take about 15 seconds to compute on a standard machine.

Theorem \ref{cor:BFCalculation_rx_and_ry} makes several minor technical assumptions. In particular, it assumes $\cov(W_1, Y)$ and $\cov(W_1, X)$ are linearly independent for simplicity. Moreover, the theorem assumes $d_1 \geq 2$, requiring at least two observed covariates in $W_1$. This is not restrictive since the purpose of this result is primarily to show that the optimization problem does not depend on the dimension of $W_1$. If $d_1 = 1$ then the breakdown frontier can instead be easily computed using equation \eqref{eq:Theorem4simpleCase} in the appendix, which only requires searching over a 3-dimensional space.

\subsubsection*{Practical Recommendations for Applied Researchers}

To assess the impact of different choices of $(\bar{r}_X, \bar{r}_Y)$ on the robustness of their results, researchers can plot sample analog estimates of the function $\bar{r}_Y^\text{bf}(\cdot, \underline{b})$ (for example, see the solid line in the right plots of figures \ref{fig:republican} and \ref{fig:cutpoor} in section \ref{sec:empirical} below). The conclusion of interest $\beta_\text{long} \geq \underline{b}$ holds for all pairs $(\bar{r}_X, \bar{r}_Y)$ below this function. In practice it is also useful to report scalar measures of robustness to omitted variables. One way to do this while also incorporating restrictions on the impact of the omitted variables on outcomes is to set $\bar{r}_X = \bar{r}_Y$. This assumption says that the maximal impact of unobservables relative to observables is the same for outcomes as it is for treatment.  Under this common maximal impact assumption, the breakdown point for the common value of the sensitivity parameter is
\[
	\bar{r}^\text{bp}(\underline{b}) \coloneqq \sup\{r \geq 0: b \geq \underline{b} \text{ for all } b \in \mathcal{B}_I(r,r)\}.
\]
Let $\bar{r}^\text{bp} = \bar{r}^\text{bp}(0)$. Our second main recommendation is that researchers report estimates of this breakdown point for some value $\underline{b}$ of interest. $\bar{r}^\text{bp}$ can be viewed as a less conservative version of $\bar{r}_X^\text{bp}$, the breakdown point that only imposes restrictions on how treatment is related to the omitted variables. In particular, note that $\bar{r}_X^\text{bp} \leq \bar{r}^\text{bp}$ always holds. The common maximal impact breakdown point $\bar{r}^\text{bp}(\underline{b})$ can be directly computed from the function $\bar{r}^\text{bf}_Y(\rx, \underline{b})$ by solving $\bar{r}^\text{bf}_Y(\rx, \underline{b}) = \rx$ numerically.

Finally, note that sharp bounds on the identified set $\mathcal{B}_I(\rx,\ry)$ can be easily computed by using a similar duality relationship between the breakdown frontier and the identified set as described in Proposition 1 of \cite{MastenPoirier2023}.

\subsection{Interpreting and Calibrating the Sensitivity Parameters}\label{sec:interpretation}

Having stated our main identification results, we next make several remarks regarding how to interpret the magnitudes of these parameters.

\subsubsection*{Which Covariates to Calibrate Against?}

The sensitivity parameters $\rx$ and $\ry$ are \textit{relative} sensitivity parameters, in the sense that they compare the importance of the unobservables $W_2$ relative to the importance of the observed covariates $W_1$. Thus far we have assumed that $W_1$ includes \emph{all} observed covariates. Here we extend the analysis to allow researchers to select which covariates to use in the sensitivity parameters. Hence the interpretation of the sensitivity parameters depends on the choice of covariates that we calibrate against. Put differently, when we say that we compare ``selection on unobservables to selection on observables,'' \emph{which} observables do we mean?

To answer this, we split the observed covariates into two groups: (1) The control covariates, which we label $W_0$, and (2) The calibration covariates, which we continue to label $W_1$. Write equation \eqref{eq:outcome} as
\[
	Y = \beta_\text{long} X + \gamma_0' W_0 + \gamma_1' \Wob + \gamma_2' \Wu + Y^{\perp X,W}
	\tag{\ref{eq:outcome}$^\prime$}
\]
where $W \coloneqq (W_0,W_1,W_2)$. Likewise, write equation \eqref{eq:XprojectionW1W2} as
\[
	X = \pi_0' W_0 + \pi_1' \Wob + \pi_2' \Wu + X^{\perp W}.
	\tag{\ref{eq:XprojectionW1W2}$^\prime$}
\]
The key difference from our earlier analysis is that, like in Assumption \ref{assump:rx}, we will continue to only compare $\pi_1$ with $\pi_2$. That is, we only compare the omitted variable to the observed variables in $W_1$; we do not use $W_0$ for calibration. A similar remark applies to \ref{assump:ry}.

This distinction between control and calibration covariates is useful because in many applications we do not necessarily think the omitted variables have similar explanatory power as \emph{all} of the observed covariates included in the model. For example, in our empirical application in section \ref{sec:empirical}, we include state fixed effects as control covariates, but we do not use them for calibration.

We next briefly describe how to generalize our identification results to account for this distinction. By the FWL theorem, a linear projection of $Y^{\perp W_0}$ onto $(1,X^{\perp W_0}, W_1^{\perp W_0}, W_2^{\perp W_0})$ has the same coefficients as equation \eqref{eq:outcome}. Likewise for a linear projection of $X^{\perp W_0}$ onto $(1,W_1^{\perp W_0}, W_2^{\perp W_0})$. Hence we can write
\begin{align*}
	Y^{\perp W_0} &= \beta_\text{long} X^{\perp W_0} + \gamma_1'W_1^{\perp W_0} + \gamma_2' \Wu^{\perp W_0} + Y^{\perp X,W}\\
	X^{\perp W_0} &= \pi_1' W_1^{\perp W_0} + \pi_2' \Wu^{\perp W_0} + X^{\perp W}.
\end{align*}
Therefore our earlier results continue to hold when $(Y,X,\Wob,\Wu)$ are replaced with ($Y^{\perp W_0}$, $X^{\perp W_0}$, $W_1^{\perp W_0}$, $\Wu^{\perp W_0}$). 

\subsubsection*{Internal Calibration of the Magnitude of $\bar{r}_X$}

Here we sketch a data-driven approach to calibrating the magnitude of the sensitivity parameter $\bar{r}_X$, and therefore for assessing the robustness of one's results to omitted variables. By analogy to $r_X = \sqrt{ \var(\pi_2'W_2)} / \sqrt{ \var(\pi_1'W_1)}$, define
\[
	\rho_k	
	\coloneqq \frac{ \sqrt{\var(\pi_{1,\text{med},k} W_{1k})}
	}{
	\sqrt{ \var(\pi_{1,\text{med},-k}'W_{1,-k} )
	}
	}
\]
for $k=1,\ldots,d_1$, where $\pi_{1,\text{med}} = (\pi_{1,\text{med},k}, \pi_{1,\text{med},-k})$ is the coefficient on the observed covariates $W_1 = (W_{1k}, W_{1,-k})$ from OLS of treatment $X$ on $(1,W_1)$. $\rho_k$ measures the importance of variable $k$ relative to all other observed covariates. Moreover, unlike $r_X$, the $\rho_k$ values are point identified from the data. We suggest that researchers compare their estimates of the breakdown point $\bar{r}_X^\text{bp}$ against estimates of the values $\rho_1,\ldots,\rho_{d_1}$. We illustrate this approach in our empirical analysis in section \ref{sec:empirical}. Similar approaches to empirically calibrating sensitivity parameters were previously proposed in \cite{HosmanHansenHolland2010}, \cite{HsuSmall2013}, \cite{MastenPoirier2018}, and \cite{CinelliHazlett2020}. Importantly, as  \cite{HosmanHansenHolland2010} discuss in detail, values like $\rho_k$ are ``reference points for speculation'' about the true, but unknown \emph{and completely unidentified} value of $r_X$. Put differently: There is no logical guarantee that $r_X$ will be smaller than any of the $\rho_k$'s. Instead, their purpose is simply to help us use data to reason about plausible values of the true, but unknown sensitivity parameter $r_X$. Finally, here we defined $\rho_k$ by comparing a single covariate against all others. More generally one could compare the impact of a group of covariates against all others; we omit this extension for brevity.

\subsection{Restricting the Controls' Endogeneity}\label{sec:endogeneity_restrictions}

Our results thus far have allowed for arbitrarily endogenous controls. In this section we present more general identification results that allow researchers to impose restrictions on the correlation between the observed covariates and the omitted variable. There are at least two reasons why researchers may be interested in imposing such restrictions: (1) First, it allows us to compare bounds obtained under exogenous controls with those obtained under arbitrarily endogenous controls. This allows us to assess the impact of endogenous controls on the robustness of one's results. (2) Second, when the observed controls are believed to be sufficiently endogenous, it may be plausible to impose a lower bound on their correlation.

To motivate our formal results, note that for scalar $W_2$ with $\var(W_2) = 1$ we can write the omitted variable bias as a function of the coefficient $\gamma_2$ on $W_2$ in the long regression outcome equation \eqref{eq:outcome} and the coefficient $\pi_2$ on $W_2$ in the selection equation \eqref{eq:XprojectionW1W2} as follows:
\begin{equation}\label{eq:ourOVBformula}
	\beta_{\text{med}} - \beta_\text{long}
	=
	\frac{\gamma_2 \pi_2 (1 - R_{\Wu \sim \Wob}^2)}{\var(X^{\perp W_1})}
\end{equation}
where $R_{\Wu \sim \Wob}^2$ denotes the population $R^2$ in a linear regression of the unobserved $\Wu$ on the observed covariates $(1,\Wob)$. Our restrictions using $\bar{r}_X$ and $\bar{r}_Y$ imply bounds on $\pi_2$ and $\gamma_2$. The remaining parameter in equation \eqref{eq:ourOVBformula} is $R_{W_2 \sim W_1}^2$. We consider the following restrictions directly on this R-squared.

\begin{Aassumption}\label{assump:corr}
$R_{W_2 \sim W_1} \in [\uc,\bar{c}]$ for known values of $[\underline{c},\bar{c}] \subseteq [0,1]$.
\end{Aassumption}

Our results in section \ref{sec:MainNewAnalysis} can be thought of as the agnostic case $[\underline{c}, \bar{c}] = [0,1]$. We first consider restrictions under Assumption \ref{assump:corr} and a bound on $r_X$. We then consider restrictions on $r_Y$ as well. We conclude this section by discussing an approach to assessing control endogeneity and to calibrating the values $\underline{c}$ and $\bar{c}$. 

\bigskip

\emph{Remark on extension to non-calibration covariates}. In section \ref{sec:interpretation} we discussed how to include covariates $W_0$ that are not used for calibration. When including these variables, the $\underline{c}$ and $\bar{c}$ variables should be interpreted as bounds on $R_{W_2 \sim W_1 \sbullet W_0}$, the square root of the R-squared from the regression of $\Wu$ on $W_1$ after partialling out $W_0$.

\subsubsection*{Identification Using the $\rx$ and $(\underline{c},\bar{c})$ Restrictions}

Let $\mathcal{B}_I(\rx,\uc,\bar{c})$ denote the identified set for $\beta_\text{long}$ under the positive definite variance assumption \ref{assump:posdefVar}, the restriction \ref{assump:rx} on $r_X$, Assumption \ref{assump:corr} on $R_{W_2 \sim W_1}^2$, and when $W_2$ is a scalar with $\var(W_2)$ normalized to 1. Let
\[
	\underline{B}(\rx, \bar{c}) \coloneqq \inf \mathcal{B}_I(\rx,\uc, \bar{c})
	\qquad \text{and} \qquad
	\overline{B}(\rx, \bar{c}) \coloneqq \sup \mathcal{B}_I(\rx,\uc, \bar{c})
\]
denote its greatest lower bound and least upper bound. Like Theorem \ref{cor:IdsetRyANDcFree}, Theorem \ref{thm:IdsetRyFree} below provides simple, closed form expressions for these sharp bounds.

Appendix equation \eqref{eq:z_X formula} defines the function $\bar{z}_X(\rx,\uc,\bar{c})$; its exact form is unimportant here. The sensitivity parameters will only affect the bounds via this function. 

\begin{theorem}\label{thm:IdsetRyFree}
Suppose the joint distribution of $(Y,X,\Wob)$ is known. Suppose \ref{assump:posdefVar}, \ref{assump:rx}, \ref{assump:noKnifeEdge}, and \ref{assump:corr} hold. Suppose $W_2$ is a scalar and normalize $\var(W_2) = 1$. Normalize $\var(X) = 1$ and $\var(W_1) = I$. If $\bar{z}_X(\rx,\uc, \bar{c})^2 < \var(X^{\perp W_1})$, then
\[
	\underline{B}(\rx,\uc, \bar{c}) = \beta_{\text{med}} - \text{dev}(\rx,\uc, \bar{c})
	\qquad \text{and} \qquad
	\overline{B}(\rx,\uc, \bar{c}) = \beta_{\text{med}} + \text{dev}(\rx,\uc,\bar{c})	
\]
where
\[
	\text{dev}(\rx,\uc, \bar{c}) \coloneqq \sqrt{\frac{\var(Y^{\perp X,W_1})}{\var(X^{\perp W_1})}} \cdot \sqrt{\frac{\bar{z}_X(\rx,\uc,\bar{c})^2 }{\var(X^{\perp W_1}) - \bar{z}_X(\rx,\uc,\bar{c})^2 } }.
\]
Otherwise, $\underline{B}(\rx,\uc,\bar{c}) = -\infty$ and $\overline{B}(\rx,\uc,\bar{c}) = +\infty$.
\end{theorem}

Theorem \ref{cor:IdsetRyANDcFree} can be obtained as a special case of Theorem \ref{thm:IdsetRyFree} by setting $\underline{c} = 0$ and $\overline{c} = 1$, and hence the two theorems have similar interpretations: Theorem \ref{thm:IdsetRyFree} characterizes the largest and smallest possible values of $\beta_\text{long}$ when some selection on unobservables is allowed and the controls are allowed to be partially but not arbitrarily endogenous. We also make no restrictions on the coefficients in the outcome equation. As before, with the exception of at most three singletons, the interval $[\underline{B}(\rx,\uc,\bar{c}), \overline{B}(\rx,\uc, \bar{c})]$ is the identified set for $\beta_\text{long}$ under these assumptions.

Earlier we saw that $\rx < 1$ is necessary for the bounds of Theorem \ref{cor:IdsetRyANDcFree} to be finite. Theorem \ref{thm:IdsetRyFree} shows that, if we are willing to make assumptions on $R^2_{W_2 \sim W_1}$ through $[\uc,\bar{c}]$, then we can allow for $\rx > 1$ while still obtaining finite bounds since $\bar{z}_X(\rx,\uc,\bar{c})$ is finite whenever $\rx < \bar{c}^{-1}$. Thus there is a trade-off between (i) the magnitude of selection on unobservables we can allow for and (ii) the magnitude of control endogeneity. We can use breakdown frontiers to summarize these trade-offs. Specifically, when $\beta_\text{med} \geq 0$, define
\[
	\bar{r}_X^\text{bf}(\underline{c},\bar{c}) 
	\coloneqq \sup \{ \bar{r}_X \geq 0 : b \geq 0 \text{ for all } b \in \mathcal{B}_I(\bar{r}_X, \underline{c}, \bar{c}) \}.
\]
Then the conclusion that $\beta_\text{long} \geq 0$ is robust to any combination $(\bar{r}_X, \underline{c}, \bar{c})$ such that $\bar{r}_X \leq \bar{r}_X^\text{bf}(\underline{c},\bar{c})$. Put differently, for any fixed $\underline{c}$ and $\bar{c}$, $\bar{r}_X^\text{bf}(\underline{c},\bar{c})$ is a breakdown point: It is the largest magnitude of selection on unobservables relative to selection on observables that we can allow for while still concluding that our parameter of interest is nonnegative.

\subsubsection*{Identification Using Restrictions on $(\rx,\ry,\uc,\bar{c})$}

Next we generalize Theorem \ref{cor:BFCalculation_rx_and_ry}. Let 
\[
	\bar{r}_Y^\text{bf}(\rx,\uc,\bar{c}, \underline{b})
	\coloneqq \sup \{ \bar{r}_Y \geq 0 : b \geq \underline{b} \text{ for all } b \in \mathcal{B}_I(\rx, \ry, \uc,\bar{c}) \}.
\]

\begin{theorem}\label{cor:BFCalculation3D}
Suppose the joint distribution of $(Y,X,\Wob)$ is known. Suppose \ref{assump:posdefVar} and \ref{assump:noKnifeEdge} hold. Suppose $W_2$ is a scalar and normalize $\var(W_2) = 1$. Normalize $\var(X) = 1$ and $\var(W_1) = I$. Suppose $\cov(W_1,Y)$ and $\cov(W_1,X)$ are linearly independent. Suppose $d_1 \geq 2$. Let $\underline{b} \in \R$ and $\rx, \ry \geq 0$.
\begin{enumerate}
\item If $\underline{b} \ge \beta_{\text{med}}$ then $\bar{r}_Y^\text{bf}(\rx,\uc,\bar{c}, \underline{b}) = 0$.

\item If $\underline{B}(\rx,\uc,\bar{c}) > \underline{b}$, then $\bar{r}_Y^\text{bf}(\rx,\uc,\bar{c}, \underline{b}) = +\infty$.
    
\item If $\underline{B}(\rx,\uc,\bar{c}) \le \underline{b} < \beta_\text{med}$, then $ \bar{r}_Y^\text{bf}(\rx,\uc,\bar{c}, \underline{b})$ equals
\begin{align*}
     \min_{(z,c_1,c_2,b) \in (-\sqrt{\var(X^{\perp W_1})},\sqrt{\var(X^{\perp W_1})}) \times \R \times \R \times (-\infty, \underline{b}]} \ \ \underline{r}_Y(z,c_1\cov(W_1,Y) + c_2 \cov(W_1,X),b)& \\
      \text{subject to } \ p(z, c_1\cov(W_1,Y) + c_2 \cov(W_1,X); \rx) \ge 0& \\
	(b - \beta_{\text{med}})^2 < \frac{\var(Y^{\perp X, W_1})}{\var(X^{\perp W_1})} \frac{z^2}{\var(X^{\perp W_1}) - z^2}& \\
	\|c_1\cov(W_1,Y) + c_2 \cov(W_1,X)\| \in [\underline{c},\bar{c}] \cap [0, 1).&
\end{align*}
\end{enumerate}
\end{theorem}

Theorem \ref{cor:BFCalculation_rx_and_ry} is the special case obtained by setting $[\uc,\bar{c}] = [0,1]$. 

\subsubsection*{Assessing Exogenous Controls and Calibrating $\underline{c}$ and $\overline{c}$}

Assumption \ref{assump:corr} is a constraint on the covariance between the observed calibration covariates and the unobserved covariate. Specifically, when including non-calibration variables, it is a bound on $R_{\Wu \sim \Wob \sbullet W_0}$. So what values of this parameter should be considered large, and what values should be considered small? One way to calibrate this parameter is to compute
\[
	c_k \coloneqq R_{W_{1k} \sim W_{1,-k} \sbullet W_0}
\]
for each covariate $k$ in $W_1$. That is, compute the square root of the population R-squared from the regression of $W_{1k}$ on the rest of the calibration covariates $W_{1,-k}$, after partialling out the control covariates $W_0$. These numbers tell us two things. First, if many of these values are nonzero and large, we may worry that the exogenous controls assumption fails. That is, if $W_2$ is in some way similar to the observed covariates $W_1$, then we might expect that $R_{\Wu \sim \Wob \sbullet W_0}$ is similar to some of the $c_k$'s. So this gives us one method for assessing the plausibility of exogenous controls. Second, we can use the magnitudes of these values to calibrate our choice of $\bar{c}$, in analysis based on Theorem \ref{cor:BFCalculation_rx_and_ry} or Theorem \ref{thm:IdsetRyFree}. For example, you could choose $\underline{c} = \min_k c_k$ and $\bar{c} = \max_k c_k$. Table \ref{table:ck_calib} in our empirical analysis shows sample analog estimates of $c_k^2$.

\subsection{Estimation and Inference}\label{sec:estimationInference}

Thus far we have described population level identification results. In practice, we only observe finite sample data. Our identification results depend on the observables $(Y,X,\Wob)$ solely through their covariance matrix. In our empirical analysis in section \ref{sec:empirical}, we apply our identification results by using sample analog estimators that replace $\var(Y,X,\Wob)$ with a consistent estimator $\widehat{\var}(Y,X,\Wob)$. For example, we let $\widehat{\beta}_\text{med}$ denote the OLS estimator of $\beta_\text{med}$, the coefficient on $X$ in the medium regression of $Y$ on $(1,X,W_1)$. We expect the corresponding asymptotic theory for estimation and inference on the bound functions to be straightforward, but for brevity we do not develop it in this paper.

\section{Causal Models}\label{sec:causalModels}

While the results in section \ref{sec:MainNewAnalysis} are stated under general conditions, empirical researchers are often interested in assessing sensitivity to omitted variables in causal models. In this section, we review how $\beta_\text{long}$ in equation \eqref{eq:outcome} can be interpreted in the context of four specific causal models. These models are based on three identification strategies: Unconfoundedness, difference-in-differences, and instrumental variables. Here we focus on simple models, but our analysis can be used anytime the causal parameter of interest can be written as the coefficient on a treatment variable in a long regression of the form in equation \eqref{eq:outcome}.

\subsection{Unconfoundedness for Linear Models}

Recall that $Y$ denotes the realized outcome, $X$ denotes treatment, $W_1$ denotes the observed covariates, and $W_2$ denotes the unobserved variables of concern. Let $Y(x)$ denote potential outcomes, where $x$ is any logically possible value of treatment. Assume this potential outcome has the following form:
\begin{equation}\label{eq:OsterOutcomeEq}
	Y(x) = \beta_c x + \gamma_1' W_1 + \gamma_2' W_2 + U
\end{equation}
where $(\beta_c,\gamma_1,\gamma_2)$ are unknown constants. The parameter of interest is $\beta_c$, the causal effect of treatment on the outcome. $U$ is an unobserved random variable. Suppose the realized outcome satisfies $Y = Y(X)$. Consider the following assumption.
\begin{itemize}
	\item[] Linear Latent Unconfoundedness: $\text{corr}(X^{\perp W_1,W_2}, U^{\perp W_1,W_2}) = 0$.
\end{itemize}
This assumption says that, after partialling out the observed covariates $W_1$ and the unobserved variables $W_2$, treatment is uncorrelated with the unobserved variable $U$. This model has two unobservables, which are treated differently via this assumption. We call $W_2$ the \emph{confounders} and $U$ the \emph{non-confounders}. $W_2$ are the unobserved variables which, when omitted, may cause bias. In contrast, as long as we adjust for $(W_1,W_2)$, omitting $U$ does not cause bias. Note that, given equation \eqref{eq:OsterOutcomeEq}, linear latent unconfoundedness can be equivalently written as $\text{corr}(X^{\perp W_1,W_2}, Y(x)^{\perp W_1,W_2}) = 0$ for all logically possible values of treatment $x$.

Linear latent unconfoundedness can be interpreted as a linear parametric version of the nonparametric latent unconfoundedness assumption
\begin{equation}\label{eq:nonparaLatentUnconf}
	Y(x) \indep X \mid (W_1,W_2)
\end{equation}
for all logically possible values of $x$.

The following result shows that, in this model, the causal effect of $X$ on $Y$ can be obtained from $\beta_\text{long}$, the coefficient on $X$ in the long regression described in equation \eqref{eq:outcome}.

\begin{proposition}\label{prop:unconfoundednessBaseline}
Consider the linear potential outcomes model \eqref{eq:OsterOutcomeEq}. Suppose linear latent unconfoundedness holds. Suppose \ref{assump:posdefVar} holds. Then $\beta_c = \beta_\text{long}$.
\end{proposition}

Since $W_2$ is unobserved, however, this result cannot be used to identify $\beta_c$. Instead, suppose we believe the no selection on unobservables assumption \ref{assump:baselineEndogControl1}. Recall that this assumption says that $\pi_2 = 0$, where $\pi_2$ is the coefficient on $W_2$ in the OLS estimand of $X$ on $(1,W_1,W_2)$. Under this assumption, we obtain the following result. Recall that $\beta_\text{med}$ denotes the coefficient on $X$ in the medium regression of $Y$ on $(1,X,W_1)$. 

\begin{corollary}\label{corr:SelectionOnObsCausal}
Suppose the assumptions of Proposition \ref{prop:unconfoundednessBaseline} hold. Suppose \ref{assump:baselineEndogControl1} holds ($\pi_2 = 0$). Then $\beta_c = \beta_\text{med}$.
\end{corollary}

The selection on observables assumption \ref{assump:baselineEndogControl1} is usually thought to be quite strong, however. Nonetheless, since $\beta_c = \beta_\text{long}$, our results in section \ref{sec:MainNewAnalysis} can be used to assess sensitivity to selection on unobservables.

\subsection{Nonparametric Unconfoundedness}

In the previous section we considered a linear parametric version of the unconfoundedness identification strategy. Here we consider a nonparametric version, which allows us to extend our analysis to models with heterogeneous treatment effects. Let $Y(x)$ denote potential outcomes, $X$ be a binary treatment, and $W \coloneqq (W_1,W_2)$ a vector of covariates. Suppose the following two assumptions hold:
\begin{itemize}
\item (Latent Unconfoundedness): $Y(x) \indep X \mid (W_1,W_2)$.

\item (Latent Overlap) $\Prob(X=1 \mid W=w) \in (0,1)$ for $w \in \supp(W)$.
\end{itemize}
For any $w \in \supp(W)$, let
\[
	\text{CATE}(w) \coloneqq \Exp[Y(1) - Y(0) \mid W=w]
\]
denote the conditional average treatment effect. Let $\beta_\text{long}$ be the coefficient in OLS of $Y$ on $(1,X,W_1,W_2)$. Assume $\Exp(X \mid W=w)$ is linear in $w$. That is, we can write
\[
	\Exp(X \mid W=w) = \pi_1' w_1 + \pi_2' w_2.
\]
Define
\[
	\beta_c
	= \Exp[ \omega(W) \text{CATE}(W)]
\]
where
\[
	\omega(w) = \frac{\var(X \mid W=w)}{\Exp[\var(X \mid W)]}.
\]
Under the assumptions above, it follows from \cite{Angrist1998} that $\beta_\text{long} = \beta_c$. Thus $\beta_\text{long}$ has a causal interpretation as a weighted average of CATE's. 

Now suppose $W_2$ is not observed. Recall that $\beta_\text{med}$ denotes the coefficient on $X$ in the medium regression of $Y$ on $(1,X,W_1)$. If $\pi_2 = 0$, then $\beta_\text{med} = \beta_\text{long}$. But if $\pi_2 \neq 0$, generally $\beta_\text{med} \neq \beta_\text{long}$. We can then use our results in section \ref{sec:MainNewAnalysis} to do sensitivity analysis for $\beta_c$.

\subsection{Difference-in-Differences}

Let $Y_t(x_t)$ denote potential outcomes at time $t$, where $x_t$ is a logically possible treatment value. For simplicity we do not consider models with dynamic effects. Also suppose $W_{2t}$ is a scalar for simplicity. Suppose
\begin{equation}\label{eq:DiDoutcomeEq}
	Y_t(x_t) = \beta_c x_t + \gamma_1' W_{1t} + \gamma_2 W_{2t} + V_t
\end{equation}
where $V_t$ is an unobserved random variable and $(\beta_c,\gamma_1,\gamma_2)$ are unknown parameters that are constant across units. The classical two way fixed effects model is a special case where
\begin{equation}\label{eq:TWFE}
	V_t = A + \delta_t + U_t.
\end{equation}
where $A$ is an unobserved random variable that is constant over time, $\delta_t$ is an unobserved constant, and $U_t$ is an unobserved random variable.

Suppose there are two time periods, $t \in \{1,2\}$. Let $Y_t = Y_t(X_t)$ denote the observed outcome at time $t$. For any time varying random variable like $Y_t$, let $\Delta Y = Y_2 - Y_1$. Then taking first differences of the observed outcomes yields
\[
	\Delta Y = \beta_c \Delta X + \gamma_1' \Delta W_1 + \gamma_2 \Delta W_2 + \Delta V.
\]
Let $\beta_\text{long}$ denote the OLS coefficient on $\Delta X$ from the long regression of $\Delta Y$ on $(1,\Delta X, \Delta W_1, \Delta W_2)$.

\begin{proposition}\label{prop:DiD}
Consider the linear potential outcome model \eqref{eq:DiDoutcomeEq}. Suppose the following exogeneity assumptions hold: $\cov(\Delta X, \Delta V) = 0$, $\cov(\Delta W_2, \Delta V) = 0$, and $\cov(\Delta W_1, \Delta V) = 0$. Then $\beta_c = \beta_\text{long}$.
\end{proposition}

The exogeneity assumption in Proposition \ref{prop:DiD} says that $\Delta V$ is uncorrelated with all components of $(\Delta X, \Delta W_1, \Delta W_2)$. A sufficient condition for this is the two way fixed effects assumption \eqref{eq:TWFE} combined with the assumption that the $U_t$ are uncorrelated with $(X_s,W_{1s}, W_{2s})$ for all $t$ and $s$. Given this exogeneity assumption, the only possible identification problem is that $\Delta W_2$ is unobserved. Hence we cannot adjust for this trend variable. If we assume, however, that treatment trends $\Delta X$ are not related to the unobserved trend $\Delta W_2$, then we can point identify $\beta_c$. Specifically, consider the linear projection of $\Delta X$ onto $(1,\Delta W_1, \Delta W_2)$:
\[
	\Delta X = \pi_1' (\Delta W_1) + \pi_2 (\Delta W_2) + (\Delta X)^{\perp \Delta W_1, \Delta W_2}.
\]
Using this equation to define $\pi_2$, we now have the following result. Here we let $\beta_\text{med}$ denote the coefficient on $\Delta X$ in the medium regression of $\Delta Y$ on $(1, \Delta X, \Delta W_1)$.

\begin{corollary}\label{corr:DiD}
Suppose the assumptions of Proposition \ref{prop:DiD} hold. Suppose \ref{assump:baselineEndogControl1} holds ($\pi_2 = 0$). Then $\beta_c = \beta_\text{med}$.
\end{corollary}

This result implies that $\beta_c$ is point identified when $\pi_2 = 0$. This assumption is a version of common trends, because it says that the unobserved trend $\Delta W_2$ is not related to the trend in treatments, $\Delta X$. Our results in section \ref{sec:MainNewAnalysis} allow us to analyze the impacts of failure of this common trends assumption on conclusions about the causal effect of $X$ on $Y$, $\beta_c$. In particular, our results allow researchers to assess the failure of common trends by comparing the impact of observed time varying covariates with the impact of unobserved time varying confounders. In this context, allowing for endogenous controls means allowing for the trend in observed covariates to correlate with the trend in the unobserved covariates. Finally, note that this approach allows researchers to assess sensitivity to common trends even when it is not possible to examine pre-trends; that is, even when there are not multiple time periods where all units are untreated.

\subsection{Instrumental Variables}

Let $Z$ be an observed variable that we want to use as an instrument. Let $Y(z)$ denote potential outcomes, where $z$ is any logical value of the instrument. Assume
\[
	Y(z) = \beta_c z + \gamma_1' W_1 + \gamma_2' W_2 + U
\]
where $U$ is an unobserved scalar random variable and $(\beta_c, \gamma_1, \gamma_2)$ are unknown constants. Thus $\beta_c$ is the causal effect of $Z$ on $Y$. In an instrumental variables analysis, this is typically called the reduced form causal effect, and $Y(z)$ are reduced form potential outcomes. Suppose $\cov(Z^{\perp W_1, W_2}, Y(z)^{\perp W_1, W_2}) = 0$, a linear parametric version of conditional independence between the instrument $Z$ and the unobserved $U$ given $(W_1,W_2)$. Then $\beta_c$ equals the OLS coefficient on $Z$ from the long regression of $Y$ on $(1,Z,W_1,W_2)$. In this model, Proposition \ref{thm:baselineEndogControl1} implies that $\beta_c$ is also obtained as the coefficient on $Z$ in the medium regression of $Y$ on $(1,Z,W_1)$, and thus is point identified. In this case, Assumption \ref{assump:baselineEndogControl1} is an instrument exogeneity assumption. Our results in section \ref{sec:MainNewAnalysis} thus allow us to analyze the impacts of instrument exogeneity failure on conclusions about the reduced form causal effect of $Z$ on $Y$, $\beta_c$.

In a typical instrumental variable analysis, the reduced form causal effect of the instrument on outcomes is not the main effect of interest. Instead, we usually care about the causal effect of a treatment variable on outcomes. The reduced form is often just an intermediate tool for learning about that causal effect. Our analysis in this paper can be used to assess the sensitivity of conclusions about this causal effect to failures of instrument exclusion or exogeneity too. This analysis is somewhat more complicated, however, and so we leave it for a separate paper. Nonetheless, empirical researchers do sometimes examine the reduced form directly to study the impact of instrument exogeneity failure. For example, see Section D7 and Table D15 of \cite{Tabellini2020}.

\section{Empirical Application: The Frontier Experience and Culture}\label{sec:empirical}

Where does culture come from? \cite{BFG2020} study the origins of people's preferences for or against government redistribution, intervention, and regulation. They provide the first systematic empirical analysis of a famous conjecture that living on the American frontier cultivated individualism and antipathy to government intervention. The idea is that life on the frontier was hard and dangerous, had little to no infrastructure, and required independence and self-reliance to survive. These features then create cultural change, in particular, leading to ``more pervasive individualism and opposition to redistribution''. Overall, \cite{BFG2020} find evidence supporting this frontier life conjecture.

\cite{BFG2020} provide a wide variety of analyses and contributions. In our analysis, we focus on a subset of their main results, concerning the long run effects of the frontier experience on modern culture. These results are based on an unconfoundedness identification strategy and use linear models. They note that ``the main threat to causal identification of $\beta$ lies in omitted variables'' and hence they strongly rely on Oster's (2019) sensitivity analysis to ``show that unobservables are unlikely to drive our results'' (page 2344). As we have discussed in the introduction, however, this method may lead to incorrect conclusions about robustness. In this section, we apply our methods to examine the impact of allowing for endogenous controls on Bazzi et al.'s empirical conclusions about the long run effect of the frontier experience on culture. Overall, while they found that all of their analyses were robust to the presence of omitted variables, we find that their analysis using questionnaire based outcomes are sometimes substantially sensitive to omitted variables, while their analysis using property tax levels and voting patterns is generally robust. We also find suggestive evidence that the controls are endogenous, which highlights the value of sensitivity analysis methods that allow for endogenous controls. We discuss all of these findings in more detail below.

\subsection{Data}

We first describe the variables and data sources. The main units of analysis are counties in the U.S., although we will also use some individual level data. The treatment $X$ is the ``total frontier experience'' (TFE). This is defined as the number of decades between 1790 and 1890 a county spent ``on the frontier''. A county is ``on the frontier'' if it had a population density less than 6 people per square mile and was within 100 km of the ``frontier line''. The frontier line is a line that divides sparse counties (less than or equal to 2 people per square mile) from less sparse counties. By definition, the frontier line changed over time in response to population patterns, but it did so unevenly, resulting in some counties being ``on the frontier'' for longer than others. Figure 3 in \cite{BFG2020} shows the spatial distribution of treatment.

The outcome variable $Y$ is a measure of modern culture. They consider 8 different outcome variables. Since data is not publicly available for all of them, we only look at 5 of these. They can be classified into two groups. The first are questionnaire based outcomes:
\begin{enumerate}
\item \emph{Cut spending on the poor}. This variable comes from the 1992 and 1996 waves of the American National Election Study (ANES), a nationally representative survey. In those waves, it asked
\begin{itemize}
\item[] ``Should federal spending be increased, decreased, or kept about the same on poor people?''
\end{itemize}
Let $Y_{1i} = 1$ if individual $i$ answered ``decreased'' and 0 otherwise.

\item \emph{Cut welfare spending}. This variable comes from the Cooperative Congressional Election Study (CCES), waves 2014 and 2016. In those waves, it asked
\begin{itemize}
\item[] ``State legislatures must make choices when making spending decisions on important state programs. Would you like your legislature to increase or decrease spending on Welfare? 1. Greatly Increase 2. Slightly Increase 3. Maintain 4. Slightly Decrease 5. Greatly Decrease.''
\end{itemize}
Let $Y_{2i} = 1$ if individual $i$ answered ``slightly decrease'' or ``greatly decrease'' and 0 otherwise.

\item \emph{Reduce debt by cutting spending}. This variable also comes from the CCES, waves 2000--2014 (biannual). It asked
\begin{itemize}
\item[] ``The federal budget deficit is approximately [\$ year specific amount] this year. If the Congress were to balance the budget it would have to consider cutting defense spending, cutting domestic spending (such as Medicare and Social Security), or raising taxes to cover the deficit. Please rank the options below from what would you most prefer that Congress do to what you would least prefer they do: Cut Defense Spending; Cut Domestic Spending; Raise Taxes.''
\end{itemize}
Let $Y_{3i} = 1$ if individual $i$ chooses ``cut domestic spending'' as a first priority, and 0 otherwise.
\end{enumerate}
These surveys also collected data on individual demographics, specifically age, gender, and race. The second group of outcome variables are based on behavior rather than questionnaire responses:
\begin{enumerate}
\setcounter{enumi}{3}

\item $Y_{4i}$ is the average effective \emph{property tax rate} in county $i$, based on data from 2010 to 2014 from the National Association of Home Builders (NAHB) data, which itself uses data from the American Community Survey (ACS) waves 2010--2014.

\item $Y_{5i}$ is the average \emph{Republican vote share} over the five presidential elections from 2000 to 2016 in county $i$, using data from Leip's Atlas of U.S. Presidential Elections.
\end{enumerate}

Next we describe the observed covariates. We partition these covariates into $W_1$ and $W_0$ by following the implementation of Oster's \citeyearpar{Oster2019} approach in \cite{BFG2020}. $W_1$, the calibration covariates which \emph{are} used to calibrate selection on unobservables, is a set of geographic and climate controls: Centroid Latitude, Centroid Longitude, Land area, Average rainfall, Average temperature, Elevation, Average potential agricultural yield, and Distance from the centroid to rivers, lakes, and the coast. $W_0$, the control covariates which are \emph{not} used to calibrate selection on unobservables, includes state fixed effects. The questionnaire based outcomes use individual level data. For those analyses, we also include age, age-squared, gender, race, and survey wave fixed effects in $W_0$. In \cite{BFG2020}, they were included in $W_1$. We instead include them in $W_0$ to keep the set of calibration covariates $W_1$ constant across the five main specifications. This allows us to directly compare the robustness of our baseline results across different specifications. We discuss the results that include state fixed effects in $W_1$ in Appendix \ref{sec:additionalEmpirical}.

\subsection{Baseline Model Results}

\cite{BFG2020} perform a variety of analyses to examine the long run effect of the frontier experience on culture. We focus on the subset of their main results for which replication data is publicly available. These are columns 1, 2, 4, 6, and 7 of their table 3. Panel A in our Table \ref{table:mainTable1} replicates those results. From columns (1)--(3) we see that individuals who live in counties with more exposure to the frontier prefer cutting spending on the poor, on welfare, and to reduce debt by spending cuts. Moreover, these point estimates are statistically significant at conventional levels. From columns (4) and (5), we see that counties with more exposure to the frontier have lower property taxes and are more likely to vote for Republicans. As \cite{BFG2020} argue, these baseline results support the conjecture that frontier life led to opposition to government intervention and redistribution.

\def\mystrut{\rule{0pt}{1.25\normalbaselineskip}}
\begin{table}[t]
\centering
\SetTblrInner[talltblr]{rowsep=0pt}
\resizebox{.98\textwidth}{!}{
  \begin{talltblr}[
    caption = {The Effect of Frontier Life on Opposition to Government Intervention and Redistribution.\label{table:mainTable1}},
    remark{Note} = {Panel A and the first row of Panel B replicate columns 1, 2, 4, 6, and 7 of table 3 in \cite*{BFG2020}, while the second row of Panel B and Panel C are new. As in \cite{BFG2020}, Panel B uses Oster's rule of thumb choice $R_{Y \sim X,W_1,W_2 \sbullet W_0}^2 = 1.3 \cdot \widehat{R}_{Y \sim X,W_1 \sbullet W_0}^2$.},
  ]{
  p{0.25\textwidth}>{\centering}
  *{3}{>{\centering \arraybackslash}p{0.15\textwidth}} |
  *{2}{>{\centering \arraybackslash}p{0.15\textwidth}}
  }
    \toprule
  
\SetCell{valign=m} \hfill Outcome:  & 
\SetCell{valign=m} Prefers Cut Public Spending on Poor & 
\SetCell{valign=m} Prefers Cut Public Spending on Welfare &
\SetCell{valign=m} Prefers Reduce Debt by Spending Cuts &
\SetCell{valign=m} County Property Tax Rate &
\SetCell{valign=m} Republican Presidential Vote Share \\[4pt]
   & (1) & (2) & (3) & (4) & (5) \\[4pt]
    \hline
  \multicolumn{5}{l}{Panel A. Baseline Results}  \\
  \hline
  
  Total Frontier Exp. 
  & 0.010 & 0.007 & 0.014 & -0.034 & 2.055  \\
  {\footnotesize \quad (standard error)}&{\footnotesize  (0.004) }&{\footnotesize  (0.003) }&{\footnotesize  (0.002) }&{\footnotesize  (0.007) }&{\footnotesize  (0.349) } \\

  Mean of Dep Variable&  0.09 &  0.40 &  0.41 &  1.02 &  60.04  \\ 
  Number of Individuals & 2,322& 53,472& 111,853& - & - \\
  Number of Counties&  95 &  1,863 &  1,963 &  2,029 &  2,036  \\[2pt]
    
    Controls: & & &  & & \\[2pt]
    \ \ Survey Wave FEs              & X & X & X & - & - \\
    \ \ Ind.\ Demographics       & X & X & X & - & - \\
    \ \ State Fixed Effects  & X  & X   & X & X & X \\
    \ \ Geographic/Climate            &  X & X & X &X  & X \\[2pt]
    \hline
  \multicolumn{5}{l}{Panel B. Sensitivity Analysis (Oster 2019)}  \\
  \hline
  $\widehat{\delta}_\text{resid}^\text{ bp}$ (incorrect) & 16.01 & 3.10 & 5.89 & -27.45 & -8.55 \\
  $\widehat{\delta}_\text{resid}^\text{ bp}$ (correct) &  2.28 &  3.05 &  2.58 &  90.7 &  -23.3  \\
  \hline
  \multicolumn{5}{l}{Panel C. Sensitivity Analysis (Our Approach)}  \\
  \hline
  
  $\widehat{\bar{r}}_X^\text{ bp}$ ($\times 100$)&  2.81 &  3.05 &  5.85 &  72.6 &  80.4  \\
  $\widehat{\bar{r}}^\text{ bp}$ ($\times 100$)&  54.1 &  74.4 &  83.9 &  91.8 &  95.9 \\
  \bottomrule
  \end{talltblr}  
}
\end{table}

\subsection{Assessing Selection on Observables}

The baseline results in Table 1 rely on a selection on observables assumption, that treatment $X$ is exogenous after adjusting for the observed covariates $(W_0,W_1)$. How plausible is this assumption? \cite{BFG2020} say
\begin{itemize}
\item[] ``The main threat to causal identification of $\beta$ lies in omitted variables correlated with both contemporary culture and TFE. We address this concern in four ways. First, we rule out confounding effects of modern population density. Second, we augment [the covariates] to remove cultural variation highlighted in prior work. \emph{Third, we show that unobservables are unlikely to drive our results.} Finally, we use an IV strategy that isolates exogenous variation in TFE due to changes in national immigration flows over time.'' (page 2344, emphasis added)
\end{itemize}
Their first two approaches continue to rely on selection on observables, and consist of including additional control variables. We focus on their third strategy: to use a formal econometric method to assess the importance of omitted variables. 

\subsubsection*{Sensitivity Analysis Based on \cite{Oster2019}} 

\cite{BFG2020} use Oster's \citeyearpar{Oster2019} to assess the impact of omitted variables on their results. That method uses a sensitivity parameter $\delta_\text{resid}$, which is informally described in Oster's appendix A (see appendix C of our companion paper \cite{DiegertMastenPoirier2025} for a detailed, formal description of this parameter). The variables $(X,W_1,W_2)$ are replaced by $(X^{\perp W_0}, W_1^{\perp W_0}, W_2^{\perp W_0})$ to adjust for the non-calibration covariates $W_0$. Finally, the auxiliary sensitivity parameter in Oster's analysis, $R_{Y \sim X,W_1,W_2 \sbullet W_0}^2$, is set to $1.3 \cdot \widehat{R}_{Y \sim X,W_1 \sbullet W_0}^2$. Using these parameters and Oster's analysis, \cite{BFG2020} report the estimated breakdown point, which is a measure of how large $\delta_\text{resid}$ must be for the estimated effect to be fully explained away by the unobservables.

The second row of Panel B of Table \ref{table:mainTable1} shows sample analog estimates of this breakdown point, which is commonly referred to as \emph{Oster's delta}. The first row of Panel B shows the values of Oster's delta as reported in table 2 of \cite{BFG2020}. These were computed incorrectly.\footnote{There are three differences between the two rows in Panel B. First, it appears to us that, rather than using the correct expression in Proposition 3 of \cite{Oster2019}, \cite{BFG2020} set the first displayed equation on page 193 of \cite{Oster2019} equal to zero and solved for $\delta_\text{resid}$. That does not give the correct breakdown point. This mistake was not unique to those authors; while replicating other results, we discovered the same mistake in several other papers published in `top 5' economics journals. Second, \cite{BFG2020} used $\widehat{R}_{Y \sim X,W_0,W_1}^2$ and $\widehat{R}_{Y \sim X,W_0}^2$ to adjust for $W_0$ rather than $\widehat{R}_{Y \sim X,W_1 \sbullet W_0}^2$ and $\widehat{R}_{Y \sim X \sbullet W_0}^2$. Third, as we discussed earlier, they include the individual demographics and survey wave fixed effects in $W_1$ rather than $W_0$. This is a valid choice, but we chose to instead include these variables in $W_0$ to keep the covariates in $W_1$ constant across columns (1)--(5) for comparability.} Based on these estimates, \cite{BFG2020} conclude:
\begin{itemize}
\item[] ``Oster (2019) suggests $| \delta | > 1$ leaves limited scope for unobservables to explain the results'' and therefore, based on their $\delta_\text{resid}^\text{bp}$ estimates, ``unobservables are unlikely to drive our results''. (page 2344)
\end{itemize}
This conclusion remains unchanged if the same robustness cutoff is applied to the correctly computed $\delta_\text{resid}^\text{bp}$ estimates.

\subsubsection*{Assessing Exogenous Controls}

In our companion paper \cite{DiegertMastenPoirier2025}, we argue that robustness conclusions based on $\delta_\text{resid}$ are particularly problematic when the controls are endogenous. Are the controls in this application plausibly exogenous, however? The answer depends on which omitted variables $W_2$ we are concerned about. \cite{BFG2020} does not specifically describe the unmeasured omitted variables of concern, nor do they discuss the plausibility of exogenous controls. However, in their extra robustness checks they consider the variables listed in Table \ref{table:extra_vars}.

\begin{table}[ht]
\caption{Additional Covariates Included by \cite{BFG2020} as Robustness Checks. \label{table:extra_vars}}
\centering
\begin{tabular}{l l}
Contemporary population density & Sex ratio \\
Conflict with Native Americans & Rainfall risk \\
Employment share in manufacturing & Portage sites \\
Mineral resources & Prevalence of slavery \\
Immigrant share & Scotch-Irish settlement \\
Timing of railroad access & Birthplace diversity \\
Ruggedness &
\end{tabular}
\end{table}

\noindent The additional omitted variables of concern might therefore be similar to these variables. Thus the question is: Are \emph{all} of the geographic/climate variables in $W_1$ uncorrelated with variables like these? This seems unlikely, especially since many of these additional variables are also geographic/climate type variables. Moreover, although this assumption is not falsifiable---since $W_2$ is unobserved---we can assess its plausibility by examining the correlation structure of the observed covariates. Specifically, we estimate the values $R_{W_{1k} \sim W_{1,-k} \sbullet W_0}^2$ for each covariate $k$ in $W_1$. For a given $k$, this is the population R-squared from the regression of $W_{1k}$ on the rest of the calibration covariates $W_{1,-k}$, after partialling out the control covariates $W_0$. We described these values in section \ref{sec:endogeneity_restrictions}. Table \ref{table:ck_calib} shows sample analog estimates of these values.

The estimates in Table \ref{table:ck_calib} show a substantial range of correlation between the observed covariates in $W_1$. Recall that the exogenous controls assumption says that each element of $W_1$ is uncorrelated with $W_2$, after partialling out $W_0$. So under that assumption, if $W_2$ was included in this table, it would have a value of zero. Therefore, if $W_2$ is a variable similar to the components of $W_1$ then we would expect exogenous controls to fail. This suggests that, in this application, robustness conclusions based on $\delta_\text{resid}$ could be incorrect. Consequently, one should instead use sensitivity analysis methods which are valid even when the controls are endogenous.

\def\mystrut{\rule{0pt}{1.25\normalbaselineskip}}
\begin{table}[th]
\caption{Correlations Between Observed Covariates. \label{table:ck_calib}}
\centering
\begin{tabular}{l c}
\toprule
$W_{1k}$ & $\widehat{R}_{W_{1k} \sim W_{1,-k} \sbullet W_0}^2$ \\[4pt]
  \hline
  \mystrut
Average temperature & 0.893 \\
Centroid Latitude & 0.876 \\
Elevation & 0.681 \\
Average potential agricultural yield & 0.648 \\
Average rainfall & 0.560 \\
Distance from centroid to the coast & 0.487 \\
Centroid Longitude & 0.434 \\
Distance from centroid to rivers & 0.135 \\
Distance from centroid to lakes & 0.100 \\
Land area & 0.098 \\
\bottomrule
\end{tabular}
\end{table}

\subsubsection*{Results from Our Sensitivity Analysis} 

Next we present the findings from the sensitivity analysis that we developed in section \ref{sec:MainNewAnalysis}. We first discuss our two main robustness summary statistics, and then we give a more detailed analysis.

\bigskip

\emph{Robustness Summary Statistics}. Our simplest result, Theorem \ref{cor:IdsetRyANDcFree}, only uses a single sensitivity parameter $\bar{r}_X$. The first row of Panel C of Table \ref{table:mainTable1} reports sample analog estimates of the breakdown point $\bar{r}_X^\text{bp}$ described in Corollary \ref{corr:breakdownPointRXonly}. This is the largest amount of selection on unobservables, as a percentage of selection on observables, allowed for until we can no longer conclude that $\beta_\text{long}$ is nonzero. Recall that, since this result allows for arbitrarily endogenous controls, Theorem \ref{cor:IdsetRyANDcFree} implies that $\bar{r}_X^\text{bp} < 1$. This does not imply that these results should always be considered non-robust. Instead, when the calibration covariates $W_1$ are a set of variables that are important for treatment selection, researchers should consider large values of $\bar{r}_X^\text{bp}$ to indicate the robustness of their baseline results. For example, in columns (4) and (5) of Panel C we see that the breakdown point estimates for the two behavior based outcomes are 72.6\% and 80.4\%. For example, for the average Republication vote share outcome, we can conclude $\beta_\text{long} > 0$ as long as selection on unobservables is at most 80.4\% as large as selection on observables. In contrast, the breakdown point estimates in columns (1)--(3) are substantially smaller: between about 3\% and 6\%. For these outcomes, we therefore only need selection on unobservables to be at least 3 to 6\% as large as selection on observables to overturn our conclusion that $\beta_\text{long} > 0$. Thus, without imposing restrictions on the impact of unobservables on outcomes or on the magnitude of control endogeneity, we find that the analysis using questionnaire based outcomes is highly sensitive to selection on unobservables. In contrast, the analysis using behavior based outcomes is quite robust to selection on unobservables.

The breakdown point $\bar{r}_X^\text{bp}$ is a conservative measure of robustness to omitted variables, because it does not restrict the impact of the omitted variables on outcomes. A less conservative measure of robustness is $\bar{r}^\text{bp}$, which restricts the omitted variables' impact on treatment and on outcomes. The second row of Panel C of Table \ref{table:mainTable1} reports sample analog estimates of this breakdown point. In columns (4) and (5) we see that the breakdown point increases to above 90\%. For example, for the average Republican vote share outcome, we can conclude $\beta_\text{long} > 0$ as long as selection on unobservables is at most 96\% as large as selection on observables, and as long as the impact of omitted variables on outcomes is at most 96\% as large as the impact of the observables on outcomes. The breakdown point estimates in columns (1)--(3) are smaller, between 54\% and 84\%. Nonetheless, these breakdown points are substantially larger than the corresponding estimates of $\bar{r}_X^\text{bp}$. For example, for the cut spending on the poor outcome, the breakdown point increases from about 3\% to 54\% once we restrict the impact of omitted variables on outcomes. Overall, this less conservative measure of robustness shows that it is possible to conclude that the analysis using questionnaire based outcomes is robust. But since there is more variation in the breakdown points for columns (1)--(3), this robustness conclusion depends more importantly on one's subjective judgment about the cutoff for robustness, unlike the analysis for behavior based outcomes.

\bigskip

\begin{table}[tb]
\centering
\def\mystrut2{\rule{0pt}{1\normalbaselineskip}}
\resizebox{0.9\textwidth}{!}{
\SetTblrInner[talltblr]{rowsep=0pt}
\begin{talltblr}[
    caption = {Calibrating $\bar{r}_X$: The Relative Impact of Each Observed Covariate.\label{table:rxbarCalibrationTable}},
    remark{Note} = {Columns (1) and (5) refer to the specifications in Table \ref{table:mainTable1}.},
]{l *{2}{>{\centering \arraybackslash}p{0.3\textwidth}}}
\toprule
\mystrut2
& \multicolumn{2}{c}{$\widehat{\rho}_k$ (\%)} \\[4pt] \cline{2-3}
\rule{0pt}{1.25\normalbaselineskip}
\SetCell{valign=m} \hfill Outcome:  & 
\SetCell{valign=m} Prefers Cut Public Spending on Poor &
\SetCell{valign=m} Republican Presidential Vote Share \\[4pt]
     \hline
 $W_{1k}$                          & (1) & (5) \\[4pt]
  \midrule
   Average potential agricultural yield &  97.0 &  118.3  \\ 
   Distance from centroid to the coast &  71.1 &  78.6  \\ 
   Centroid Longitude &  38.7 &  49.9  \\ 
   Average temperature &  68.3 &  37.3  \\ 
   Average rainfall &  68.9 &  29.3  \\ 
   Centroid Latitude &  4.8 &  26.9  \\ 
   Distance from centroid to rivers &  9.6 &  25.0  \\ 
   Land area &  8.1 &  22.5  \\ 
   Elevation &  7.8 &  20.2  \\ 
   Distance from centroid to lakes &  67.3 &  12.1  \\ 
   
\bottomrule
\end{talltblr}
}
\end{table}

\emph{Calibrating $\bar{r}_X$}. Table \ref{table:rxbarCalibrationTable} shows estimates of the calibration parameters $\rho_k$ that we discussed in section \ref{sec:interpretation}. Although the treatment variable is the same across all specifications in Table \ref{table:mainTable1}, the estimation datasets vary. Consequently, the estimates of $\rho_k$ vary across specifications as well. Here we only show the estimates corresponding to the specifications in columns (1) and (5) for brevity. First consider column (5), the results for average Republican vote share. Here we see that $\widehat{\rho}_k$ is smaller than the $\bar{r}_X^\text{bp}$ breakdown point estimate of 80.4\% for all but one covariate value. This suggests that the results for average Republican vote share is quite robust to selection on unobservables, assuming we view the calibration parameter estimates as plausible values of the unknown ratio $r_X$. In contrast, consider column (1), the results for the cut spending on the poor outcome variable. In this case, the $\bar{r}_X^\text{bp}$ breakdown point estimate of 2.81\% is smaller than \emph{all} the calibration parameters $\widehat{\rho}_k$. Even if we use the less conservative breakdown point $\widehat{\bar{r}}^\text{bp} = 54.1$\%, this value is still smaller than half of the calibration parameters. This suggests that the results for cut spending on the poor are sensitive to selection on unobservables, again assuming the calibration parameter estimates are plausible values for the unknown $r_X$. Overall, our calibration analysis confirms the findings discussed above: The conclusions using the behavior based outcomes are substantially more robust than the conclusions using questionnaire based outcomes.

\bigskip

\begin{figure}[t]
\caption{Sensitivity Analysis for \emph{Average Republican Vote Share}. Left: Bounds on $\beta_\text{long}$ as a function of $\bar{r}_X$. The solid line sets $\bar{r}_Y = \infty$ while the dashed line restricts $\bar{r}_Y = \bar{r}_X$. The tick marks on the horizontal axis at zero show the estimated calibration parameters from Table \ref{table:rxbarCalibrationTable}. Right: Thick solid line shows an estimate of the breakdown frontier $\bar{r}_Y^\text{bf}(\bar{r}_X, 0)$, which allows for arbitrarily endogenous controls. The lighter lines show estimated breakdown frontiers for $\underline{c} = 0$ and various values of $\overline{c} < 1$. \label{fig:republican}}
\centering
\includegraphics[width=0.475\textwidth]{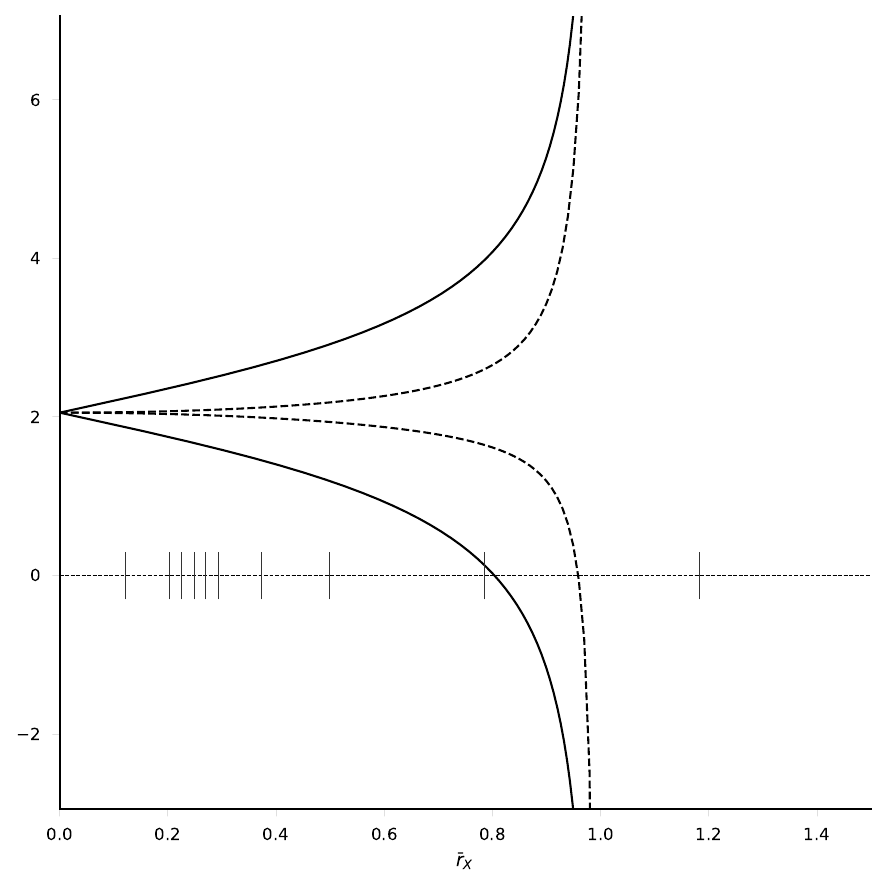}
\hspace{0.005\textwidth}
\includegraphics[width=0.475\textwidth]{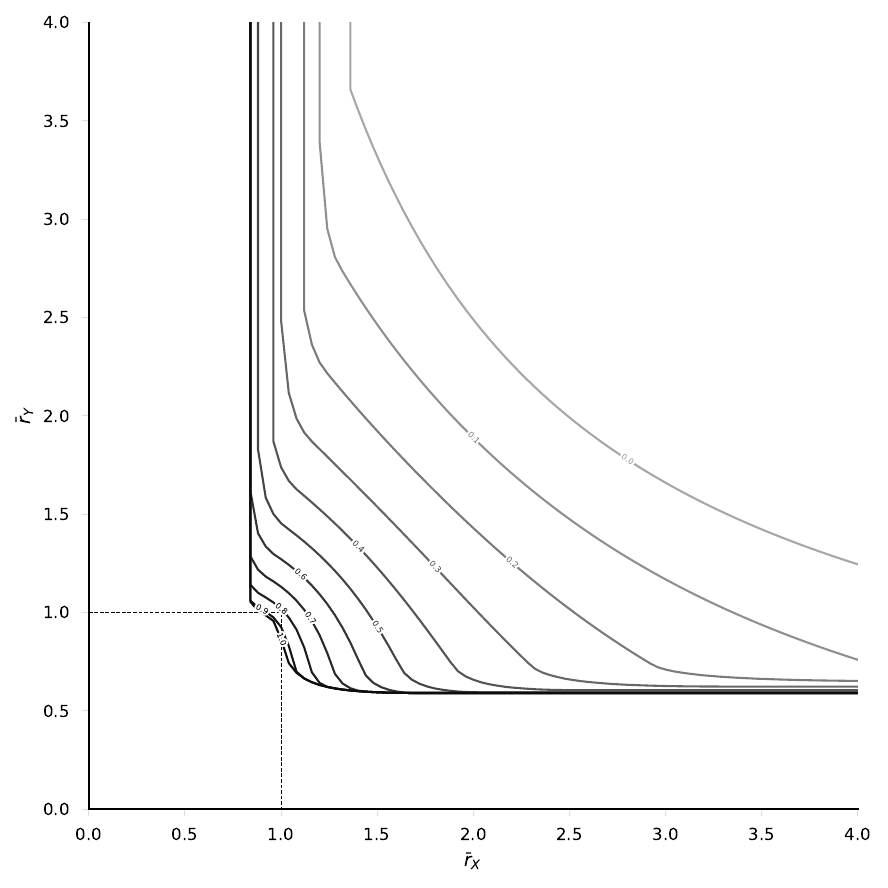}
\end{figure}

\emph{Beyond the Summary Statistics}. Our results also allow researchers to explore the impact of a wide range of assumptions on the omitted variables. We present this analysis next. For brevity we discuss just one questionnaire based outcome, cut spending on the poor, and one behavior based outcome, average Republican vote share. Figure \ref{fig:republican} shows the results for average Republican vote share. The solid lines in the left plot show the estimated identified set for $\beta_\text{long}$ as a function of $\bar{r}_X$, allowing for arbitrarily endogenous controls and no restrictions on the outcome equation. This is the set described by Theorem \ref{cor:IdsetRyANDcFree}. The horizontal intercept is the estimated breakdown point $\widehat{\bar{r}}_X^\text{ bp} = 80.4\%$, as reported in Panel C, column (5) of Table \ref{table:mainTable1}. The dashed lines show the estimated identified set once we impose the common maximal impact assumption $\bar{r}_Y = \bar{r}_X$. The horizontal intercept is the estimated breakdown point $\widehat{r}^\text{ bp} = 96\%$. For all values of $\bar{r}_X$, we see that imposing the restriction $\bar{r}_Y = \bar{r}_X$ shrinks the size of the identified set substantially. The figure also shows the estimated calibration parameters from Table \ref{table:rxbarCalibrationTable} as tick marks on the horizontal line at zero.

Next we consider robustness to choices of $\bar{r}_Y \neq \bar{r}_X$. To show this, we plot the estimated breakdown frontier $\bar{r}_Y^\text{bf}(\bar{r}_X, 0)$ that we showed how to compute in Theorem \ref{cor:BFCalculation_rx_and_ry}. This is shown as the solid black line in the right plot of figure \ref{fig:republican}. For any value of $(\bar{r}_X, \bar{r}_Y)$ below this line we can conclude that $\beta_\text{long} > 0$. For example, $(\bar{r}_X, \bar{r}_Y) = (2, 0.5)$ lies below this line. This means that our conclusion that $\beta_\text{long} > 0$ is robust to unobservables up to twice as strong as the observables for their impact on treatment, and up to half as strong as the observables for their impact on outcomes.

In section \ref{sec:endogeneity_restrictions} we also extended our results to allow researchers to restrict the magnitude of control endogeneity as well. These restrictions shift the breakdown frontier away from zero, as shown by the gray lines in the right plot of figure \ref{fig:republican}. These plots correspond to the restriction $R_{W_2 \sim W_1 \sbullet W_0} \in [0,\bar{c}]$ for the various displayed values of $\bar{c}$. For example, if we set $\bar{c} = 0.7$ then the point $\bar{r}_X = \bar{r}_Y = 110\%$ is now below the breakdown frontier. Hence with this restriction on control endogeneity, our conclusion $\beta_\text{long} > 0$ is robust to unobservables up to 1.1 times as strong as unobservables in for their impact on treatment and on outcomes. Note that $0.7^2 = 0.49$ is around the middle of the distribution of values in Table \ref{table:ck_calib}, and hence might be considered a moderate or slightly conservative value of the magnitude of control endogeneity. If we impose exogenous controls ($\bar{c} = 0$) then we can allow the impact of the omitted variable on outcomes to be 200\% as large as the observables and the impact of the omitted variable on treatment to be up to about 240\% as large as the observables, and yet still conclude that $\beta_\text{long} > 0$.

\begin{figure}[t]
\caption{Sensitivity Analysis for \emph{Cut Spending on Poor}. Left: Bounds on $\beta_\text{long}$ as a function of $\bar{r}_X$. The solid line sets $\bar{r}_Y = \infty$ while the dashed line restricts $\bar{r}_Y = \bar{r}_X$. The tick marks on the horizontal axis at zero show the estimated calibration parameters from Table \ref{table:rxbarCalibrationTable}. Right: Thick solid line shows an estimate of the breakdown frontier $\bar{r}_Y^\text{bf}(\bar{r}_X, 0)$, which allows for arbitrarily endogenous controls. The lighter lines show estimated breakdown frontiers for $\underline{c} = 0$ and various values of $\overline{c} < 1$. \label{fig:cutpoor}}
\centering
\includegraphics[width=0.475\textwidth]{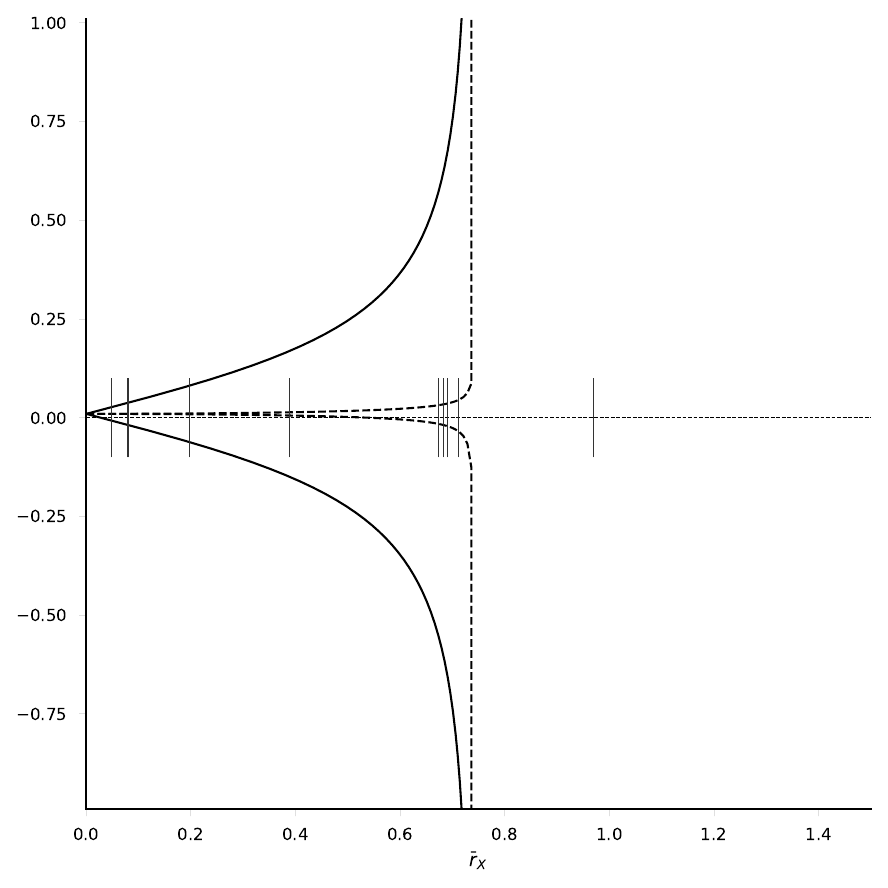}
\hspace{0.005\textwidth}
\includegraphics[width=0.475\textwidth]{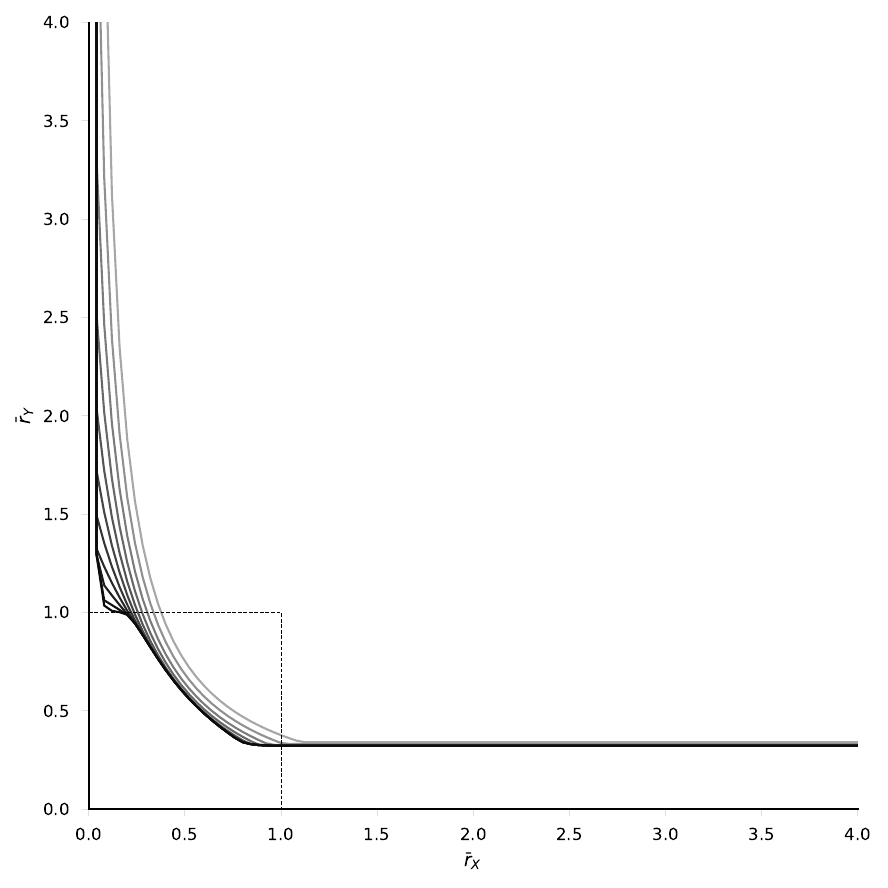}
\end{figure}

These findings suggest that the empirical conclusions for average Republican vote share are quite robust to failures of the selection on observables assumption. In contrast, we next consider the analysis for the cut spending on the poor outcome. Figure \ref{fig:cutpoor} shows the results. The solid lines in the left plot show the estimated identified sets for $\beta_\text{long}$ as a function of $\bar{r}_X$ on the horizontal axis. The horizontal intercept gives an estimated value for $\bar{r}_X^\text{bp}$ of 2.81\%, as reported in Panel C, column (1) of Table \ref{table:mainTable1}. The dashed lines show the estimated identified sets once we impose the common maximal impact assumption $\bar{r}_Y = \bar{r}_X$. This assumption substantially shrinks the identified set, and increases the estimated breakdown point up to 54.1\%. This point can also be seen as the intersection between a 45 degree line and the solid black line on the right plot of figure \ref{fig:cutpoor}, which shows the estimated breakdown frontier $\bar{r}_Y^\text{bf}(\bar{r}_X, 0)$. Compared to figure \ref{fig:republican}, these breakdown frontiers are all shifted substantially toward the origin, suggesting that the conclusion $\beta_\text{long} > 0$ is relatively less robust for this outcome variable.

Overall, there are some cases where the results for cutting spending on the poor could be considered robust, especially if one is willing to restrict the impact of omitted variables on outcomes. But there are also cases where the results could be considered sensitive. In contrast, the results for average Republican vote share are robust across a wide range of relaxations of the baseline model. Similar findings hold for the other three outcome variables: The three results using the questionnaire based outcomes tend to be more sensitive to assumptions on the unobservables than the two results using behavior based outcomes.

\subsection{Empirical Conclusions}

Overall, a sensitivity analysis based on our new methods leads to a richer set of empirical conclusions about the robustness of the long run effect of the frontier experience on modern culture than those originally obtained by \cite{BFG2020}. We found that their analysis using questionnaire based outcomes is generally more sensitive to the presence of omitted variables than their analysis using property tax levels and voting patterns. This has several empirical implications. 

First, the questionnaire based outcomes are the most easily interpretable as measures of opposition to redistribution, regulation, and preferences for small government. In contrast, it is less clear that property taxes and Republican presidential vote share alone should be interpreted as direct measures of opposition to redistribution. So the fact that the questionnaire based outcomes are more sensitive to the presence of omitted variables suggests that Bazzi et al.'s overall conclusion in support of the ``frontier thesis'' might be considered more tentative than previously stated. Second, it suggests that the impact of frontier life may occur primarily through broader behavior based channels like elections, rather than individuals' more specific policy preferences and behavior in their personal lives. It may be useful to explore this difference in future empirical work.

Finally, note that \cite{BFG2020} perform a wide variety of additional supporting analyses that we have not examined here. In particular, their figure 5 considers another set of outcome variables: Republican vote share in each election from 1900 to 2016. In contrast, our analysis above looked only at one election outcome: the average Republican vote share over the five elections from 2000 to 2016. They use these additional baseline estimates along with a qualitative discussion of the evolution of Republican party policies over time to argue that the average Republican vote share outcome between 2000--2016 can be interpreted as a measure of opposition to redistribution. It would be interesting to also apply our methods to these additional analyses.

\section{Conclusion}\label{sec:conclusion}

This paper develops a new method to assess the sensitivity of regression results to omitted variables that does not require assuming the controls are exogenous. Our approach maintains a key aspect of the popular methods of \cite{AltonjiElderTaber2005} and \cite{Oster2019} which let researchers calibrate the magnitude of the sensitivity parameter by comparing a measure of selection on observables with a measure of selection on unobservables, while simultaneously resolving several limitations of those earlier methods. Our results are also simple to implement in practice, via the accompanying Stata package \texttt{regsensitivity}. Finally, in our empirical application to Bazzi et al.'s \citeyearpar{BFG2020} study of the impact of frontier life on modern culture, we showed that allowing for endogenous controls does matter in practice, leading to richer empirical conclusions than those obtained in \cite{BFG2020}.

\bibliographystyle{econometrica}
\bibliography{BadControls_paper}

\appendix
\section{Sketch of Main Identification Proof}\label{sec:mainResultProofIntuition}

To provide intuition for our first main identification result, Theorem \ref{cor:IdsetRyANDcFree}, we give a brief sketch derivation of its proof here. Suppose for simplicity that $W_2$ is a scalar. Using properties of correlations and linear projections, it can be shown that the omitted variable bias can be written as
\begin{equation}\label{eq:appendixIntuitionOVB}
	| \beta_\text{med} - \beta_\text{long} | 
	=  \sqrt{\frac{\var(Y^{\perp X,W_1})}{\var(X^{\perp W_1})} \cdot \frac{R^2_{X \sim W_2 \sbullet W_1}}{1 - R^2_{X\sim W_2 \sbullet W_1}} \cdot R^2_{Y \sim W_2 \sbullet X, W_1}}.
\end{equation}
Versions of this decomposition appear in \cite{HosmanHansenHolland2010} and \cite{CinelliHazlett2020}, for example. This expression depends on two unknown partial R-squareds: $R^2_{X\sim W_2 \sbullet W_1}$ and $R^2_{Y \sim W_2 \sbullet X, W_1}$. Now suppose we impose Assumption \ref{assump:rx}, $r_X \leq \rx$. We can write 
\begin{equation}\label{eq:rxIntuition}
	r_X^2 
	= \frac{R^2_{X \sim W_1,W_2} - R^2_{X \sim W_1}}{R^2_{X \sim W_1,W_2} - R^2_{X \sim W_2}} 
	= \frac{R_{X \sim W_2 \sbullet W_1}^2(1 - R^2_{X \sim W_1})}{R_{X \sim W_2 \sbullet W_1}^2(1 - R^2_{X \sim W_1}) + R^2_{X \sim W_1} - R^2_{X \sim W_2}}
\end{equation}
and hence the assumption that $r_X \leq \bar{r}_X$ \emph{does} restrict the value of $R_{X \sim W_2 \sbullet W_1}^2$. For example, setting $\bar{r}_X = 0$ implies $R^2_{X \sim W_2 \sbullet W_1} = 0$. Equation \eqref{eq:rxIntuition} does not depend on $R^2_{Y \sim W_2 \sbullet X,W_1}$, however. Since equation \eqref{eq:rxIntuition} is monotonically increasing in $R^2_{Y \sim W_2 \sbullet X,W_1}$, we can therefore set it to 1 to obtain the worst case bounds (this can be done independently from $R^2_{X \sim W_2 \sbullet W_1}$ by Lemma \ref{lem:unconstrainedR2_vector}). Similarly, equation \eqref{eq:rxIntuition} is monotonically increasing in $R_{X \sim W_2 \sbullet W_1}^2$. So it suffices to find the largest possible value of this R-squared that is consistent with the assumption that $r_X \leq \bar{r}_X$. To do this, notice that equation \eqref{eq:rxIntuition} also depends on the unknown $R^2_{X \sim W_2}$. This term does not enter the OVB equation \eqref{eq:appendixIntuitionOVB} and hence we can treat it as a free parameter and simply pick the largest value of $R^2_{X \sim W_2 \sbullet W_1} \in [0,1)$ consistent with some value of $R^2_{X \sim W_2} \in [0,1]$ and with the observed data and the constraint that equation \eqref{eq:rxIntuition} is not larger than $\bar{r}_X$. The solution is $R^2_{X \sim W_2} = 0$ and 
\begin{align*}
	R^2_{X \sim W_2 \sbullet W_1}
	&= \begin{cases}
		\frac{\rx^2 R^2_{X \sim W_1}}{(1 - R^2_{X \sim W_1})(1 - \rx^2)}
		&\text{ if } \rx^2 < 1 - R^2_{X \sim W_1} \\
		1
		&\text{ if } \rx^2 \geq 1 - R^2_{X \sim W_1}
		\end{cases}\\[0.5em]
	&= \min\left\{\frac{\rx^2 R^2_{X \sim W_1}}{(1 - R^2_{X \sim W_1})(1 - \rx^2)},1\right\}.
\end{align*}
See Lemmas \ref{lem:unconstrainedR2_vector}, \ref{lem:marg_R2_XonW2_vector}, and \ref{lem:max_abstract_vector} for formal details. Substituting this and  $R^2_{Y \sim W_2 \sbullet X,W_1} = 1$ into \eqref{eq:appendixIntuitionOVB} gives
\begin{align*}
	| \beta_\text{med} - \beta_\text{long} |
	&=  \sqrt{\dfrac{\var(Y^{\perp X,W_1})}{\var(X^{\perp W_1})} \cdot \frac{\min\left\{\frac{\rx^2 R^2_{X \sim W_1}}{(1 - R^2_{X \sim W_1})(1 - \rx^2)},1\right\}}{1 - \min\left\{\frac{\rx^2 R^2_{X \sim W_1}}{(1 - R^2_{X \sim W_1})(1 - \rx^2)},1\right\}} \cdot 1}\\
	&= \begin{cases}
		\sqrt{\dfrac{\var(Y^{\perp X,W_1})}{\var(X^{\perp W_1})} \cdot \dfrac{\rx^2 R^2_{X \sim W_1}}{1 - R^2_{X \sim W_1} - \rx^2}} &\text{ if } \rx^2 < 1 - R^2_{X \sim W_1}\\
		+\infty &\text{ if } \rx^2 \geq 1 - R^2_{X \sim W_1}.
		\end{cases}
\end{align*}
This is the largest magnitude of omitted variable bias under the restriction that $r_X \leq \rx$, which therefore gives us the bounds in Theorem \ref{cor:IdsetRyANDcFree}. This argument can be modified to account for constraints on $r_Y$ as in Theorem \ref{cor:BFCalculation_rx_and_ry} or constraints on the correlation between $W_1$ and $W_2$ as in Theorems \ref{thm:IdsetRyFree} and \ref{cor:BFCalculation3D}. We omit this discussion for brevity; see the online appendices for full proofs.

\section{Additional Empirical Analysis}\label{sec:additionalEmpirical}

\subsubsection*{The Effect of the Choice of Calibration Covariates}

\begin{figure}[t]
\caption{Effect of Calibration Covariates on Analysis For Republican Vote Share. See body text for discussion.\label{fig:calibrationCovars}}
\centering
\includegraphics[width=0.475\textwidth]{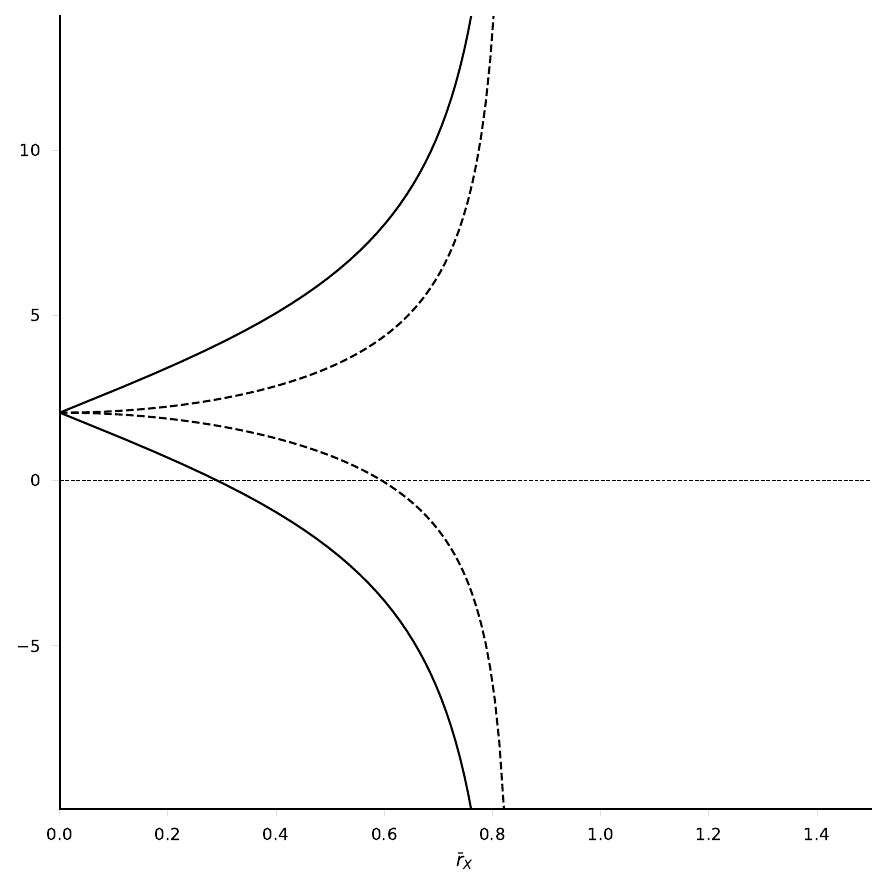}
\hspace{0.005\textwidth}
\includegraphics[width=0.475\textwidth]{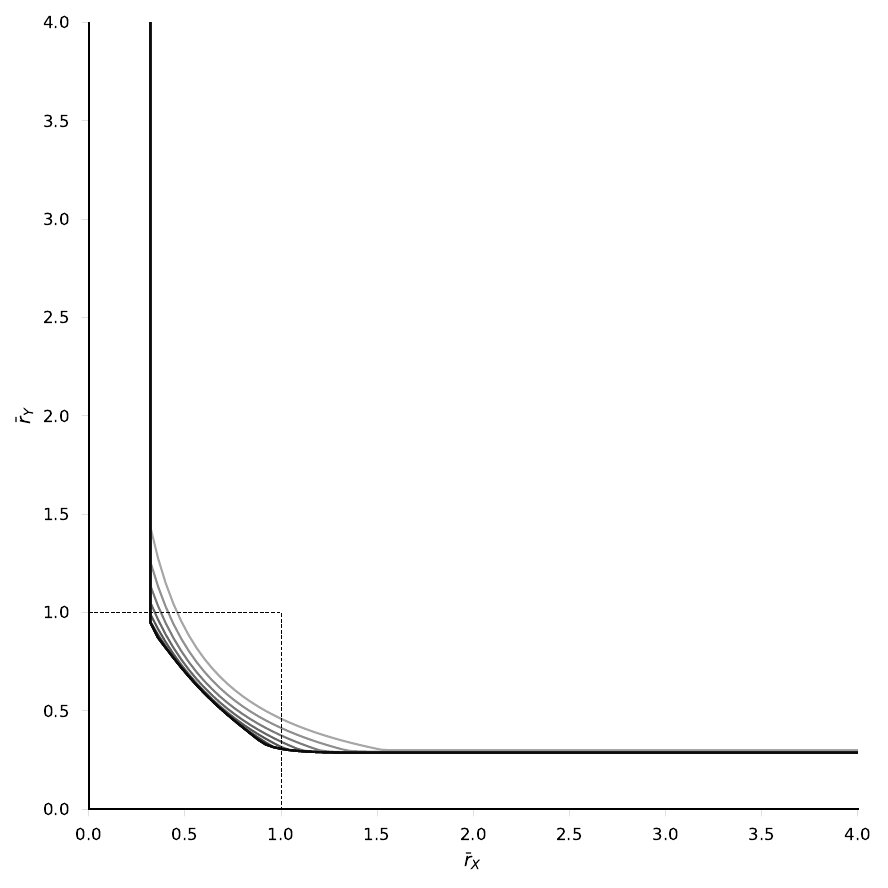}
\end{figure}

In section \ref{sec:interpretation} we discussed the importance of choosing which variables to calibrate against (the variables in $W_1$) versus which variables to use as controls only (the variables in $W_0$). Here we briefly illustrate this in our empirical application. The results in Table \ref{table:mainTable1} and figures \ref{fig:republican} and \ref{fig:cutpoor} all include state fixed effects as controls, but do not use them for calibration; that is, these variables are in $W_0$. Next we consider the impact of instead putting them into $W_1$ and calibrating the magnitude of selection on unobservables against them, in addition to the geographic and climate controls already in $W_1$.

Figure \ref{fig:calibrationCovars} shows plots corresponding to figure \ref{fig:republican}, but now also using state fixed effects for calibration. We first see that the identified sets for $\beta_\text{long}$ (left plot) are larger, for any fixed $\bar{r}_X$. This makes sense because the \emph{interpretation} of $\bar{r}_X$ has changed with the change in calibration controls. In particular, the breakdown point $\bar{r}_X^\text{bp}$ is now about 30\%, whereas previously it was about 80\%. By including state fixed effects---which have a large amount of explanatory power---in our calibration controls, we have increased the magnitude of selection on observables. Holding selection on unobservables fixed, this implies that $r$ must decrease. This discussion reiterates the point that the magnitude of $\bar{r}_X$ must always be interpreted as dependent on the set of calibration controls. For example, our finding in figure \ref{fig:calibrationCovars} that the estimated $\bar{r}_X^\text{bp}$ is about 30\% should not be interpreted as saying that the results are sensitive; in fact, an effect about 30\% as large as these calibration covariates is substantially large, and so it may be that we do not expect the omitted variable to have such a large additional impact.

The right plot in figure \ref{fig:calibrationCovars} shows the estimated breakdown frontiers. The frontiers have all shifted inward, compared to the right plot of figure \ref{fig:republican} which did not use state fixed effects for calibration. Consequently, a superficial reading of this plot may suggest that the results for average Republican vote share are no longer robust. However, as we just emphasized in our discussion of the left plot, by including state fixed effects in the calibration covariates $W_1$, we are changing the meaning of all sensitivity parameters. Since the expanded set of calibration covariates has substantial explanatory power, even a relaxation like $(\bar{r}_Y, \bar{r}_X) = (50\%, 50\%)$---which is below the breakdown frontier and hence allows us to conclude that $\beta_\text{long}$ is positive---could be considered to be a large impact of omitted variables. So these figures do not change our overall conclusions about the robustness of the analysis for average Republican vote share.

\section{The Identified Set For $\beta_\text{long}$ With Fixed $(\bar{r}_X, \bar{r}_Y, \underline{c}, \bar{c}$)}\label{sec:generalIdentSet}

In this appendix we characterize the identified set for $\beta_\text{long}$, the coefficient on $X$ in the long regression of $Y$ on $(1,X,W_1,W_2)$, under assumptions \ref{assump:rx}--\ref{assump:corr} hold and when $W_2$ is a scalar. That is, we use information from all three sensitivity parameters to learn about $\beta_\text{long}$ (Theorem \ref{thm:mainMultivariateResult} below). This result is a preliminary step in deriving our main results in sections \ref{subsec:identrXonly} and \ref{sec:identrYtoo}.

Here and throughout the appendix we let $\| a \|_V \coloneqq \sqrt{a'Va}$ denote the weighted Euclidean norm, for a positive definite matrix $V$. Let $\| \cdot \|$ denote the unweighted Euclidean norm. Let $\Sigma_\text{obs} \coloneqq \var(W_1)$. Define
\begin{align}
	\mathcal{B}(\tilde{r}_X,\tilde{r}_Y,c)
	\coloneqq
	\{ b \in \R : \text{Equations \eqref{eq:IDseteq1}--\eqref{eq:IDseteq6} hold for some } (p_1,g_1) \in \R^{d_1} \times \R^{d_1} \} \label{eq:condIDsetdefTEMP}
\end{align}
where these equations are
\begin{align}
	\cov(Y,X) &= b \var(X) + g_1'(\var(W_1) + c\tilde{r}_X' + \tilde{r}_Yc' + \tilde{r}_Y \tilde{r}_X')p_1 \label{eq:IDseteq1}\\
	\cov(Y,W_1) &= b \cov(X,W_1) + g_1'(\var(W_1) + \tilde{r}_Yc')\label{eq:IDseteq2}\\
	\cov(X,W_1) &= p_1'(\var(W_1) + \tilde{r}_X c')\label{eq:IDseteq3}\\
	\var(Y) &> b^2\var(X) + g_1'\left(\var(\Wob) + \tilde{r}_Y\tilde{r}_Y' + 2\tilde{r}_Yc'\right)g_1 \notag\\
	&\quad  + 2b g_1'(\var(W_1) + c\tilde{r}_X' + \tilde{r}_Yc' + \tilde{r}_Y \tilde{r}_X')p_1\label{eq:IDseteq4}\\
	\var(X) &> p_1'(\var(\Wob) + 2\tilde{r}_Xc' + \tilde{r}_X \tilde{r}_X')p_1\label{eq:IDseteq5}\\
	1 &> c'\var(W_1)^{-1}c.\label{eq:IDseteq6}
\end{align}

\begin{theorem}\label{thm:mainMultivariateResult}
Suppose the joint distribution of $(Y,X,\Wob)$ is known. Suppose \ref{assump:posdefVar} holds. Suppose \ref{assump:rx}--\ref{assump:corr} hold. Suppose $W_2$ is a scalar with $\var(W_2)$ normalized to 1. Let $\mathcal{B}_I(\rx,\ry,\underline{c},\bar{c})$ denote the identified set for $\beta_\text{long}$ under these assumptions. Then
\begin{align}
	\mathcal{B}_I(\rx,\ry,\underline{c},\bar{c})
	= \bigcup_{(\tilde{r}_X,\tilde{r}_Y,c) : \| \tilde{r}_X \|_{\Sigma_{\text{obs}}^{-1}} \leq \rx, \| \tilde{r}_Y \|_{\Sigma_{\text{obs}}^{-1}} \leq \ry, \| c \|_{\Sigma_{\text{obs}}^{-1}} \in [\underline{c}, \bar{c}]} \mathcal{B}(\tilde{r}_X,\tilde{r}_Y,c). \label{eq:IDsetdef}
\end{align}
\end{theorem}

\makeatletter\@input{DMP2aux.tex}\makeatother
\end{document}


\maketitle
\begin{abstract}
This document contains appendices to accompany the main text. Appendix \ref{sec:proofsForIntuitionAppendix} provides formal details to accompany the sketch of Appendix \ref{sec:mainResultProofIntuition}. Appendix \ref{sec:modelProofs} contains proofs for Section \ref{sec:model}, and Appendix \ref{sec:proofsForMainNewAnalysis} contains all remaining proofs for the main paper.
\end{abstract}

\singlespacing

\appendix
\allowdisplaybreaks

\setcounter{section}{3}

\section{Formal Details and Proofs for Appendix \ref{sec:mainResultProofIntuition}}\label{sec:proofsForIntuitionAppendix}

Here we give formal details regarding the sketch of the main identification proof in Appendix \ref{sec:mainResultProofIntuition}. We give lemmas used in this intuition section and their proofs. We also provide an alternate proof of Theorem \ref{cor:IdsetRyANDcFree} that uses only these lemmas.

\subsection{Supporting Results}

Let $R_{A \sim B}$ denote the vector of correlations between components of a random vector $A$ and a random variable $B$. Note that $R_{B \sim A}^2 = R_{A \sim B}' \corr(A)^{-1} R_{A \sim B}$ where $\corr(A)$ is the matrix of correlations between the components of $A$. Let $\Omega_\text{obs}$ denote the correlation matrix of $W_1$.  

We first show that a scalar unobserved confounder $W_2$ can simultaneously attain arbitrary values of the following (partial) correlations: $(R_{W_1 \sim W_2}, R_{X \sim W_2\sbullet W_1},R_{Y \sim W_2\sbullet X,W_1})$. This allows us to vary these values independently of each other when searching for dgps that maximize the magnitude of the omitted variable bias.

\begin{lemma}\label{lem:unconstrainedR2_vector}
Suppose the joint distribution of $(Y,X,W_1)$ is known. Suppose that Assumption \ref{assump:posdefVar} holds, and that $W_2$ is a random variable with finite variance. Then the identified set for
\[
	(R_{W_1 \sim W_2}, R_{X \sim W_2\sbullet W_1},R_{Y \sim W_2\sbullet X,W_1}) \in \R^{d_1 + 1 + 1}
\]
contains 
\[
	\{r \in (-1,1)^{d_1} : r'\Omega_\text{obs}^{-1}r < 1\} \times (-1,1) \times (-1,1).
\]
\end{lemma}

The next lemma shows that we can write the R-squared between the variable of interest $X$ and the unobserved confounder $W_2$ as a function of some of the three R-squareds from the previous lemma, which are unconstrained.

\begin{lemma}\label{lem:marg_R2_XonW2_vector}
Suppose $R_{W_1 \sim W_2}, R_{W_1 \sim X} \in \{r \in (-1,1)^{d_1} : r'\Omega_\text{obs}^{-1}r < 1\}$ and $R_{X \sim W_2 \sbullet W_1} \in [-1,1]$ are fixed. Then
\[
	R_{X \sim W_2}
	= R_{X \sim W_2 \sbullet W_1} \sqrt{(1 - R^2_{X \sim W_1})(1 - R^2_{W_2 \sim W_1})} + R_{W_1 \sim X}' \Omega_\text{obs}^{-1} R_{W_1 \sim W_2}.
\]
\end{lemma}

We use the final lemma to calculate the maximum value of $R^2_{X \sim W_2 \sbullet W_1}$ that is consistent with the constraint $|r_X| \leq \rx$, for a given value of $R_{X \sim W_1}$.

\begin{lemma}\label{lem:max_abstract_vector}
Fix $R_{W_1 \sim X} \in \R^{d_1}$ such that $R^2_{X \sim W_1} = R_{W_1 \sim X}'\Omega_\text{obs}^{-1}R_{W_1 \sim X} \in (0, 1)$, and fix $\rx \geq 0$. Let 
\[
	f(r_1,r_2) \coloneqq \frac{r_2^2 (1-R^2_{X \sim W_1})}{r_2^2 (1-R^2_{X \sim W_1}) + R^2_{X \sim W_1} - (r_2 \sqrt{(1 - R^2_{X \sim W_1})(1-r_1'\Omega_\text{obs}^{-1}r_1)} + R_{W_1 \sim X}'\Omega_\text{obs}^{-1}r_1)^2}
\]
for $(r_1,r_2) \in \{ \tilde{r}_1 \in \R^{d_1}: \tilde{r}_1'\Omega_\text{obs}^{-1} \tilde{r}_1 < 1\} \times (-1,1)$. Then
\[
	\sup_{(r_1,r_2) \in \{ \tilde{r}_1 \in \R^{d_1}: \tilde{r}_1'\Omega_\text{obs}^{-1} \tilde{r}_1 < 1\} \times (-1,1) : f(r_1,r_2) \leq \rx^2} 
	r_2^2 
	=
	\begin{cases}
		\frac{\rx^2 R^2_{X \sim W_1}}{(1- R^2_{X \sim W_1})(1 - \rx^2)}
		&\text{ if } \rx^2 < 1 - R^2_{X \sim W_1}\\
	        1
	        &\text{ if } \rx^2 \geq 1 - R^2_{X \sim W_1}.
	 \end{cases} 
\]
\end{lemma}

Combining the previous lemmas with equation \eqref{eq:appendixIntuitionOVB} yields an alternative derivation of Theorem \ref{cor:IdsetRyANDcFree}. We conclude this section by formalizing the intuition of Appendix \ref{sec:mainResultProofIntuition} via this alternate proof. This proof is shorter than the main one and arguably provides better intuition for the structure of the bounds. However, we keep the main proof since it can be generalized to prove theorems \ref{cor:BFCalculation_rx_and_ry} and \ref{thm:IdsetRyFree}.

\begin{proof}[Alternate Proof]
By equation \eqref{eq:appendixIntuitionOVB}, we can write the omitted variable bias as
\[
	| \text{OVB} |
	= \sqrt{\frac{\var(Y^{\perp X,W_1})}{\var(X^{\perp W_1})}} \cdot  \frac{| R_{X \sim W_2\sbullet W_1}|}{\sqrt{1 - R^2_{X \sim W_2\sbullet W_1}}}  \cdot |R_{Y \sim W_2\sbullet X,W_1} |.
\]
We will find a dgp that maximizes this expression subject to the constraint $| r_X | \leq \rx$. By Lemma \ref{lem:unconstrainedR2_vector}, for any small $\varepsilon_Y > 0$, we can set $R_{Y \sim W_2 \sbullet X,W_1} = 1 - \varepsilon_Y$ independently from $R_{X \sim W_2 \sbullet W_1}$ and $R_{W_1 \sim W_2}$. By Lemma \ref{lem:marg_R2_XonW2_vector}, the constraint $| r_X | \leq \rx$ is equivalent to the constraint
\begin{align*}
	\rx^2 &\geq \frac{R^2_{X \sim W_1,W_2} - R^2_{X \sim W_1}}{R^2_{X \sim W_1,W_2} - R^2_{X \sim W_2}}\\
	&= \frac{R_{X \sim W_2\sbullet W_1}^2(1 - R^2_{X \sim W_1})}{R_{X \sim W_2\sbullet W_1}^2(1 - R^2_{X \sim W_1}) + R^2_{X \sim W_1} - R^2_{X \sim W_2}}\\
	&= f(R_{W_1 \sim W_2}, R_{X \sim W_2\sbullet W_1})
\end{align*}	
so long as $(R_{X \sim W_2 \sbullet W_1},R_{W_1 \sim W_2}) \in (-1,1) \times \{ \tilde{r}_1 \in (-1,1)^{d_1}: \tilde{r}_1'\Omega_\text{obs}^{-1} \tilde{r}_1 < 1\}$. By Lemma \ref{lem:max_abstract_vector}, the maximum value of $R_{X \sim W_2 \sbullet W_1}^2$ subject to that constraint is given by $\rx^2 R^2_{X \sim W_1} / ((1 - R^2_{X \sim W_1})(1 - \rx^2))$ for $\rx^2 \leq 1 - R^2_{X \sim W_1}$ and $1$ for $\rx^2 > 1 - R^2_{X \sim W_1}$. By Lemma \ref{lem:unconstrainedR2_vector} we can choose $| R_{X \sim W_2 \sbullet W_1} | = \min\left\{\sqrt{\frac{\rx^2 R^2_{X \sim W_1}}{(1 - R^2_{X \sim W_1})(1 - \rx^2)}},1 - \varepsilon_X \right\}$, where $\varepsilon_X  > 0$ is arbitrarily small. Substituting our choices for $R_{X \sim W_2 \sbullet W_1}$ and $R_{Y \sim W_2 \sbullet X,W_1}$ into the OVB expression yields
\begin{align*}
	|\text{OVB}|
	&\leq \sqrt{\frac{\var(Y^{\perp X,W_1})}{\var(X^{\perp W_1})}} \cdot  \frac{\min\left\{\sqrt{\frac{\rx^2 R^2_{X \sim W_1}}{(1 - R^2_{X \sim W_1})(1 - \rx^2)}},1 - \varepsilon_X \right\}}{\sqrt{1 - \min\left\{\frac{\rx^2 R^2_{X \sim W_1}}{(1 - R^2_{X \sim W_1})(1 - \rx^2)},(1 - \varepsilon_X )^2\right\}}}  \cdot (1 - \varepsilon_Y).
\end{align*}
We obtain the bounds of Theorem \ref{cor:IdsetRyANDcFree} by letting $\varepsilon_X ,\varepsilon_Y \to 0$.
\end{proof}

\subsection{Proof of Supporting Lemmas}

\begin{proof}[Proof of Lemma \ref{lem:unconstrainedR2_vector}]
Before proceeding with the proof, note that
\[
	\Omega_\text{obs}
	= \textbf{S}^{-1}\Sigma_\text{obs} \textbf{S}^{-1}
\]
where $\textbf{S} \coloneqq \text{diag}(\SD(W_{1,1}),\ldots, \SD(W_{1,d_1}))$, a diagonal matrix with the standard deviations of the components of $W_1$ along its diagonal, and $\Sigma_\text{obs} \coloneqq \var(W_1)$.

Fix $(v_2,R_W,R_X,R_Y) \in (0,\infty) \times \{\tilde{r}_1 \in (-1,1)^{d_1} : \tilde{r}_1'\Omega_\text{obs}^{-1} \tilde{r}_1 < 1\} \times (-1,1) \times (-1,1)$. We will construct a joint distribution of $(Y,X,W_1,W_2)$ that is consistent with the known distribution of $(Y,X,W_1)$ and with $(\var(W_2),R_{W_1 \sim W_2}, R_{X \sim W_2\sbullet W_1},R_{Y \sim W_2\sbullet X,W_1}) = (v_2,R_W,R_X,R_Y)$. (For generality, here we also show that any value of $\var(W_2) \in (0,\infty)$ is attainable.) For this construction, let $U$ be unit-variance random variable that is uncorrelated with $(Y,X,W_1)$. Let
\[
	W_2 \coloneqq a' W_1 + b X^{\perp W_1} + c Y^{\perp X,W_1} + \sigma_U U
\]
where
\begin{align*}
	a &\coloneqq \Sigma_\text{obs}^{-1} \textbf{S} R_W v_2^{1/2}\\
	b &\coloneqq R_X  \sqrt{v_2(1 - R_W'\Omega_\text{obs}^{-1}R_W})/\SD(X^{\perp W_1})\\
	c &\coloneqq R_Y \sqrt{v_2(1 - R_W'\Omega_\text{obs}^{-1}R_W)(1 - R_X^2)}/\SD(Y^{\perp X,W_1})\\
	\sigma_U &\coloneqq \sqrt{v_2(1 - R_W'\Omega_\text{obs}^{-1}R_W)(1 - R_X^2)(1-R^2_Y)}.
\end{align*}
$(a,b,c,\sigma_U)$ are all well-defined since $\var(Y,X,W_1)$ is positive definite, $R_W'\Omega_\text{obs}^{-1}R_W < 1$, and $R_X, R_Y \in (-1,1)$.
	
We now compute the implied values of $(\var(W_2),R_{W_1 \sim W_2}, R_{X \sim W_2\sbullet W_1},R_{Y \sim W_2\sbullet X,W_1})$ and verify that they coincide with $(v_2,R_W,R_X,R_Y)$:
	\begin{align*}
		\var(W_2) &= a'\var(W_1)a + b^2 \var(X^{\perp W_1}) + c^2 \var(Y^{\perp X,W_1}) + \sigma^2_U\\
		&= v_2 R_W'\Omega_\text{obs}^{-1}R_W  + v_2 R_X^2(1 - R_W'\Omega_\text{obs}^{-1}R_W ) + v_2 R^2_Y (1-R_W'\Omega_\text{obs}^{-1}R_W )(1 - R^2_X)\\
		&\qquad + v_2(1 - R_W'\Omega_\text{obs}^{-1}R_W )(1 - R_X^2)(1-R^2_Y)\\
		&= v_2,
	\end{align*}
	\begin{align*}
		R_{W_1 \sim W_2} &\coloneqq \corr(W_1,W_2)\\
		&= \textbf{S}^{-1}v_2^{-1/2}\cov(W_1,a'W_1 + b X^{\perp W_1} + c Y^{\perp X,W_1} + U)\\
		&= \textbf{S}^{-1}v_2^{-1/2}\var(W_1)\Sigma_\text{obs}^{-1} \textbf{S} R_W v_2^{1/2}\\
		&= R_W,
	\end{align*}
	\begin{align*}
		R_{X \sim W_2 \sbullet W_1} &= \frac{\cov(W_2,X^{\perp W_1})}{\SD(W_2^{\perp W_1})\SD(X^{\perp W_1})}\\
		&= b \SD(X^{\perp W_1})/\sqrt{v_2 (1 - R_{W_2 \sim W_1}^2)}\\
		&= b \SD(X^{\perp W_1})/\sqrt{v_2 (1 - R_W'\Omega_\text{obs}^{-1}R_W)}\\
		&= R_X,
	\end{align*}
	\begin{align*}
		R_{Y \sim W_2 \sbullet X,W_1} &= \frac{\cov(W_2,Y^{\perp X,W_1})}{\SD(W_2^{\perp X,W_1})\SD(Y^{\perp X, W_1})}\\
		&= c \SD(Y^{\perp X,W_1})/\sqrt{v_2 (1 - R^2_{W_2 \sim X,W_1})}\\
		&= c \SD(Y^{\perp X,W_1})/\sqrt{v_2 (1 - R_{W_2 \sim W_1}^2)(1 - R_{X \sim W_2 \sbullet W_1}^2)}\\
		&= c \SD(Y^{\perp X,W_1})/\sqrt{v_2 (1 - R_W'\Omega_\text{obs}^{-1}R_W )(1 - R_X^2)}\\
		&= R_Y.
	\end{align*}
	We conclude the proof by noting that $(v_2,R_W,R_X,R_Y)$ was an arbitrary element of $(0,\infty) \times \{c \in (-1,1)^{d_1} : c'\Omega_\text{obs}^{-1}c < 1\} \times (-1,1) \times (-1,1)$.
\end{proof}

\begin{proof}[Proof of Lemma \ref{lem:marg_R2_XonW2_vector}]
We have
\begin{align*}
	R_{X \sim W_2 \sbullet W_1}
	&\coloneqq \corr(X^{\perp W_1},W_2^{\perp W_1})\\
	&= \frac{\cov(X,W_2^{\perp W_1})}{\SD(X^{\perp W_1}) \SD(W_2^{\perp W_1})}\\
	&= \frac{\cov(X,W_2 - \cov(W_2,W_1)\Sigma_\text{obs}^{-1}W_1)}{\SD(X^{\perp W_1}) \SD(W_2^{\perp W_1})}\\
	&= \frac{\cov(X,W_2) - \cov(W_2,W_1)\Sigma_\text{obs}^{-1}\cov(W_1,X)}{\SD(X^{\perp W_1}) \SD(W_2^{\perp W_1})}\\
	&= \frac{\corr(X,W_2)\SD(X)\SD(W_2) - \corr(W_2,W_1)\textbf{S}\Sigma_\text{obs}^{-1}\textbf{S}\corr(W_1,X)\SD(X)\SD(W_2)}{\SD(X)\sqrt{1 - R^2_{X \sim W_1}} \SD(W_2)\sqrt{1 - R^2_{W_2 \sim W_1}}}\\
	&= \frac{R_{X \sim W_2} - R_{W_1 \sim W_2}'\Omega_\text{obs}^{-1}R_{W_1 \sim X}}{\sqrt{1 - R^2_{X \sim W_1}}\sqrt{1 - R^2_{W_2 \sim W_1}}}.
\end{align*}
Rearrange this equation to obtain the lemma's conclusion.
\end{proof}

\begin{proof}[Proof of Lemma \ref{lem:max_abstract_vector}]
Notice that $r_1$ only affects the constraint set. Therefore, for any feasible choice $(r_1,r_2)$, the value $(r_1^*(r_2), r_2)$ will also be feasible, where
\[
	r_1^*(r_2) \coloneqq \inf_{\tilde{r}_1 \in (-1,1)^{d_1} : \tilde{r}_1' \Omega_\text{obs} \tilde{r}_1 < 1} \; f(\tilde{r}_1, r_2),
\]
and these two choices lead to the same value of the objective function. Hence it suffices to solve
\begin{equation}\label{eq:concentratedOutSup}
	\sup_{r_2 \in (-1,1) : f(r_1^*(r_2),r_2) \leq \bar{r}_X } \; r_2^2.
\end{equation}
That is, we concentrate out $r_1$. First we derive $r_1^*(r_2)$. By two applications of the Cauchy–Schwarz inequality,
\begin{align*}
	&\left(r_2 \sqrt{(1-R^2_{X \sim W_1})(1-r_1'\Omega_\text{obs}^{-1}r_1)} + R_{W_1 \sim X}'\Omega_\text{obs}^{-1}r_1\right)^2\\
	&= \left(\left(r_2\sqrt{1-R^2_{X \sim W_1}}\right) \left(\sqrt{1-r_1'\Omega_\text{obs}^{-1}r_1}\right) + \left(\sqrt{R^2_{X \sim W_1}}\right)
		\left( \tfrac{R_{W_1 \sim X}'\Omega_\text{obs}^{-1}r_1}{\sqrt{R^2_{X \sim W_1}}}\right) \right)^2\\
	&\leq \left[\left(r_2\sqrt{1-R^2_{X \sim W_1}}\right)^2 + \left(\sqrt{R^2_{X \sim W_1}}\right)^2\right] 
	\times \left[\left(\sqrt{1-r_1'\Omega_\text{obs}^{-1}r_1}\right)^2 + \left(\tfrac{R_{W_1 \sim X}'\Omega_\text{obs}^{-1}r_1}{\sqrt{R^2_{X \sim W_1}}}\right)^2\right]\\
	&= \left[r_2^2(1-R^2_{X \sim W_1}) + R^2_{X \sim W_1}\right] \times \left[(1-r_1'\Omega_\text{obs}^{-1}r_1) + \frac{(R_{W_1 \sim X}'\Omega_\text{obs}^{-1}r_1)^2}{R^2_{X \sim W_1}}\right]\\
	&\leq \left[r_2^2(1-R^2_{X \sim W_1}) + R^2_{X \sim W_1}\right] \times \left[(1-r_1'\Omega_\text{obs}^{-1}r_1) + \frac{(R_{W_1 \sim X}'\Omega_\text{obs}^{-1}R_{W_1 \sim X})(r_1'\Omega_\text{obs}^{-1}r_1)}{R^2_{X \sim W_1}}\right]\\
	&= r_2^2(1-R^2_{X \sim W_1}) + R^2_{X \sim W_1},
\end{align*}
which implies that the denominator of $f(r_1,r_2)$ is non-negative. Since $r_1$ only enters the denominator, for any value of $r_2 \in (-1, 1)$, $f(r_1,r_2)$ is minimized at a value of $r_1$ satisfying
\[
	r_2 \sqrt{(1 - R^2_{X \sim W_1})(1-r_1'\Omega_\text{obs}^{-1}r_1)} + R_{W_1 \sim X}'\Omega_\text{obs}^{-1}r_1 = 0.
\]
This can be attained by setting
\[
	r_1^*(r_2) = -R_{W_1 \sim X}r_2\sqrt{\frac{1 - R^2_{X \sim W_1}}{R^2_{X \sim W_1}(R^2_{X \sim W_1} + r_2^2(1 - R^2_{X \sim W_1}))}} 
\]
which is feasible since $r_1^*(r_2)'\Omega^{-1}r_1^*(r_2) = r_2^2(1 - R^2_{X \sim W_1})/(R^2_{X \sim W_1} + r_2^2(1 - R^2_{X \sim W_1})) < 1$. Equation \eqref{eq:concentratedOutSup} is therefore simply maximization of $r_2^2$ subject to
\[
	f(r_1^*(r_2),r_2) = \frac{r_2^2(1 - R^2_{X \sim W_1})}{r_2^2(1 - R^2_{X \sim W_1}) + R^2_{X \sim W_1}} \le \rx^2.
\]
For $\rx^2 < 1$, the bound can be rearranged as
\begin{equation}\label{eq:upper_bound_x2}
	r_2^2 \le \frac{\rx^2 R^2_{X \sim W_1}}{(1 - R^2_{X \sim W_1})(1 - \rx^2)}
\end{equation}
At $\rx^2 = 1 - R^2_{X \sim W_1}$, the bound becomes $r_2^2 \le 1$, and the right hand side of \eqref{eq:upper_bound_x2} is strictly increasing in $\rx^2$, so for $\rx^2 \ge 1 - R^2_{X \sim W_1}$, $r_2^2$ is maximized at $1$. For values $\rx^2 < 1 - R_{X \sim W_1}^2$, the solution obtains by setting $r_2^2$ equal to the upper bound of equation \eqref{eq:upper_bound_x2}.
\end{proof}

\section{Proofs for Section \ref{sec:model}}\label{sec:modelProofs}

In the following proofs, let $\Lin(A \mid B) \coloneqq \Exp(AB)\Exp(BB')^{-1}B$ denote the linear projection of random vector $A$ on random column vector $B$. Note that $\Lin(A \mid 1,B) = \cov(A,B)\var(B)^{-1} B + k$ where $k \coloneqq \Exp(A) - \cov(A,B) \var(B)^{-1} \Exp(B)$ is constant. Finally, recall that $A^{\perp B} \coloneqq A - \cov(A,B) \var(B)^{-1} B$.

\begin{proof}[Proof of Proposition \ref{thm:baselineEndogControl1}]
\noindent\emph{\textbf{Part 1}}. We first show that $\beta_\text{long} = \beta_\text{med}$. 
By equation \eqref{eq:XprojectionW1W2} and assumption \ref{assump:baselineEndogControl1} ($\pi_2 = 0$), we have $X^{\perp W} = X - \pi_1'\Wob$. This yields $\cov(X,W_1) = \pi_1'\var(W_1)$. By definition of projection residuals, we then have
\begin{align*}
	X^{\perp W_1} = X - \cov(X,W_1)\var(W_1)^{-1}W_1 = X - \pi_1'W_1.
\end{align*}
Thus $X^{\perp W_1} = X^{\perp W}$. By the FWL theorem,
\[
	\beta_\text{long} = \cov(Y, X^{\perp W})/\var(X^{\perp W})
	\qquad \text{and} \qquad
	\beta_\text{med} = \cov(Y, X^{\perp \Wob})/\var(X^{\perp \Wob}).
\]
Since $X^{\perp W_1} = X^{\perp W}$, $\beta_\text{long} = \beta_\text{med}$.

\bigskip

\noindent\emph{\textbf{Part 2}}. Next we'll show that the identified set for $\gamma_1$ is $\R^{\dob}$. To show this, fix $g_1 \in \R^{\dob}$. We will construct a joint distribution of $(W_2,W_1,X,Y)$ that is consistent with (a) the distribution of observables $(Y,X,\Wob)$, (b) assumptions \ref{assump:posdefVar} and \ref{assump:baselineEndogControl1}, and (c) equation \eqref{eq:outcome} holding with $\gamma_1 = g_1$. We first consider the scalar $W_2$ case, $d_2 = 1$.

\bigskip

\noindent\textbf{Part 2(i): Construction of $(W_2,W_1,X,Y)$ joint distribution}.

\medskip

Before constructing this distribution, we begin by finding values $g_2 \in \R$ and $c \in \R^{d_1}$ such that
\begin{align}
	\cov(\Wob,Y) &=  \beta_\text{long}\cov(\Wob,X) + \var(\Wob)g_1 + g_2 c. \label{eq:thm1eq1}
\end{align}
For any fixed $g_1$, there always exists multiple values of $(g_2,c)$ satisfying this equation.
We will select specific solutions $(g_2,c)$ below.

For a given $g_1$ and $(g_2,c)$ satisfying \eqref{eq:thm1eq1}, define the $(1+\dob+2) \times (1+\dob+2)$ matrix 
\begin{align*}
	\Sigma(g_2,c)
	&= 
	\begin{pmatrix}
		1 & c' & \pi_1' c & \beta_\text{long} \pi_1' c + g_1' c + g_2\\
		c & \var(\Wob) & \cov(W_1,X) & \cov(\Wob,Y)\\
		\pi_1' c & \cov(X,W_1) & \var(X) & \cov(X,Y)\\
		\beta_\text{long} \pi_1' c + g_1' c + g_2 & \cov(Y,\Wob) & \cov(Y,X) & \var(Y)
	\end{pmatrix}.
\end{align*}
Note that the bottom right $(\dob+2)\times(\dob+2)$ block corresponds to $\var(W_1,X,Y)$, which is known. Overall, we will think of this matrix as a covariance matrix for $(W_2,W_1,X,Y)$. Specifically, we will show that for a fixed $g_1 \in \R^{\dob}$, and for a pair of values $(g_2^*, c^*)$ that satisfy equation \eqref{eq:thm1eq1}, the matrix $\Sigma(g_2^*,c^*)$ is positive definite. Then we will show how to use this matrix to construct $\Wu$ in a way that is consistent with the above equations and assumptions. 

To check whether a matrix is positive definite, we use Sylvester's criterion, which states that a matrix is positive definite if and only if all its leading principal minors have positive determinants. First, we show that the $(1+\dob) \times (1+\dob)$ matrix 
\[
	\Sigma_2(c) = \begin{pmatrix}
	1 & c^{ \prime}\\
	c & \var(\Wob)
	\end{pmatrix}
\]
is positive definite. By symmetry, $\Sigma_2(c)$ is positive definite if and only if
\[\begin{pmatrix}
	\var(\Wob) & c\\
	c^{ \prime} & 1
\end{pmatrix}\]
is positive definite. Therefore, since $\var(\Wob)$ is positive definite (by \ref{assump:posdefVar}), $\Sigma_2(c)$ is positive definite if and only if its determinant is positive. By the determinant formula for partitioned matrices, it equals
\[
	\det\begin{pmatrix}
	\var(\Wob) & c\\
	c^{ \prime} & 1
	\end{pmatrix}
	= \det(\var(\Wob)) \cdot \left(1 - c^{ \prime} \var(\Wob)^{-1} c\right).
\]
$\det(\var(\Wob)) > 0$ by \ref{assump:posdefVar}. By equation \eqref{eq:thm1eq1}, $\det(\Sigma_2(c))$ is positive if
\begin{align*}
	c^{ \prime}\var(\Wob)^{-1}c &= g_2^{-2} \|\cov(Y,\Wob) - \beta_\text{long}\cov(X,\Wob) - g_1'\var(\Wob)\|^2_{\var(\Wob)^{-1}} < 1.
\end{align*}
Thus the first $(1 + \dob)$ leading principal minors are positive definite when $g_2^2 > L(g_1)$, where 
\begin{align*}
	L(g_1) &= \|\cov(Y,\Wob) - \beta_\text{long}\cov(X,\Wob) - g_1'\var(\Wob)\|^2_{\var(\Wob)^{-1}}.
\end{align*}

Second, define the $(2+\dob)\times (2+\dob)$ matrix
\[
	\Sigma_3(c)
	=
	\begin{pmatrix}
		1 & c^{ \prime} & \pi_1' c \\
		c & \var(\Wob) & \cov(\Wob,X) \\
		\pi_1' c & \cov(X,\Wob) & \var(X) 
	\end{pmatrix}.
\]
This is the next leading principal minor of ${\Sigma}(g_2, c)$. We will show its determinant is also positive when $g_2^2 > L(g_1)$. By the partitioned matrix determinant formula,
\begin{align*}
\det(\Sigma_3(c))
	 &= \det(\Sigma_2(c)) \cdot \left(\var(X) - 
	\begin{pmatrix}
	\pi_1'c & \cov(X,\Wob)
	\end{pmatrix}
	\Sigma_2(c)^{-1}
	\begin{pmatrix}
	\pi_1'c \\
	\cov(X,\Wob)
	\end{pmatrix}
	\right).
\end{align*}
We have already shown that $\det(\Sigma_2(c)) > 0$ when $g_2^2 > L(g_1)$. Therefore, when $g_2^2 > L(g_1)$,
\begin{align*}
&\var(X) - \begin{pmatrix}
	\pi_1'c & \cov(X,\Wob)
\end{pmatrix}\Sigma_2(c)^{-1}\begin{pmatrix}
	\pi_1'c \\
	\cov(X,\Wob)
\end{pmatrix}\\[0.5em]
&=\var(X) - \frac{1}{1 - c^{ \prime}\var(\Wob)^{-1}c} \begin{pmatrix}
	\pi_1'c & \cov(X,\Wob)
\end{pmatrix} \\
&\qquad \cdot 
\begin{pmatrix}
1 & -c^{ \prime}\var(\Wob)^{-1}\\
 -\var(\Wob)^{-1}c & \var(\Wob)^{-1} (1 - c^{ \prime}\var(\Wob)^{-1}c) + \var(\Wob)^{-1}c c^{ \prime}\var(\Wob)^{-1}
\end{pmatrix}\begin{pmatrix}
	\pi_1'c \\
	\cov(\Wob,X)
\end{pmatrix} \\[0.5em]
	&= \var(X) - \cov(X,\Wob)\var(\Wob)^{-1}\cov(\Wob,X) \\
	&= \var(X^{\perp \Wob}) >0.
\end{align*}
The first line follows by the partitioned inverse formula and by $g_2^2 > L(g_1)$ (which we showed is equivalent to $c^{ \prime}\var(\Wob)^{-1}c < 1$). The last line follows by \ref{assump:posdefVar}.

Finally, we compute $\det(\Sigma(g_2,c))$. By the partitioned matrix determinant formula,
\begin{align*}
	&\det(\Sigma(g_2,c)) = \det (\Sigma_3(c)) \cdot \left(\var(Y) - \left\| \begin{pmatrix}
		\beta_\text{long} \pi_1' c + g_1' c + g_2 \\
		\cov(\Wob,Y) \\
		\cov(Y,X)
	\end{pmatrix}\right\|_{\Sigma_3(c)^{-1}}\right).
\end{align*}
We have already shown that $\det (\Sigma_3(c)) > 0$ when $g_2^2 > L(g_1)$. We finish by showing that the second term in the expression for $\det(\Sigma(g_2,c))$ is positive so long as $(g_1,g_2)$ satisfies one additional constraint. Specifically, we compute
\begin{align}
	&\var(Y) - \begin{pmatrix}
		\beta_\text{long} \pi_1' c + g_1' c + g_2 & \cov(Y,X) & \cov(Y,\Wob) 
	\end{pmatrix}\Sigma_3(c)^{-1}\begin{pmatrix}
		\beta_\text{long} \pi_1' c + g_1' c + g_2 \\
		\cov(Y,X)\\
		\cov(\Wob,Y)		
	\end{pmatrix} \notag \\
	&= \var(Y) - \beta_\text{long}^2 \var(X) + g_1' \var(\Wob)g_1 + 2\beta_\text{long} g_1' \cov(\Wob,X) \notag\\
	&\qquad - 2\beta_\text{long}\pi_1' \cov(\Wob,Y) - 2g_1' \cov(\Wob,Y) + 2\beta_\text{long}^2 \pi_1'\cov(\Wob,X) - g_2^2 \label{eq:thm1eq2}\\
	&=: U(g_1) -g_2^{ 2}. \notag
\end{align}
Thus this determinant is positive if $g_2^2 < U(g_1)$. 

Putting all of these results together, by Sylvester's criterion, we have shown that $\Sigma(g_2,c)$ is positive definite whenever $g_2$ satisfies $L(g_1) < g_2^2 < U(g_1)$. To show there exists such a $g_2 \in \R$, we show that $0 \leq L(g_1) < U(g_1)$ for all $g_1 \in \R$. To see this, first note that $L(g_1)\geq 0$ for all $g_1 \in \R$. To see that $L(g_1) < U(g_1)$ for all $g_1 \in \R$, note that
\begin{align*}
	&U(g_1) - L(g_1)\\
	&= \var(Y) - \beta_\text{long}^2 \var(X) + g_1' \var(\Wob)g_1 + 2\beta_\text{long} g_1' \cov(\Wob,X) - 2(\beta_\text{long}\pi_1 + g_1)' \cov(\Wob,Y)\\
	&\qquad  + 2\beta_\text{long}^2 \pi_1'\cov(\Wob,X) - \|\cov(Y,\Wob) - \beta_\text{long}\cov(X,\Wob) - g_1'\var(\Wob)\|^2_{\var(\Wob)^{-1}}\\[0.5em]
	&= \var(Y) - \beta_\text{long}^2 \var(X)   + \beta_\text{long}^2 \pi_1'\cov(\Wob,X) - \cov(Y,\Wob)\var(\Wob)^{-1}\cov(\Wob,Y)\\
	&= \var(Y^{\perp W_1}) - \frac{\cov(Y^{\perp W_1},X^{\perp W_1})^2}{\var(X^{\perp W_1})^2}\left(\var(X) - \cov(X,W_1)\var(W_1)^{-1}\cov(W_1,X)\right)\\
	&= \var(X^{\perp \Wob})^{-1}\left(\var(Y^{\perp \Wob})\var(X^{\perp \Wob})  - \cov(Y^{\perp \Wob},X^{\perp \Wob})^2 \right) >0.
\end{align*}
The second equality follows by distributing terms and using our equation for $\pi_1$. The third equality follows from $\beta_\text{long} = \beta_\text{med} = \cov(Y^{\perp W_1},X^{\perp W_1})/\var(X^{\perp W_1})$. The inequality follows from the Cauchy-Schwarz inequality and \ref{assump:posdefVar}, which ensures that the inequality is strict. Also note that $\var(X^{\perp \Wob}) > 0$ by \ref{assump:posdefVar}.

Therefore, let $g_2^*$ satisfy $g_2^{\ast 2} \in (L(g_1),U(g_1))$ and
\[
	c^* = (g_2^*)^{-1}(\cov(W_1,Y) - \beta_\text{long} \cov(W_1,X) - \var(W_1)g_1).
\]
Then $\Sigma(g_2^*,c^*)$ is positive definite and hence a valid covariance matrix.

\bigskip

We next use this matrix to construct the unobservable $\Wu$. Since $\Sigma(g_2^*,c^*)$ and $\var(W_1,X,Y)$ are positive definite, the partitioned inverse matrix formula yields that
\begin{align*}
	v \coloneqq 1 - \begin{pmatrix}
	c^{*\prime} & \pi_1'c^* & \beta_\text{long} \pi_1'c^* + g_1'c^* + g_2^*
	\end{pmatrix}\var(W_1,X,Y)^{-1}\begin{pmatrix}
	c^* \\ \pi_1'c^* \\ \beta_\text{long} \pi_1'c^* + g_1'c^* + g_2^*
	\end{pmatrix} > 0.
\end{align*}
Let $\widetilde{W_2}$ be a unit-variance random variable that is uncorrelated with $(W_1,X,Y)$ and let
\begin{align}\label{eq:W2construction}
	W_2 &= \begin{pmatrix}
	c^{*\prime} & \pi_1'c^* & \beta_\text{long} \pi_1'c^* + g_1'c^* + g_2^*
	\end{pmatrix}\var(W_1,X,Y)^{-1}\begin{pmatrix}
		W_1\\
		X\\
		Y
	\end{pmatrix} + v^{1/2} \widetilde{W}_2.
\end{align}
We can verify that $\var(W_2,W_1,X,Y) = \Sigma(g_2^*,c^*)$. Thus we have constructed a joint distribution of $(W_2,W_1,X,Y)$. By construction, this distribution is consistent with the distribution of $(Y,X,\Wob)$.

\bigskip

\noindent\textbf{Part 2(ii): This distribution is consistent with \ref{assump:posdefVar} and \ref{assump:baselineEndogControl1} and with equation \eqref{eq:outcome} where $\gamma_1 = g_1$}.

\bigskip

\noindent \emph{Checking \ref{assump:posdefVar}}. We showed that $\var(W_2, W_1, X, Y) = \Sigma(g_2^*, c^*)$ is positive definite. Thus \ref{assump:posdefVar} holds.

\bigskip

\noindent\emph{Checking that equation \eqref{eq:outcome} holds with $\gamma_1 = g_1$}. By construction, equation \eqref{eq:thm1eq1} holds at $(g_1, g_2^*, c^*)$:
\[
	\cov(\Wob,Y) = \beta_\text{long} \cov(\Wob,X) + \var(\Wob)g_1 + g_2^*c^*.
\]
Substituting this in our expression for $\beta_\text{long}$, we obtain
\begin{align*}
	\cov(Y,X)
	&= \beta_\text{long} \var(X^{\perp \Wob}) + \cov(Y,\Wob)\pi_1\\
	&=\beta_\text{long} \var(X) + g_1'\var(\Wob)\pi_1 + g_2^* \pi_1'c^*.
\end{align*}
Directly from $\Sigma(g_2^*,c^*)$, we have that $\cov(Y,W_2) = \beta_\text{long}\pi_1' c^* + g_1' c^* + g_2^*$. We use these to calculate
\begin{align}
	Y^{\perp X,W}& = Y - \begin{pmatrix}
	\cov(Y,\Wu)  & \cov(Y,X)& \cov(Y,\Wob)
	\end{pmatrix}\begin{pmatrix}
		1  & \pi_1' c^* & c^{* \prime} \\
		\pi_1' c^* & \var(X) & \cov(X,\Wob) \\
		c^* & \cov(\Wob,X) & \var(W_1) 
	\end{pmatrix}^{-1} \begin{pmatrix}
		\Wu\\
		X\\
		W_1
	\end{pmatrix} \notag\\
	&= Y - (\beta_\text{long} X + g_1' \Wob + g_2^* \Wu)  \label{eq:thm1eq3}
	\end{align}
This implies equation \eqref{eq:outcome} holds with $\gamma_1 = g_1$.

\bigskip

\emph{Checking \ref{assump:baselineEndogControl1}}. From the definition of $\Sigma(g_2^*,c^*)$, we have that $\cov(X,W_2) = \pi_1'c^*$. We also have $\cov(X,W_1) = \pi_1' \var(\Wob)$. Using this, we next show that $\pi_2 = 0$. We have
\begin{align*}
	X^{\perp W} &= X - \begin{pmatrix}
	\cov(X,\Wu) & \cov(X,\Wob) 
	\end{pmatrix}\begin{pmatrix}
		1 & c^{* \prime} \\
		c^* & \var(\Wob) 
	\end{pmatrix}^{-1} \begin{pmatrix}
		\Wu\\
		\Wob
	\end{pmatrix} 
	= X - (0 \cdot \Wu + \pi_1' \Wob).
\end{align*}
Thus we see that equation \eqref{eq:XprojectionW1W2} holds with $\pi_2 = 0$.

\bigskip

\noindent\textbf{Part 2(iii): Generalization to $d_2 > 1$}.
If $W_2$ is multivariate, the previous proof can still be applied to show the identified set for $\gamma_1$ is $\R^{d_1}$. To see this, we need to exhibit a joint distribution of $(W_2,W_1,X,Y)$ that satisfies the requirements of Part 2(ii). This can be achieved by defining $W_2^{(1)}$, the first element of $W_2$, to be the one defined in equation \eqref{eq:W2construction}, and letting the remaining $d_2 - 1$ elements all be uncorrelated with $(W_2^{(1)},W_1,X,Y)$, and each other. Then $Y^{\perp X,W_1,W_2} = Y^{\perp X,W_1,W_2^{(1)}}$ and $X^{\perp W_1,W_2} = X^{\perp W_1,W_2^{(1)}}$. Therefore $d_2 > 1$ leads to the same identified set as $d_2 = 1$. This concludes the proof. 
\end{proof}

\section{Proofs for Sections \ref{sec:MainNewAnalysis} and \ref{sec:endogeneity_restrictions}}\label{sec:proofsForMainNewAnalysis}

\subsection{Proof of Alternative Expressions for $r_X$}

We first show the validity of equations \eqref{eq:rx_interpretation_scalarW_1W_2} and \eqref{eq:rx_interpretation_vectorW_1W_2}. We begin by showing equation \eqref{eq:rx_interpretation_vectorW_1W_2}. We have that
	\begin{align*}
		&\var(Y)(R^2_{X \sim W_1,W_2} - R^2_{X \sim \pi_1'W_1}) \\
		&= \var(\pi_1' W_1 + \pi_2' W_2) - \frac{\cov(X,\pi_1'W_1)^2}{\var(\pi_1'W_1)}\\
		&= \var(\pi_1' W_1 + \pi_2' W_2) - \frac{\cov(\pi_1' W_1 + \pi_2' W_2,\pi_1'W_1)^2}{\var(\pi_1'W_1)}\\
		&= \var(\pi_2'W_2)\left(1 - \frac{\cov(\pi_2'W_2,W_1)^2}{\var(\pi_1'W_1)\var(\pi_2'W_2)}\right)\\
		&= \var(\pi_2'W_2)\left(1 - R^2_{\pi_1'W_1 \sim \pi_2'W_2}\right).
	\end{align*}
A symmetric argument reveals that
\[
	\var(Y)(R^2_{X \sim W_1,W_2} - R^2_{X \sim \pi_1'W_1}) = \var(\pi_1'W_1)\left(1 - R^2_{\pi_1'W_1 \sim \pi_2'W_2}\right).
\]
Taking their ratio yields equation \eqref{eq:rx_interpretation_vectorW_1W_2}. 
	
	We now show equation \eqref{eq:rx_interpretation_vectorW_1W_2} implies equation \eqref{eq:rx_interpretation_scalarW_1W_2} when $W_1$ and $W_2$ are scalar. First note that, for $j \in \{1,2\}$, $R^2_{X \sim \pi_jW_j} = R^2_{X \sim W_j}$ whenever $\pi_j \neq 0$. Therefore, \eqref{eq:rx_interpretation_scalarW_1W_2} holds whenever $\pi_1 \neq 0$ and $\pi_2 \neq 0$. We now consider the cases where $\pi_1$ or $\pi_2$ are zero. If $\pi_1 = 0$, then both sides of  \eqref{eq:rx_interpretation_scalarW_1W_2} are undefined, which we rule out. If $\pi_2 = 0$, then $r_X = 0$ and $R^2_{X \sim W_1,W_2} = R^2_{X \sim W_1}$. Moreover, $R^2_{X \sim W_1,W_2} - R^2_{X \sim \pi_2 W_2} = R^2_{X \sim W_1} - R^2_{X \sim 0} = R^2_{X \sim W_1} > 0$. Therefore, equation \eqref{eq:rx_interpretation_scalarW_1W_2} holds in this case as well.

\subsection{Useful Definitions and Lemmas}

\textbf{Important Notation Remark}: In this section we use $r_X$ and $r_Y$ to denote related but different objects than the selection ratios defined in the main paper.

We impose the normalizations $\var(X) = 1$ and $\var(W_1) = I$ throughout this section. We also let $W_2 \in \R$ and $\var(W_2) = 1$. We use the notation $\sigma_{A,B} = \cov(A,B)$, and let $(k_0,k_1,k_2) = (\var(X^{\perp W_1}), \cov(Y^{\perp W_1},X^{\perp W_1}), \var(Y^{\perp W_1}))$. We begin by providing a simplified characterization of the set $\mathcal{B}(r_X,r_Y,c)$.

\begin{lemma}\label{lemma:IDsetSimplification}
Let $\var(Y, X, W_1)$ be positive definite. Then $b \in \mathcal{B}(r_X,r_Y,c)$ if and only if the following six equations hold for some $(p_1,g_1) \in \R^{d_1}\times \R^{d_1}$:
\begin{align}
	(p_1'r_X)(g_1'r_Y)(1 - \|c\|^2) &= k_1 - b k_0 \label{eq:IDseteqsimp1}\\
	(I + cr_Y')g_1 &= \sigma_{W_1,Y} - b \sigma_{W_1,X} \label{eq:IDseteqsimp2}\\
	(I + cr_X')p_1 &= \sigma_{W_1,X} \label{eq:IDseteqsimp3}\\
	(g_1'r_Y)^2(1 - \|c\|^2) &< k_0(\beta_\text{med} - b)^2 + \var(Y^{\perp X,W_1}) \label{eq:IDseteqsimp4}\\
	(p_1'r_X)^2(1 - \|c\|^2) &< k_0 \label{eq:IDseteqsimp5} \\
	\|c\|^2 &< 1.\label{eq:IDseteqsimp6}
\end{align}
\end{lemma}

\begin{proof}
	First note that equations \eqref{eq:IDseteq2}--\eqref{eq:IDseteq3} are equivalent to  \eqref{eq:IDseteqsimp2}--\eqref{eq:IDseteqsimp3} by applying the normalization $\var(W_1) = I$. Second, we can see that \eqref{eq:IDseteq1} is equivalent to \eqref{eq:IDseteqsimp1} by
\begin{align*}
    \sigma_{XY} - b &= p_1'(I + r_Xc' + cr_Y' + r_Xr_Y')g_1 \\
    &= p_1'(I + r_Xc')(I + cr_Y')g_1 + (p_1'r_X)(g_1'r_Y)(1-\|c\|^2)\\
    &= \sigma_{W_1,X}'(\sigma_{W_1,Y} - b \sigma_{W_1,X}) + (p_1'r_X)(g_1'r_Y)(1 - \|c\|^2).
\end{align*}
Third, we show that \eqref{eq:IDseteq4} is equivalent to \eqref{eq:IDseteqsimp4} since \eqref{eq:IDseteq4} can be written as
\begin{align*}
	\sigma_Y^2&> b^2 + g_1'\left(I + r_Yr_Y' + 2r_Yc'\right)g_1  + 2b g_1'(I + cr_X' + r_Yc' + r_Y r_X')p_1\\
	&= b^2 + \|g_1'(I+ r_Yc')\|^2 + (1-\|c\|^2)(g_1'r_Y)^2 + 2b(\sigma_{XY} - b)\\
	&= b^2 + \| \sigma_{W_1,Y} \|^2 + b^2 \| \sigma_{W_1,X} \|^2 - 2 b \sigma_{W_1,X}' \sigma_{W_1,Y} + (1-\|c\|^2)(g_1'r_Y)^2 + 2b(\sigma_{XY} - b).
\end{align*}
This inequality is equivalent to
\begin{align*}
	(1-\|c\|^2)(g_1'r_Y)^2 &< \sigma_Y^2- b^2 - \| \sigma_{W_1,Y} \|^2 - b^2 \| \sigma_{W_1,X} \|^2 + 2 b \sigma_{W_1,X}' \sigma_{W_1,Y} - 2b(\sigma_{XY} - b)\\
	&= k_0(\beta_\text{med} - b)^2 + k_0\left(\frac{k_2}{k_0} - \beta_\text{med}^2\right).
\end{align*}
This inequality, in turn, is equivalent to equation \eqref{eq:IDseteqsimp4} since, by properties of projections, $k_2 - k_0\beta_\text{med}^2 = \var(Y^{\perp W_1}) - \frac{\cov(Y^{\perp W_1},X^{\perp W_1})^2}{\var(X^{\perp W_1})} = \var(Y^{\perp X,W_1})$.

Fourth, \eqref{eq:IDseteq5} is equivalent to \eqref{eq:IDseteqsimp5} since $p_1'(I + r_Xr_X' + 2cr_X')p_1 = \|\sigma_{W_1,X}\|^2 + (p_1'r_X)^2(1 - \|c\|^2)$ and by rearranging terms. Finally, \eqref{eq:IDseteq6} is equivalent to \eqref{eq:IDseteqsimp6} by $\var(W_1) = I$.
\end{proof}

Our previous lemma characterized $\mathcal{B}(r_X, r_Y, c)$. The next lemma characterizes every element of this set which is not equal to $\beta_\text{med}$. By treating the element $\beta_\text{med}$ separately, we obtain an even simpler characterization of the elements in $\mathcal{B}(r_X, r_Y, c)$. Let $\text{devsq}(z) \coloneqq \frac{\var(Y^{\perp X, W_1})}{\var(X^{\perp W_1})} \frac{z^2}{\var(X^{\perp W_1}) - z^2}$.

\begin{lemma}\label{lemma:IDsetAlt}
Let $\var(Y, X, W_1)$ be positive definite. Let $b \ne \beta_\text{med}$. Then $b \in
  \mathcal{B}(r_X,r_Y,c)$ if and only if there exists
  $z \in \mathbb{R} \setminus \{0\}$ such that the following hold:
  \begin{align}
    k_1 - b k_0 &= r_Y'\left(z \sqrt{1 - \|c\|^2}(\sigma_{W_1,Y} - b\sigma_{W_1,X}) - (k_1 -bk_0)c\right) \label{eq:IDsetEqAltA}  \\
    z &= r_X'\left(\sqrt{1 - \|c\|^2}\sigma_{W_1,X} - cz\right) \label{eq:IDsetEqAltB} \\
    z^2 &< k_0 \label{eq:IDsetIneqZx} \\
    (b - \beta_\text{med})^2 &< \text{devsq}(z)  \label{eq:IDsetIneqBeta}\\
    \|c\|^2 &< 1.\label{eq:IDsetIneq_c}
\end{align}
\end{lemma}

\begin{proof}[Proof of Lemma \ref{lemma:IDsetAlt}]

\bigskip
  
\textbf{Step 1 ($\Rightarrow$).} First we show that $b \in \mathcal{B}(r_X,r_Y,c)$ implies \eqref{eq:IDsetEqAltA}--\eqref{eq:IDsetIneq_c} hold for some $z \in \mathbb{R} \setminus \{0\}$.

\bigskip

By Lemma \ref{lemma:IDsetSimplification}, $b \in \mathcal{B}(r_X,r_Y,c)$ is equivalent to the existence of $(p_1,g_1) \in \R^{d_1} \times \R^{d_1}$ such that equations \eqref{eq:IDseteqsimp1}--\eqref{eq:IDseteqsimp6} hold. Fix such $(p_1,g_1) \in \R^{d_1} \times \R^{d_1}$ and let $z = p_1'r_X \sqrt{1 - \|c\|^2}$.

First we show that $z \neq 0$. By equation \eqref{eq:IDseteqsimp6}, $\sqrt{1-\|c\|^2} \neq 0$. By equation \eqref{eq:IDseteqsimp1} and $b \neq \beta_\text{med} = k_1/k_0$, $p_1'r_X \neq 0$. Therefore $z \neq 0$.

Second, we show that \eqref{eq:IDseteqsimp1} implies \eqref{eq:IDsetEqAltA}. Multiplying both sides of \eqref{eq:IDseteqsimp1} by $(1 + r_Y'c)$, we have
\begin{align*}
	(k_1 - bk_0)(1 + r_Y'c) &= (p_1'r_X)(1 - \|c\|^2)r_Y'(g_1(1 + r_Y'c)) \\
    &= r_Y'(\sigma_{W_1,Y} - b \sigma_{W_1,X})(p_1'r_X)(1 - \|c\|^2) \\
    &= r_Y'(\sigma_{W_1,Y} - b \sigma_{W_1,X})z\sqrt{1 - \|c\|^2}.
\end{align*}
The second line follows from \eqref{eq:IDseteqsimp2} and the last line follows by the definition of $z$. Rearranging the last equality yields \eqref{eq:IDsetEqAltA}.

Third, we show that \eqref{eq:IDsetEqAltB} holds for this choice of $z$. Note that $p_1 = \sigma_{W_1,X} - c(p_1'r_X)$. Thus
\begin{align*}
	z &= r_X'p_1 \sqrt{1 - \|c\|^2}  = r_X'(\sigma_{W_1,X} - c(p_1'r_X))\sqrt{1 - \|c\|^2}   = r_X'(\sigma_{W_1,X} \sqrt{1 - \|c\|^2} - cz).
\end{align*}
Fourth, equation \eqref{eq:IDseteq5} implies \eqref{eq:IDsetIneqZx} by substituting $p_1'r_X \sqrt{1 - \|c\|^2} = z$, which gives $z^2 < k_0$. Fifth, we show equation \eqref{eq:IDsetIneqBeta} holds. From equation \eqref{eq:IDseteqsimp4}, we can write
\begin{align*}
    k_0(\beta_\text{{med}} - b)^2 + \var(Y^{\perp X,W_1}) &> (g_1'r_Y)^2(1 - \|c\|^2) = \frac{(k_1 - bk_0)^2}{(p_1'r_X)^2(1 - \|c\|^2)} = \frac{k_0^2(\beta_\text{{med}} - b)^2}{z^2}
\end{align*}
where the first equality follows from \eqref{eq:IDseteqsimp1}. Rearranging this inequality and recalling that $k_0 > z^2$, we obtain \eqref{eq:IDsetIneqBeta}. Finally, note that equations \eqref{eq:IDseteqsimp6} and \eqref{eq:IDsetIneq_c} are identical.

\bigskip

\noindent\textbf{Step 2 ($\Leftarrow$).} We show that if equations \eqref{eq:IDsetEqAltA}--\eqref{eq:IDsetIneq_c} hold for some $z \in \mathbb{R} \setminus \{0\}$ then $b \in \mathcal{B}(r_X,r_Y,c)$.
   
Let
\begin{align*}
	p_1 &= \sigma_{W_1,X} - c\frac{z}{\sqrt{1 - \|c\|^2}}  \qquad \text{and} \qquad   g_1 = \sigma_{W_1,Y} - b \sigma_{W_1,X} - c \frac{k_1 - b k_0}{z\sqrt{1 - \|c\|^2}}.
\end{align*}
First, we verify \eqref{eq:IDseteqsimp1}. By the definition of $p_1$ and equation \eqref{eq:IDsetEqAltB}, $p_1'r_X \sqrt{1 - \|c\|^2} = z$. Similarly, by the definition of $g_1$ and equation \eqref{eq:IDsetEqAltA}, $g_1'r_Y \sqrt{1 - \|c\|^2} = (k_1 - b k_0)/z$. Therefore $(p_1'r_X)(g_1'r_Y)(1-\|c\|^2) = z  \cdot (k_1 - bk_0)/z = k_1 - bk_0$ hence equation \eqref{eq:IDseteqsimp1} holds.

Second, using equation \eqref{eq:IDsetEqAltA}, we verify \eqref{eq:IDseteqsimp2}:
\begin{align*}
	(I + cr_Y')g_1 &= (I + cr_Y')\left(\sigma_{W_1,Y} - b \sigma_{W_1,X} - c \frac{k_1 - b k_0}{z\sqrt{1 - \|c\|^2}}\right) \\
	&= \sigma_{W_1,Y} - b \sigma_{W_1,X} + c\left((\sigma_{W_1,Y} - b \sigma_{W_1,X})'r_Y\right) - c \frac{(1 + r_Y'c)(k_1 - b k_0)}{z\sqrt{1 - \|c\|^2}} \\
	&= \sigma_{W_1,Y} - b \sigma_{W_1,X}.        
\end{align*}

Third, using equation \eqref{eq:IDsetEqAltB}, we verify \eqref{eq:IDseteqsimp3}:
\begin{align*}
	(I + cr_X')p_1 &= \sigma_{W_1,X} + c(r_X'\sigma_{W_1,X}) - c \frac{(1 + r_X'c)z}{\sqrt{1 - \|c\|^2}} = \sigma_{W_1,X} + c(r_X'\sigma_{W_1,X}) - c  (r_X'\sigma_{W_1,X} ) = \sigma_{W_1,X}.
\end{align*}

Fourth, since $z^2 < k_0$ by inequality \eqref{eq:IDsetIneqZx}, inequality \eqref{eq:IDsetIneqBeta} yields after rearrangement $(k_1 - bk_0)^2k_0 < z^2(k_1 - b k_0)^2 + z^2(k_2k_0 - k_1^2)$. Using this inequality we find that
\begin{align*}
	(g_1'r_Y)^2(1-\|c\|^2) &= \frac{k_0(k_1 - b k_0)^2}{k_0 z^2}< \frac{z^2(k_1 - b k_0)^2 + z^2(k_2k_0 - k_1^2)}{k_0 z^2}
	= k_0(b - \beta_\text{{med}})^2 + \var(Y^{\perp X,W_1}).
\end{align*}
Therefore inequality \eqref{eq:IDseteqsimp4} holds. 

Finally, we see that $(p_1'r_X)^2(1-\|c\|^2) = z^2 < k_0$ by equation \eqref{eq:IDsetIneqZx}. Therefore inequality \eqref{eq:IDseteqsimp5} holds. Note that equations \eqref{eq:IDseteqsimp6} and \eqref{eq:IDsetIneq_c} are identical. We have shown that equations \eqref{eq:IDseteqsimp1}--\eqref{eq:IDseteqsimp6} hold for given values of $(p_1,g_1)$. Hence $b \in \mathcal{B}(r_X,r_Y,c)$ by Lemma \ref{lemma:IDsetSimplification}.
\end{proof}

For the next several lemmas, we use the function
\[
	z_X(r_X,c) \coloneqq \frac{r_X'\sigma_{W_1,X} \sqrt{1 - \|c\|^2}}{1 + r_X'c}.
\]
This function is well defined for all $(r_X,c)$ such that $r_X'c \ne -1$. When $r_X'c \neq -1$, equation \eqref{eq:IDsetEqAltB} is equivalent to $z = z_X(r_X,c)$. The following lemma examines the set $\mathcal{B}(r_X, r_Y, c)$ when $r_Y = 0$.
 
\begin{lemma}\label{lemma:IDsetAltCase2}
Let $\var(Y, X, W_1)$ be positive definite. Let $\|c\| < 1$. Then
\[
	\mathcal{B}(r_X,0,c) = \{\beta_{\text{med}}\}
\]
if and only if there exists a $p_1 \in \R^{d_1}$ such that equations \eqref{eq:IDseteqsimp3} and \eqref{eq:IDseteqsimp5} hold. In particular, $\mathcal{B}(r_X,0,c) = \{\beta_\text{med}\}$ for all $(r_X,c)$ such that $r_X'c \ne -1$ and $z_X(r_X,c)^2 < k_0$.
\end{lemma} 

\begin{proof}[Proof of Lemma \ref{lemma:IDsetAltCase2}]
($\Rightarrow$) Suppose $\mathcal{B}(r_X,0,c) = \{\beta_\text{med}\}$. By Lemma \ref{lemma:IDsetSimplification}, there exists a $p_1 \in \R^{d_1}$ such that \eqref{eq:IDseteqsimp3} and \eqref{eq:IDseteqsimp5} hold. 

($\Leftarrow$) Conversely, let $p_1 \in \R^{d_1}$ be such that \eqref{eq:IDseteqsimp3} and \eqref{eq:IDseteqsimp5} hold. Then, letting $b = \beta_\text{med}$ and $g_1 = \sigma_{W_1,Y} - \beta_\text{med}\sigma_{W_1,X}$, we can see that equations \eqref{eq:IDseteqsimp1}--\eqref{eq:IDseteqsimp3} and \eqref{eq:IDseteqsimp5}--\eqref{eq:IDseteqsimp6} hold. Under these choices, equation \eqref{eq:IDseteqsimp4} is equivalent to $0 < \var(Y^{\perp X,W_1})$, which holds the positive definiteness of $\var(Y,X,W_1)$. Thus equations \eqref{eq:IDseteqsimp1}--\eqref{eq:IDseteqsimp6} hold, and $\beta_\text{med} \in \mathcal{B}(r_X,0,c)$ by Lemma \ref{lemma:IDsetSimplification}. Lastly, note that any $b \in \mathcal{B}(r_X,0,c)$ must satisfy equation \eqref{eq:IDseteqsimp1}. Setting $r_Y = 0$ in this equation implies that such a $b$ must equal $\beta_\text{med}$. Altogether we have shown that $\{\beta_\text{med}\} = \mathcal{B}(r_X,0,c)$.

We conclude by proving the final sentence of the lemma. Suppose $(r_X,c)$ are such that $r_X'c \neq -1$ and $z_X(r_X,c)^2 < k_0$. Then, since $I + r_Xc'$ is invertible when $r_X'c \neq -1$ (\citealt{AbadirMagnus2005}, exercise 4.28 with $A = I$), we can let $p_1 = (I + cr_X')^{-1}\sigma_{W_1,X}$. Then $r_X'\sigma_{W_1,X} = r_X' (I + cr_X')p_1 =  r_X'p_1(1 + c'r_X)$, and therefore $r_X'p_1 = (1 + r_X'c)^{-1}r_X'\sigma_{W_1,X}$. Thus
\begin{align*}
	(r_X'p_1)^2(1-\|c\|^2) &= (1 + r_X'c)^{-2}(r_X'\sigma_{W_1,X})^2(1-\|c\|^2) = z_X(r_X,c)^2 < k_0.
\end{align*}
Hence equation \eqref{eq:IDseteqsimp5} holds. By the first part of the proof we thus have $\mathcal{B}(r_X,0,c) = \{ \beta_\text{med} \}$.
\end{proof}

Let
\[
	A(\rx,\underline{c},\bar{c}) \coloneqq \{(r_X,c, b, z) \in \mathbb{R}^{d_1} \times \mathbb{R}^{d_1} \times \R \times \mathbb{R}\setminus\{0\}: \|r_X\| \leq \rx, \|c\| \in [\underline{c},\bar{c}], \text{\eqref{eq:IDsetEqAltB}--\eqref{eq:IDsetIneq_c} hold} \}.
\]
This is the set of $(r_X, c, b, z)$ that satisfies four of the five equations in the characterization of $\mathcal{B}(r_X, r_Y, c)$ given in Lemma \ref{lemma:IDsetAlt}. Our next lemma provides an analytical characterization of the smallest value of $\| r_Y \|$ such that the fifth and final equation (numbered \eqref{eq:IDsetEqAltA}) in the characterization of Lemma \ref{lemma:IDsetAlt} holds for that $r_Y$. Recall that the function $\underline{r}_Y(z,c,b)$ is defined just prior to the statement of Theorem \ref{cor:BFCalculation3D} in section \ref{sec:identrYtoo}.

\begin{lemma}
\label{lemma:IDsetRy_claim1}
Let $\var(Y, X, W_1)$ be positive definite. Fix $(r_X,c, b,z) \in A(\rx,\underline{c},\bar{c})$. Then
\[
	 \inf \left\{\|r_Y\|: r_Y \in \mathbb{R}^{d_1},  k_1 - b k_0 = r_Y'\left(z \sqrt{1 - \|c\|^2}(\sigma_{W_1,Y} - b\sigma_{W_1,X}) - (k_1 -bk_0)c\right)\right\} = \underline{r}_Y(z,c,b).
\]
\end{lemma}

\begin{proof}[Proof of Lemma \ref{lemma:IDsetRy_claim1}]
Recall that the equation in the infimum is equation \eqref{eq:IDsetEqAltA}. And recall
\begin{align*}
	\mathcal{D} &= \mathbb{R} \times \{c \in \mathbb{R}^{d_1}: \|c\| < 1\} \times \mathbb{R}\\
	\mathcal{D}^0 &= \{(z,c,b) \in \mathcal{D}: z\sqrt{1 - \|c\|^2}(\sigma_{W_1,Y} - b\sigma_{W_1,X}) - (k_1 - k_0b)c \ne 0 \}.
\end{align*}

We compute the infimum in the lemma by considering three different cases.

\medskip

\noindent\emph{\textbf{Case 1.}} $b = \beta_\text{med}$. In this case, $b = \beta_\text{med}$ implies $k_1 - bk_0 = 0$. So when $r_Y = 0$, equation \eqref{eq:IDsetEqAltA} holds for all $(z,c)$. Thus the infimum equals 0.

\medskip

\noindent\emph{\textbf{Case 2.}} $b \neq \beta_{\text{med}}$ and $(z, c, b) \in \mathcal{D} \setminus \mathcal{D}^0$. In this case, $b \neq \beta_\text{med}$ implies $k_1 - bk_0 \neq 0$. Also, $(z,c,b) \notin \mathcal{D}^0$ implies $z\sqrt{1 - \|c\|^2}(\sigma_{W_1,Y} - b\sigma_{W_1,X}) - (k_1 -bk_0)c = 0$. So there are no $r_Y$ for which equation \eqref{eq:IDsetEqAltA} holds. Thus the infimum of $\|r_Y\|$ is $+\infty$.

\medskip

\noindent \emph{\textbf{Case 3.}} $b \neq \beta_{\text{med}}$ and $(z, c, b) \in  \mathcal{D}^0$. Applying the Cauchy-Schwarz inequality to equation \eqref{eq:IDsetEqAltA}, we find that 
\begin{align*}
	|k_1 - bk_0|
	&= \left|r_Y' \left(z\sqrt{1 - \|c\|^2}(\sigma_{W_1,Y} - b\sigma_{W_1,X}) - c(k_1 -bk_0)\right)\right|\\
	&\leq \|r_Y\|\left\|z\sqrt{1 - \|c\|^2}(\sigma_{W_1,Y} - b\sigma_{W_1,X}) - c(k_1 - bk_0)\right\|.
\end{align*}
Rearranging this inequality gives
\[
	\|r_Y\| \geq \left\|z\sqrt{1 - \|c\|^2}(\sigma_{W_1,Y} - b\sigma_{W_1,X}) - (k_1 - k_0b)c\right\|^{-1}|k_1 - k_0b|,
\]
where the denominator is nonzero since $(z, c, b) \in \mathcal{D}^0$. This lower bound is attained by letting $r_Y = r_Y^*(z,c,b)$ where
\begin{align*}
	r^*_Y(z,c,b) &\coloneqq (k_1 - b k_0)\frac{z\sqrt{1 - \|c\|^2}(\sigma_{W_1,Y} - b\sigma_{W_1,X}) - (k_1 - bk_0)c}  {\left\|z\sqrt{1 - \|c\|^2}(\sigma_{W_1,Y} - b\sigma_{W_1,X}) - (k_1 - bk_0)c\right\|^2}.
\end{align*}
By construction, equation \eqref{eq:IDsetEqAltA} is satisfied when $r_Y = r_Y^*(z,c,b)$.
Since it attains the lower bound,
\[
	\inf\left\{ \|r_Y\|: r_Y \in \mathbb{R}^{d_1}, \eqref{eq:IDsetEqAltA} \text{ holds for } (r_Y, z, c, b)\right\} = \|r_Y^{*}(z,c,b)\|
\]
Finally, note that $\|r^*_Y(z,c,b)\| = \underline{r}_Y(z,c,b)$.
\end{proof}

To the state the next lemma, define the following bounds for $b$:
\[
	\underline{b}(z) \coloneqq \beta_{\text{med}} - \sqrt{\text{devsq}(z)}
	\qquad \text{and} \qquad
	\overline{b}(z) \coloneqq \beta_{\text{med}} + \sqrt{\text{devsq}(z)}.
\]

\begin{lemma}
\label{lemma:IDsetRy_claim2}
Let $\var(Y, X, W_1)$ be positive definite. Let $\sigma_{W_1,Y} \ne \sigma_{X,Y}\sigma_{W_1,X}$ and $\| c \| < 1$. For any $(r_X,c)$ such that $r_X'c \ne -1$ and $0 < z_X(r_X,c)^2 < k_0$, we have
\[
    \bigcup_{r_Y \in \mathbb{R}^{d_1}}\mathcal{B}(r_X,r_Y,c)
    = \left(\underline{b}(z_X(r_X,c)), \overline{b}(z_X(r_X,c))\right) \setminus \mathcal{B}^0(z_X(r_X,c),c)
\]
where $\mathcal{B}^0(z_X(r_X,c),c)$ is a set containing at most one point.
\end{lemma}

\begin{proof}[Proof of Lemma \ref{lemma:IDsetRy_claim2}]
Recall that by Lemma \ref{lemma:IDsetAlt}, for $b \ne \beta_\text{med}$, $b \in \mathcal{B}(r_X,r_Y,c)$ if and only if \eqref{eq:IDsetEqAltA}--\eqref{eq:IDsetIneq_c} hold for some $z \in \mathbb{R}\setminus \{0\}$. We will show two characterizations:
\begin{enumerate}[wide, labelwidth=!, labelindent=0pt]
\item First, given $r_X'c \ne -1$, $z_X(r_X,c)^2 < k_0$, and $\| c \| < 1$, \eqref{eq:IDsetEqAltB} and \eqref{eq:IDsetIneqZx} hold if and only if $z = z_X(r_X,c)$. This can be obtained directly from equation \eqref{eq:IDsetEqAltB} and $1 + r_X'c \neq 0$.

\item Next, assuming $z = z_X(r_X,c)$, \eqref{eq:IDsetIneqBeta} can be written as $(b - \beta_\text{med})^2 < \text{devsq}(z_X(r_X,c))$, which is equivalent to $b \in \big( \underline{b}(z_X(r_X,c)),\overline{b}(z_X(r_X,c)) \big)$.
\end{enumerate}

Therefore, $b \neq \beta_\text{med}$ satisfies $b \in \mathcal{B}(r_X,r_Y,c)$ if and only if $b \in \big( \underline{b}(z_X(r_X,c)),\overline{b}(z_X(r_X,c)) \big)$  and $k_1 - bk_0 = r_Y' \left(z_X(r_X,c)\sqrt{1 - \|c\|^2}(\sigma_{W_1,Y} - b\sigma_{W_1,X}) - c(k_1 -bk_0)\right)$. 

We now compute $\cup_{r_Y \in \mathbb{R}^{d_1}}\mathcal{B}(r_X,r_Y,c)$. Focusing first on the case where $b \neq \beta_\text{med}$, we obtain
\begin{align*}
	&\bigcup_{r_Y \in \mathbb{R}^{d_1}}\mathcal{B}(r_X,r_Y,c) \setminus \{ \beta_\text{med} \} \\
	&= \bigcup_{r_Y \in \mathbb{R}^{d_1}} \bigg\{b \in \left(\underline{b}(z_X(r_X,c)),\overline{b}(z_X(r_X,c))\right) \setminus \{\beta_\text{med}\}:   \\
    &\qquad\qquad\qquad  k_1 - bk_0 =  r_Y' \left(z_X(r_X,c)\sqrt{1 - \|c\|^2}(\sigma_{W_1,Y} - b\sigma_{W_1,X}) - c(k_1 -bk_0)\right)\bigg\}\\
    &= \left\{ b \in \left(\underline{b}(z_X(r_X,c)),\overline{b}(z_X(r_X,c))\right)\setminus \{ \beta_\text{med} \}: (z_X(r_X,c), c, b) \in \mathcal{D}^0\right\}\\
    &= \left(\underline{b}(z_X(r_X,c)),\overline{b}(z_X(r_X,c))\right) \setminus (\{b \in \R: (z_X(r_X,c),c,b) \in \mathcal{D} \setminus \mathcal{D}^0\}\cup \{\beta_\text{med}\})\\
    &=: \left(\underline{b}(z_X(r_X,c)),\overline{b}(z_X(r_X,c))\right) \setminus ( \mathcal{B}^0(z_X(r_X,c),c) \cup \{\beta_\text{med}\}).
\end{align*}
The second equality follows since, as we showed in the proof of Lemma \ref{lemma:IDsetRy_claim1}, for any $(r_X,c,b,z)$ such that $b \neq \beta_\text{med}$ and such that \eqref{eq:IDsetEqAltB}--\eqref{eq:IDsetIneq_c} hold, $\{r_Y \in \R^{d_1}: \eqref{eq:IDsetEqAltA} \text{ holds for } (r_Y,c,b,z) \} \neq \emptyset$ if and only if $(z, c, b) \in \mathcal{D}^0$. In the last line we defined
\[
	\mathcal{B}^0(z_X(r_X, c),c) \coloneqq \{b \in \R: (z_X(r_X,c),c,b) \in \mathcal{D} \setminus \mathcal{D}^0\}.
\]
We next show this set is either a singleton or empty. Recall that $(z,c,b) \in \mathcal{D} \setminus \mathcal{D}^0$ if and only if
\begin{align*}
	0 &= z\sqrt{1 - \|c\|^2}(\sigma_{W_1,Y} - b \sigma_{W_1,X}) - (k_1 - bk_0)c\\
	&= \left(z\sqrt{1 - \|c\|^2}\sigma_{W_1,Y} - k_1c\right) - b \left(z\sqrt{1 - \|c\|^2}\sigma_{W_1,X} - k_0 c\right)\\
	&=: A_0(z,c) - b A_1(z,c).
\end{align*}
For a given $(z,c)$ this is system of linear equations in $b$. Thus it has at most one solution unless the vectors $A_0(z,c)$ and $A_1(z,c)$ satisfy $A_0(z,c) = A_1(z,c) = 0$. This happens if and only if
\begin{align*}
	z\sqrt{1 - \|c\|^2}\sigma_{W_1,Y} &= k_1c \qquad \text{and} \qquad    z\sqrt{1 - \|c\|^2}\sigma_{W_1,X} = k_0c.
\end{align*}
Since $z \neq 0$ in this set and $\|c\| < 1$, we have $z \sqrt{1 - \|c\|^2} \neq 0$. Thus $A_0(z,c) = A_1(z,c) = 0$ implies  $\sigma_{W_1,Y}$ is proportional to $\sigma_{W_1,X}$, which is ruled out by $\sigma_{W_1,Y} \neq \sigma_{X,Y}\sigma_{W_1,X}$. This shows that there is at most one value $b$ such that $(z_X(r_X,c),c,b) \in \mathcal{D} \setminus \mathcal{D}^0$. 

To complete the proof, recall that by Lemma \ref{lemma:IDsetAltCase2}, $\{ \beta_\text{med} \} = \mathcal{B}(r_X,0,c)$ since $r_X'c \ne -1$ and $z_X(r_X,c)^2 < k_0$. Therefore $\beta_\text{med} \in \bigcup_{r_Y \in \mathbb{R}^{d_1}}\mathcal{B}(r_X,r_Y,c)$ and the lemma's statement holds.
\end{proof}

Our next lemma shows that the following function is a bound for $z_X(\cdot)$:
\begin{align}\label{eq:z_X formula}
	\bar{z}_X(\rx,\underline{c},\bar{c}) &\coloneqq \begin{cases}
		\frac{\rx\|\sigma_{W_1,X}\|\sqrt{1 - \max\{\min\{\rx,\bar{c}\},\underline{c}\}^2}}{1 - \rx\max\{\min\{\rx,\bar{c}\},\underline{c}\}} &\text{ if } \rx\bar{c} < 1\\
		+\infty &\text{ if } \rx\bar{c} \geq 1.
		\end{cases}
\end{align}

\begin{lemma}\label{lem:maxZx}
Suppose $\sigma_{W_1,X} \neq 0$, $\rx \geq 0$, and $0 \leq \underline{c} \leq \overline{c} \leq 1$ where $\underline{c} < 1$. Then
\[
	\sup\{|z_X(r_X, c)|: \|r_X\| \le \rx, \|c\| \in[\underline{c}, \bar{c}], \| c \| \neq 1 \} = \bar{z}_X(\rx,\underline{c},\bar{c}).
\]
\end{lemma}

\begin{proof}[Proof of Lemma \ref{lem:maxZx}]
Recall that, by definition,
\[
	|z_X(r_X,c)| = \frac{|r_X'\sigma_{W_1,X}|\sqrt{1 - \|c\|^2}}{|1 + r_X'c|}.
\]
We prove the result by considering two cases.

\bigskip

\noindent\textbf{Case 1}: Suppose $\bar{c}\rx \geq 1$. In this case, by definition, $\bar{z}_X(\rx,\uc, \bar{c})^2 = + \infty$. It therefore suffices to show that there exists a sequence $\{ (r_X^{(n)}, c^{(n)}) : \| r_X^{(n)} \| \leq \rx, \| c^{(n)} \| \in [\underline{c},\bar{c}], \| c^{(n)} \| \neq 1 \}$ such that $z_X(r_X,c)^2$ diverges to $+\infty$ along this sequence. When $\underline{c} < \overline{c}$, define the sequences
\[
	r_X^{(n)} = -\frac{n-1}{n}\frac{1}{\bar{c}\|\sigma_{W_1,X}\|}\sigma_{W_1,X}
	\qquad \text{and} \qquad
	c^{(n)} = \frac{n-1}{n} \frac{\bar{c}}{\|\sigma_{W_1,X}\|}\sigma_{W_1,X}.
\]
for all $n$. First we check that these sequences satisfies the norm constraints:
\[
	\|r_X^{(n)}\| = \left\|-\frac{n-1}{n}\frac{1}{\bar{c}\|\sigma_{W_1,X}\|}\sigma_{W_1,X}\right\| =  \frac{n-1}{n \bar{c}} \leq \frac{(n-1)\rx}{n} < \rx
\]
since $\bar{c} \rx \geq 1$, and $\|c^{(n)}\| = \left\| \frac{n-1}{n} \frac{\bar{c}}{\|\sigma_{W_1,X}\|}\sigma_{W_1,X}\right\| = \frac{n-1}{n} \cdot \bar{c} < \bar{c}$. In particular, note that $\| c^{(n)} \| \neq 1$ and $\|c^{(n)}\| \geq \underline{c}$ for all sufficiently large $n$. Next we evaluate the function $| z_X(\cdot,\cdot) |$ along this sequence:
\begin{align*}
	|z_X(r_X^{(n)},c^{(n)})|
	&= \frac{\frac{n-1}{n}\frac{\|\sigma_{W_1,X}\|}{\bar{c}}\sqrt{1 - \bar{c}^2 \left( \frac{n-1}{n} \right)^2 }}{1 - \left( \frac{n-1}{n} \right)^2 }
	= \frac{\| \sigma_{W_1,X} \|}{\bar{c}} \sqrt{ \frac{(n-1)^2 (n^2 - \bar{c}^2 (n-1)^2)}{(2n-1)^2} }.
\end{align*}
Since $\|\sigma_{W_1,X}\| \neq 0$ by $\sigma_{W_1,X} \neq 0$, this expression is nonzero for all $n$. The term inside the square root goes to $+\infty$ by L'Hospital's rule. Thus $\lim_{n \rightarrow \infty} | z_X(r_X^{(n)}, c^{(n)}) | = + \infty$. When $\underline{c} = \overline{c}$ we can use the same argument, replacing the $c^{(n)}$ sequence by $c^{(n)} = \bar{c}\sigma_{W_1,X}/\|\sigma_{W_1,X}\|$, noting that $\underline{c} = \overline{c}$ implies $\overline{c}<1$ by our lemma's assumptions.

\bigskip

\noindent \textbf{Case 2}: Suppose $\bar{c}\rx < 1$. Consider $(r_X, c)$ such that $\| r_X \| \leq \rx$ and $\| c \| \leq \bar{c}$. By the Cauchy Schwarz inequality, $|r_X'\sigma_{W_1,X}| \le \|r_X\|\|\sigma_{W_1,X}\|$ and $1 + r_X'c \geq 1 - \|r_X\|\|c\| \geq 1 - \rx \bar{c} > 0$.
Therefore
\begin{align*}
	|z_X(r_X, c)| &= \frac{|r_X'\sigma_{W_1,X}|\sqrt{1 - \|c\|^2}}{|1 + r_X'c|}\leq \frac{\|r_X\|\|\sigma_{W_1,X}\|\sqrt{1 - \|c\|^2}}{1- \|r_X\|\|c\|} =: f(\|r_X\|,\|c\|).
\end{align*}
This implies that
\begin{align*}
	\sup\{|z_X(r_X,c)|: \|r_X\| \leq \rx, \|c\| \in [\underline{c},\bar{c}] \} &\leq \sup\{f(\|r_X\|,\|c\|): \|r_X\| \leq \rx, \|c\| \in [\underline{c},\bar{c}] \}\\
	&= \sup\{f(a_X,a_c): (a_X,a_c) \in [0,\rx]\times [\underline{c},\bar{c}]\}.
\end{align*}
Next we show that $\sup\{f(a_X,a_c): (a_X,a_c) \in [0,\rx]\times [\underline{c},\bar{c}]\} = \bar{z}(\rx,\underline{c},\bar{c})$. To see this, first note that $f(a_X,a_c)$ is continuous on $[0,\rx]\times [\underline{c},\bar{c}]$, a compact set. $f(a_X,a_c)$ is nondecreasing in $a_X$ for any $a_c \in [\underline{c},\bar{c}]$. Hence $f(a_X,a_c) \leq f(\rx,a_c)$ for any  $(a_X,a_c) \in [0,\rx]\times [\underline{c},\bar{c}]$.

To maximize $f(\rx,a_c)$ over $a_c \in [\underline{c},\bar{c}]$, consider the derivative of $f(\rx,a_c)$ with respect to $a_c$:
\begin{align*}
	\frac{\partial f(\rx, a_c)}{\partial a_c}
	&= \rx\|\sigma_{W_1, X}\| \frac{-a_c (1 - \rx a_c) + \rx (1 - a_c^2)}{\sqrt{1 - a_c^2}(1 - \rx a_c)^2} = \frac{\rx\|\sigma_{W_1,X}\|(\rx - a_c)}{\sqrt{1 - a_c^2}(1 - \rx a_c)^2}.
\end{align*}
Examining this derivative when $\rx > 0$, $a_c = \rx$ maximizes $f(\rx, a_c)$ when $\rx \in [\underline{c},\bar{c}]$, the upper boundary value $a_c = \bar{c}$ maximizes it when $\rx > \bar{c}$, and the lower boundary value $a_c = \underline{c}$ maximizes it when $\rx < \underline{c}$. Hence $\argmax\{f(\rx, a_c): a_c \in [0, \bar{c}]\} = \max\{\min\{\rx,\bar{c}\},\underline{c}\}$. If $\rx = 0$, $f(\rx,a_c)$ is constant and is trivially maximized at $a_c = \max\{\min\{\rx,\bar{c}\},\underline{c}\}$.

Putting the previous few steps together gives
\begin{align*}
	\sup\{|z_X(r_X,c)|: \|r_X\| \leq \rx, \|c\| \in [\underline{c},\bar{c}] \} & 
\leq f(\rx,\max\{\min\{\rx,\bar{c}\},\underline{c}\})
	= \bar{z}_X(\rx,\underline{c},\bar{c}).
\end{align*}

We conclude the proof by showing that the inequality in fact holds with equality. We do this by finding $(r_X^*,c^*)$ such that $\|r_X^*\| \leq \rx$, $\|c^*\| \in [\underline{c},\bar{c}]$, and such that $|z_X(r_X^*,c^*)| = \bar{z}_X(\rx,\underline{c},\bar{c})$. Let
\[
	r_X^* = \frac{\rx}{\|\sigma_{W_1,X}\|}\sigma_{W_1,X}
	\qquad \text{and} \qquad
	c^* = -\frac{\max\{\min\{\rx,\bar{c}\},\underline{c}\}}{\|\sigma_{W_1,X}\|}\sigma_{W_1,X}.
\]
Note that $\|r_X^*\| = \rx \leq \rx$ and $\|c^*\| = \max\{\min\{\rx,\bar{c}\},\underline{c}\} \in [\underline{c},\bar{c}]$.  Moreover,
\begin{align*}
	|z_X(r_X^*,c^*)| &= \frac{|r_X^{*\prime} \sigma_{W_1,X}|\sqrt{1 - \|c^*\|^2}}{|1 + r_X^{*\prime}c^*|}
	= \frac{\rx\|\sigma_{W_1,X}\|\sqrt{1 - \max\{\min\{\rx,\bar{c}\},\underline{c}\}^2}}{1 - \rx\max\{\min\{\rx,\bar{c}\},\underline{c}\}}
	= \bar{z}_X(\rx,\underline{c}, \bar{c}).
\end{align*}
Finally note that $\|c^*\| \neq 1$ since $\|c^*\| = \max\{\min\{\rx,\bar{c}\},\underline{c}\}$ and $\rx\bar{c} < 1$.
\end{proof}

We use the next two lemmas to prove our third main result, Theorem \ref{cor:BFCalculation3D}. Recall that the function $p(z,c;\rx)$ is defined just prior to the statement of Theorem \ref{cor:BFCalculation3D} in section \ref{sec:identrYtoo}.

\begin{lemma} \label{lem:rxbound}  
Assume $\sigma_{W_1,X} \ne 0$. Fix $(z, c, \rx)$ such that $z^2 < k_0$, $\|c\| < 1$, $\rx \geq 0$.  Then $p(z,c;\rx) \geq 0$ if and only if there exists an $r_X \in \R^{d_1}$ with $\|r_X\| \le \rx$ such that
\[
	z = r_X'(\sigma_{W_1,X}\sqrt{1 - \|c\|^2} - cz).
	\tag{\ref{eq:IDsetEqAltB}}
\]
\end{lemma}

\begin{proof}[Proof of Lemma \ref{lem:rxbound}]
Let $(z, c, \rx) \in \mathbb{R} \times \mathbb{R}^{d_1} \times \mathbb{R}_{\geq 0}$ be such that $\|c\| < 1$ and $z^2 < k_0$. 

\medskip

\noindent ($\Rightarrow$) Suppose $p(z,c;\rx) \geq 0$. We will construct an $r_X \in \R^{d_1}$ with $\| r_X \| \leq \rx$ and such that equation \eqref{eq:IDsetEqAltB} holds. We consider two cases.
\begin{enumerate}[wide, labelwidth=!, labelindent=0pt]
\item Suppose $z = 0$. Let $r_X = 0$. Then $\|r_X\| \leq \rx$ and $z = r_X'(\sigma_{W_1,X}\sqrt{1 - \|c\|^2} - cz)$ holds trivially.

\item Suppose $z \neq 0$. Let $r_X = \|\sigma_{W_1,X}\sqrt{1 - \|c\|^2} - zc \|^{-2}z(\sigma_{W_1,X} \sqrt{1 - \|c\|^2} - zc)$. $r_X$ is well defined since $\|\sigma_{W_1,X}\sqrt{1 - \|c\|^2} - zc\|^2 = \left(p(z,c;\rx) + z^2\right) / \rx^2 >0$ by $z \neq 0$ and $p(z,c;\rx) \geq 0$. Also note that $p(z,c;\rx) \geq 0$ and $z \neq 0$ together imply that $\rx > 0$. Next, $\|r_X\| \leq \rx$ follows from
\begin{align*}
	\|r_X\|^2 &= \|\sigma_{W_1,X}\sqrt{1 - \|c\|^2} - zc\|^{-2}z^2\\
	&= \|\sigma_{W_1,X}\sqrt{1 - \|c\|^2} - zc\|^{-2}\left(-p(z,c;\rx) + \rx^2\|\sigma_{W_1,X}\sqrt{1 - \|c\|^2} - zc\|^{2}\right)\\
	&\leq \rx^2
\end{align*}
and $p(z,c;\rx) \geq 0$. Finally, we can verify that
\begin{align*}
	&r_X' (\sigma_{W_1,X}\sqrt{1 - \|c\|^2} - cz) = \frac{z(\sigma_{W_1,X} \sqrt{1 - \|c\|^2} - zc)' (\sigma_{W_1,X} \sqrt{1 - \|c\|^2} - zc)}{\|\sigma_{W_1,X}\sqrt{1 - \|c\|^2} - zc\|^2} = z.
\end{align*}
\end{enumerate}

\noindent ($\Leftarrow$) Suppose there exists an $r_X$ such that $\|r_X\| \le \rx$ and $z = r_X' (\sigma_{W_1,X}\sqrt{1 - \|c\|^2} - cz)$. Then
\begin{align*}
	z^2 = (r_X' (\sigma_{W_1,X}\sqrt{1 - \|c\|^2} - cz))^2 \leq \|r_X\|^2\|\sigma_{W_1,X}\sqrt{1 - \|c\|^2} - zc\|^2 \leq \rx^2\|\sigma_{W_1,X}\sqrt{1 - \|c\|^2} - zc\|^2.
\end{align*}
The first inequality follows from Cauchy-Schwarz and the second from $\|r_X\| \le \rx$. Subtracting $z^2$ from both sides yields $p(z,c;\rx)\geq 0$.
\end{proof}

The next lemma allows us to search over $c \in \vspan\{\sigma_{W_1,X}, \sigma_{W_1,Y}\}$ rather than $\mathbb{R}^{d_1}$, reducing the dimension of the minimization problem by $d_1 - 2$.

\begin{lemma}\label{lem:c2dim}  
Suppose $(\sigma_{W_1,X},\sigma_{W_1,Y})$ are linearly independent and $d_1 \geq 2$. Let $b \ne \beta_\text{med}$, $z \in \mathbb{R}\setminus\{0\}$, $\rx \geq 0$, and $[\underline{c},\bar{c}] \subseteq [0, 1]$. Then
\begin{align*}
	&\inf\{\underline{r}_Y(z,c,b): p(z,c;\rx) \geq 0, c \in \R^{d_1}, \|c\| \in [\underline{c},\bar{c}], \|c\| < 1\} \\
	&= \inf\{\underline{r}_Y(z,c,b): p(z,c;\rx) \geq 0, c \in \vspan\{\sigma_{W_1,X}, \sigma_{W_1,Y}\}, \|c\| \in [\underline{c},\bar{c}], \|c\| < 1\}.
\end{align*}  
\end{lemma}

\begin{proof}[Proof of Lemma \ref{lem:c2dim}]
If $d_1 = 2$ the lemma follows immediately because $\R^{d_1} = \vspan\{\sigma_{W_1,X}, \sigma_{W_1, Y}\}$. So suppose $d_1 > 2$. We break this proof in two parts.

\medskip

\noindent\textbf{Part 1:} We first show that for any $a \in [\underline{c},\bar{c}]$ with $a<1$,
\[
  	\argmin\{\underline{r}_Y(z,c,b): p(z,c;\rx) \geq 0, c \in \R^{d_1}, \|c\| = a\} \subseteq \vspan\{ \sigma_{W_1,X}, \sigma_{W_1,Y}\}.
\]
We prove this by contrapositive: we fix $c_\text{outside} \in \mathbb{R}^{d_1} \setminus \vspan\{\sigma_{W_1,X}, \sigma_{W_1,Y}\}$ and show $c_\text{outside} \notin \argmin\{\underline{r}_Y(z,c,b): p(z,c;\rx) \geq 0, c \in \R^{d_1}, \|c\| = a\}$. Note that such a point exists since $d_1 > 2 = \text{dim}(\vspan\{\sigma_{W_1,X}, \sigma_{W_1,Y} \})$. Let
\[
	B(a,z,\rx) = \{c \in \R^{d_1} : p(z, c; \rx) \geq 0, \|c\| = a\}
\]
denote the constraint set. The result holds immediately if $c_\text{outside} \notin B(a,z,\rx)$, so suppose $c_\text{outside} \in B(a,z,\rx)$. It therefore suffices to construct a $c_\text{better} \in \vspan\{\sigma_{W_1,X}, \sigma_{W_1,Y}\}$ with $c_\text{better} \in B(a,z,\rx)$ and $\underline{r}_Y(z,c_\text{better},b) < \underline{r}_Y(z,c_\text{outside},b)$.

By Gram-Schmidt orthogonalization, the vectors
\[
	v_1 = \frac{ \sigma_{W_1, X} }{ \| \sigma_{W_1, X}\| }
	\qquad \text{and} \qquad
	v_2 = \frac{
	\|\sigma_{W_1,X}\|^2 \sigma_{W_1, Y} - (\sigma_{W_1, Y}'\sigma_{W_1, X}) \sigma_{W_1, X}
	}{
	\|\sigma_{W_1, X}\|\sqrt{\|\sigma_{W_1, Y}\|^2 - (\sigma_{W_1, X}'\sigma_{W_1, Y})^2}
	}
\]
form an orthonormal basis for the subspace $\text{span}\{\sigma_{W_1, X}, \sigma_{W_1, Y}\}$, since we assumed $\sigma_{W_1,X}$ and $\sigma_{W_1,Y}$ are linearly independent. Since we can write $\R^{d_1} = \text{span}\{v_1, v_2\} \oplus \text{span}\{v_1, v_2\}^\perp$, $c_\text{outside}$ can be uniquely represented as $c_\text{outside} = a_1 v_1 + a_2 v_2 + a_3 v_3$ where $(a_1, a_2, a_3)$ are such that $a_1^2 + a_2^2 + a_3^2 = a^2$, $v_3 \in \text{span}\{v_1, v_2\}^\perp$, and $\|v_3\| = 1$. Since $c_\text{outside} \notin \vspan\{\sigma_{W_1,X}, \sigma_{W_1,Y}\}$, $a_3 \neq 0$. Define
\[
	c^+ \coloneqq a_1 v_1 + \left( \sqrt{a_2^2 + a_3^2} \right) v_2
	\qquad \text{and} \qquad
	c^- \coloneqq a_1 v_1 - \left( \sqrt{a_2^2 + a_3^2} \right) v_2.
\]
These points satisfy $c^+, c^- \in \text{span} \{\sigma_{W_1, X}, \sigma_{W_1, Y}\}$ and $\|c^+\| = \|c^-\| = \|c_\text{outside}\| = a$. In fact, we will show that one of $c^+$ or $c^-$ is the desired point $c_\text{better}$. To do this, we have to show that (i) $c^+, c^- \in B(a,z,\rx)$ and (ii) $\min\{\underline{r}_Y(z,c^+,b),\underline{r}_Y(z,c^-,b)\} < \underline{r}_Y(z,c_\text{outside},b)$.

\medskip

\noindent \emph{Step (i)}. Here we show that $c^+$ and $c^-$ are in the constraint set $B(a,z,\rx)$. We already showed $\|c^+\| = \|c^-\| = a$ so we now show $p(z, c; \rx) \geq 0$ for $c= c^+,c^-$. To do this, write
\begin{align*}
	p(z, c; \rx)
	&= z^2(\|c\|^2 \rx^2 - 1) - 2z \rx^2 (\sigma_{W_1,X}'c) \sqrt{1 - \|c\|^2} + \rx^2 \|\sigma_{W_1,X}\|^2 (1 - \|c\|^2),
\end{align*}
which follows by expanding $p(z, c; \rx)$. Since $\| c^+ \| = \| c^- \| = \| c_\text{outside} \| = a$, $v_1'\sigma_{W_1,X} = \|\sigma_{W_1,X}\|$, and $v_2'\sigma_{W_1,X} = v_3'\sigma_{W_1, X} = 0$, evaluating this function at any of $c \in \{c_\text{outside}, c^+,c^-\}$ returns the same value:
\begin{align*}
	p(z, c; \rx)
	&= z^2(a^2 \rx^2 - 1) - 2a_1 z  \rx^2 \|\sigma_{W_1,X}\| \sqrt{1 - a^2} + \rx^2 \|\sigma_{W_1,X}\|^2 (1 - a^2).
\end{align*}
Since $c_\text{outside} \in B(a,z,\rx)$, $p(z, c_\text{outside}; \rx) \geq 0$. Hence we must also have $p(z, c^+; \rx) \geq 0$ and $p(z, c^-; \rx) \geq 0$. Thus $c^+, c^- \in B(a,z,\rx)$.

\medskip

\noindent \emph{Step (ii)}. Next we show that either $\underline{r}_Y(z, c^+, b) < \underline{r}_Y(z, c_\text{outside}, b)$ or $\underline{r}_Y(z, c^-, b) < \underline{r}_Y(z, c_\text{outside}, b)$.

\medskip

First, we have
\[
	z\sqrt{1 - \| c_\text{outside} \|^2}(\sigma_{W_1,X} - b\sigma_{W_1,Y}) - (k_1 - k_0b)c_\text{outside} \neq 0.
\]
To see this, note that this quantity is a linear combination of $(\sigma_{W_1,X},\sigma_{W_1,Y},c_\text{outside})$. Its coefficients are nonzero because $b \neq \beta_\text{med}$, $z \neq 0$, and $\| c_\text{outside} \| = a$ where $a < 1$. Moreover, $\sigma_{W_1,X}$ and $\sigma_{W_1,Y}$ are linearly independent, and $c_\text{outside}$ is not in $\vspan\{\sigma_{W_1,X},\sigma_{W_1,Y}\}$. This linear independence therefore implies that the linear combination cannot be zero. Thus $(z,c_\text{outside},b) \in \mathcal{D}^0$ by definition of $\mathcal{D}^0$. By definition of $\underline{r}_Y(z,c,b)$, we therefore have $0 < \underline{r}_Y(z, c_\text{outside}, b) < \infty$. Consequently, $\underline{r}_Y(z,c_\text{outside},b)^{-1}$ is well defined.

For any $c$ with $\| c \| = a$, expanding the norm in $\underline{r}_Y(z,c, b)$ gives
\begin{align*}
	&\underline{r}_Y(z,c,b)^{-2} \\
	&= \frac{z^2(1 -a^2)\|\sigma_{W_1, Y} - b \sigma_{W_1, X}\|^2 - 2 z \left((\sigma_{W_1, Y} - b \sigma_{W_1, X})'c \right) \sqrt{1 - a^2}(k_1 - b k_0) + a^2(k_1 - b k_0)^2}{(k_1 - b k_0)^2}.
\end{align*}
Therefore 
\begin{align*}
	& \underline{r}_Y(z, c^+, b)^{-2} - \underline{r}_Y(z, c_\text{outside}, b)^{-2} \\
	&= 2z \sqrt{1 - a^2}(k_1 - b k_0)^{-1} (\sigma_{W_1, Y} - b \sigma_{W_1, X})'(c_\text{outside} - c^+)\\
	&= 2z \sqrt{1 - a^2}(k_1 - b k_0)^{-1} (\sigma_{W_1, Y} - b \sigma_{W_1, X})'(a_2v_2 + a_3 v_3  -\sqrt{a_2^2 + a_3^2} v_2)\\
	&= 2z \sqrt{1 - a^2}(k_1 - b k_0)^{-1} \left((\sigma_{W_1, Y} - b \sigma_{W_1, X})'v_2\right) (a_2 -\sqrt{a_2^2 + a_3^2}).
\end{align*}
The last line follows since $v_3 \in \vspan\{\sigma_{W_1,X},\sigma_{W_1,Y}\}^\perp$. Likewise,
\begin{align*}
  	 \underline{r}_Y(z, c^-, b)^{-2} - \underline{r}_Y(z, c_\text{outside}	, b)^{-2} 
	&= 2z \sqrt{1 - a^2}(k_1 - b k_0)^{-1} \left((\sigma_{W_1, Y} - b \sigma_{W_1, X})'v_2\right) (a_2 + \sqrt{a_2^2 + a_3^2}).
\end{align*}
Recall that $z\neq 0$ by assumption, $a^2 < 1$, and $k_1 \neq b k_0$ by $b \neq \beta_\text{med}$. Since $a_3 \neq 0$,
\[
	a_2  - \sqrt{a_2^2 + a_3^2} < 0 < a_2  + \sqrt{a_2^2 + a_3^2}.
\]
Therefore $\underline{r}_Y(z, c^-, b)^{-2} - \underline{r}_Y(z, c_\text{outside}, b)^{-2}$ and  $\underline{r}_Y(z, c^+, b)^{-2} - \underline{r}_Y(z, c_\text{outside}, b)^{-2}$ are of opposite signs. This implies $\min\{\underline{r}_Y(z,c^+,b),\underline{r}_Y(z,c^-,b)\} < \underline{r}_Y(z,c_\text{outside},b)$.

\medskip

\noindent\textbf{Part 2:} We finish by showing that the result of Part 1 implies the lemma's claim. We have
\begin{align*}
	&\inf\{\underline{r}_Y(z, c, b): p(z,c;\rx) \geq 0, c \in \R^{d_1}, \|c\| \in [\underline{c},\bar{c}], \|c\| < 1\} \\
	&= \inf\{\inf\{\underline{r}_Y(z, c, b): p(z,c;\rx) \geq 0, c \in \R^{d_1}, \|c\| = a\}: a \in [\underline{c},\bar{c}], a < 1\}\\
	&= \inf\{\inf\{\underline{r}_Y(z, c, b): p(z,c;\rx) \geq 0, c \in \vspan\{\sigma_{W_1,X}, \sigma_{W_1,Y}\}, \|c\| = a\}: a \in [\underline{c},\bar{c}], a < 1\}\\
	&= \inf\{\underline{r}_Y(z, c, b): p(z,c;\rx) \geq 0, c \in \vspan\{\sigma_{W_1,X}, \sigma_{W_1,Y}\}, \|c\| \in [\underline{c},\bar{c}], \|c\| < 1\}.
\end{align*}
The first and third equality follow from a sequential minimization. The second equality follows from follows from part 1 of this proof. 
\end{proof}

\subsection{Proofs of Main Results}\label{sec:proofsOfMainResults}

We prove Theorem \ref{thm:IdsetRyFree} first since Theorem \ref{cor:IdsetRyANDcFree} will be shown as the special case where $[\underline{c},\bar{c}] = [0,1]$.

\begin{proof}[Proof of Theorem \ref{thm:IdsetRyFree}]
Let $\mathcal{R}(\rx,\underline{c},\bar{c}) \coloneqq \{(r_X,c) \in \mathbb{R}^{d_1}\times \R^{d_1}: \|r_X\| \leq \rx, \|c\| \in [\underline{c},\bar{c}], \| c \| \neq 1 \}$. Note that
\begin{align*}
	\mathcal{B}_I(\rx, \underline{c},\bar{c})
	&= \bigcup_{(r_X,r_Y,c): \|r_X\| \leq \rx, \|c\| \in [\underline{c},\bar{c}]} \mathcal{B}(r_X,r_Y,c) 
	= \bigcup_{(r_X,c) \in \mathcal{R}(\rx,\underline{c},\bar{c}), r_Y \in \R^{d_1}} \mathcal{B}(r_X,r_Y,c).
\end{align*}
The first equality holds by Theorem \ref{thm:mainMultivariateResult} and the second follows since $\|c\|<1$ holds whenever $b \in \mathcal{B}(r_X,r_Y,c)$ by definition. We consider the two cases specified in the theorem statement separately. 

\bigskip

\noindent\textbf{Case 1.} $\bar{z}_X(\rx,\underline{c},\bar{c})^2 < k_0$.  

\bigskip

By Lemma \ref{lem:maxZx}, $z_X(r_X,c)^2 \leq \bar{z}_X(\rx,\uc,\bar{c})^2$ for all $(r_X,c) \in \mathcal{R}(\rx,\uc,\bar{c})$. Combining this fact with $\bar{z}_X(\rx,\uc, \bar{c})^2 < k_0 < \infty$ and the fact that the denominator of $z_X(r_X,c)^2$ is $(1+r_X'c)^2$, we have $r_X'c \neq -1$ for all $(r_X,c) \in \mathcal{R}(\rx,\uc,\bar{c})$. Partition $\mathcal{R}(\rx,\uc,\bar{c}) = \mathcal{R}^{= 0}(\rx,\uc,\bar{c}) \cup \mathcal{R}^{\neq 0}(\rx,\uc,\bar{c})$ where
  \begin{align*}
  \mathcal{R}^{= 0}(\rx,\uc, \bar{c}) &\coloneqq \{(r_X,c) \in \mathcal{R}(\rx,\uc,\bar{c}) : z_X(r_X,c) = 0\} \\
  \mathcal{R}^{\neq 0}(\rx,\uc,\bar{c}) &\coloneqq \{(r_X,c) \in \mathcal{R}(\rx,\uc,\bar{c}) : z_X(r_X,c) \neq 0\}.
\end{align*}
This partition lets us write
\[
	\mathcal{B}_I(\rx,\uc, \bar{c}) =
	\left\{\bigcup_{(r_X,c) \in \mathcal{R}^{= 0}(\rx,\uc,\bar{c}), r_Y \in \mathbb{R}^{d_1}}\mathcal{B}(r_X,r_Y,c)\right\} \bigcup \left\{\bigcup_{(r_X,c) \in \mathcal{R}^{\neq 0}(\rx,\uc,\bar{c}), r_Y \in \mathbb{R}^{d_1}}\mathcal{B}(r_X,r_Y,c)\right\}.
\]
Next we characterize each of these unions.
\begin{enumerate}[wide, labelwidth=!, labelindent=0pt]
\item First suppose $b$ be in the union on the left. Then there is a $(r_X,c) \in \mathcal{R}^{= 0}(\rx,\uc,\bar{c})$ and $r_Y \in \R^{d_1}$ such that $b \in \mathcal{B}(r_X,r_Y,c)$. By the definition of $z_X(r_X,c)$ and by $\|c\|^2 < 1$, this implies $r_X'\sigma_{W_1,X} = 0$. Using equation \eqref{eq:IDseteqsimp3} and $1 + r_X'c \neq 0$, we get that $r_X'p_1 = 0$. Using equation \eqref{eq:IDseteqsimp1}, this implies $b = \beta_\text{med}$. Thus we have shown that any element in the union on the left must be $\beta_\text{med}$. This union is nonempty since $(0,\iota_{d_1}/\|\iota_{d_1}\| \cdot \uc) \in \mathcal{R}^{=0}(\rx,\uc,\bar{c})$ and we can show $\beta_\text{med} \in \mathcal{B}(0,0,\iota_{d_1}/\|\iota_{d_1}\| \cdot \uc)$ by selecting $p_1 = \sigma_{W_1,X}$ and $g_1 = \sigma_{W_1,Y} - \beta_\text{med} \sigma_{W_1,X}$ and by applying Lemma \ref{lemma:IDsetSimplification}. Thus we conclude that
$
	\left\{\bigcup_{(r_X,c) \in \mathcal{R}^{= 0}(\rx,\uc,\bar{c}), r_Y \in \mathbb{R}^{d_1}}\mathcal{B}(r_X,r_Y,c)\right\}
	= \{\beta_\text{med}\}.
$

\item Next consider $(r_X,c) \in \mathcal{R}^{\neq 0}(\rx,\uc,\bar{c})$. By Lemma \ref{lem:maxZx}, $0 < z_X(r_X,c)^2 \le \bar{z}_X(\rx,\uc,\bar{c})^2 < k_0$ for all $(r_X,c) \in \mathcal{R}^{\neq 0}(\rx,\uc,\bar{c})$. So Lemma \ref{lemma:IDsetRy_claim2} shows that, for all $(r_X,c) \in \mathcal{R}^{\neq 0}(\rx,\uc,\bar{c})$,
\[
    \bigcup_{r_Y \in \mathbb{R}^{d_1}}\mathcal{B}(r_X,r_Y,c) = \left(\underline{b}(z_X(r_X,c)), \overline{b}(z_X(r_X,c))\right) \setminus \mathcal{B}^0(z_X(r_X,c),c).
\]
\end{enumerate}
Using these characterizations of the two partition sets gives
\begin{align*}
	\mathcal{B}_I(\rx,\uc, \bar{c})
	&= \{\beta_\text{med}\} \cup \bigcup_{(r_X,c) \in \mathcal{R}^{\neq 0}(\rx,\uc,\bar{c})}\left(\underline{b}(z_X(r_X,c)), \overline{b}(z_X(r_X,c))\right) \setminus \mathcal{B}^0(z_X(r_X,c),c)\\
	&= \{\beta_\text{med}\} \cup \bigcup_{(r_X,c) \in \mathcal{R}(\rx,\uc,\bar{c})}\left(\underline{b}(z_X(r_X,c)), \overline{b}(z_X(r_X,c))\right) \setminus \mathcal{B}^0(z_X(r_X,c),c).
\end{align*}
The second line follows since $\left(\underline{b}(z_X(r_X,c)), \overline{b}(z_X(r_X,c))\right) = \emptyset$ when $z_X(r_X,c) = 0$. We can now derive our desired expression for $\underline{B}(\rx,\uc,\bar{c})$ as follows:
\begin{align*}
    \underline{B}(\rx,\uc, \bar{c})
    &= \inf \mathcal{B}_I(\rx,\uc, \bar{c}) \\
    &= \inf \left[ \{ \beta_\text{med} \} \cup \left( \bigcup_{(r_X,c) \in \mathcal{R}(\rx,\uc,\bar{c})}\left(\underline{b}(z_X(r_X,c)), \overline{b}(z_X(r_X,c))\right) \setminus \mathcal{B}^0(z_X(r_X,c),c) \right) \right] \\
    &= \inf\{\underline{b}(z_X(r_X,c)): (r_X,c) \in \mathcal{R}(\rx,\uc,\bar{c})\} \\
    &= \beta_{\text{med}} - \sup\left\{\sqrt{\text{devsq}(z_X(r_X,c))}: \|r_X\| \leq \rx, \|c\| \in[\uc,\bar{c}], \| c \| \neq 1 \right\}.
\end{align*}
The first line follows by definition. The second line follows by our above derivations. In the third line we used the fact that $\underline{b}(z) \leq \beta_\text{med}$ for all $z \geq 0$. The fourth line follows by the definition of $\underline{b}(\cdot)$. Next, recall that $\text{devsq}(z) = z^2\left(k_2/k_0 - \beta_{\text{med}}^2\right) / (k_0 - z^2)$.
By $k_2 k_0 > k_1^2$ (from Cauchy-Schwarz and assumption \ref{assump:posdefVar}), $\text{devsq}(z)$ is increasing in $|z|$ over all $z^2 < k_0$. Therefore, since $z_X(r_X,c)^2 < k_0$ for all $(r_X,c) \in \mathcal{R}(\rx,\uc,\bar{c})$, Lemma \ref{lem:maxZx} implies that the infimum of $\underline{b}(z_X(r_X,c))$ is attained at $\underline{B}(\rx, \uc,\bar{c}) = \beta_{\text{med}} - \sqrt{\text{devsq}(\bar{z}_X(\rx,\uc,\bar{c}))}$. The analysis for $\overline{B}(\rx, \uc,\bar{c})$ is analogous.

\bigskip

\noindent\textbf{Case 2.} $\bar{z}_X(\rx, \uc, \bar{c})^2 \geq k_0$.

\bigskip

Let $\{ \rx^{(n)} \}$ be a sequence such that $\rx^{(n)} \leq \rx$ and $\bar{z}_X(\rx^{(n)},\uc, \bar{c})^2 < k_0$ for all $n \in \mathbb{N}$. Then
\begin{align*}
	\overline{B}(\rx,\uc, \bar{c})
	&\geq \overline{B}(\rx^{(n)},\uc, \bar{c})
	= \beta_{\text{med}} + \sqrt{\frac{\bar{z}_X(\rx^{(n)},\uc, \bar{c})^2(k_2/k_0 - \beta_{\text{med}}^2)}{k_0 - \bar{z}_X(\rx^{(n)},\uc, \bar{c})^2}}
\end{align*}
for all $n \in \mathbb{N}$. The equality follows by our analysis in Case 1, since $\bar{z}_X(\rx^{(n)},\uc, \bar{c})^2 < k_0$. Suppose $\bar{z}_X(\rx^{(n)},\uc, \bar{c})^2 \nearrow k_0$ as $n \rightarrow \infty$. Then
\[
	\lim_{n \rightarrow \infty}
	\left( \beta_{\text{med}} + \sqrt{\frac{\bar{z}_X(\rx^{(n)},\uc, \bar{c})^2(k_2/k_0 - \beta_{\text{med}}^2)}{k_0 - \bar{z}_X(\rx^{(n)},\uc, \bar{c})^2}} \right)
	= +\infty.
\]
This follows from $\lim_{z^2 \nearrow k_0} z^2 / (k_0 - z^2) = +\infty$
and from $k_0 > 0$. This would be sufficient to obtain $\overline{B}(\rx,\uc, \bar{c}) = +\infty$. So all that remains to be shown is that there exists such a sequence $\{ \rx^{(n)} \}$ with $\rx^{(n)} \leq \rx$ and $\bar{z}_X(\rx^{(n)}, \bar{c})^2 \nearrow k_0$ as $n \rightarrow \infty$. We consider two cases.
\begin{enumerate}[wide, labelwidth=!, labelindent=0pt]
\item Suppose $\uc = \bar{c} = 0$. Then $\bar{z}_X(\rx^{(n)},\uc, \bar{c})^2 = (\rx^{(n)})^2 \| \sigma_{W_1, X} \|^2$. Recall that $\| \sigma_{W_1,X} \| \neq 0$ by \ref{assump:noKnifeEdge}. Since $\rx^2 \| \sigma_{W_1, X} \|^2 \geq k_0$, we can find a sequence $\rx^{(n)}$ such that $\rx^{(n)} \leq \rx$ and $k_0 - (\rx^{(n)})^2 \| \sigma_{W_1, X} \|^2 \searrow 0$.

\item Suppose $\bar{c} > 0$. To ensure that $\bar{z}_X(\rx^{(n)},\uc, \bar{c})^2 < k_0$ for all $n \in \mathbb{N}$, we need that $\rx^{(n)} \bar{c} < 1$. Thus our sequence must satisfy $\rx^{(n)} < 1/\bar{c}$. Next notice that $\bar{z}_X(\rx^{(n)},\uc, \bar{c})^2$ is continuous in $\rx^{(n)}$ over the set $[0,1/\bar{c})$. Moreover, its range on this set is $[0,\infty)$. This follows from continuity combined with the boundary values $\bar{z}_X(0,\uc,\bar{c}) = 0$ and
\begin{align*}
	\lim_{\rx \nearrow 1/\bar{c}} \bar{z}_X(\rx,\uc,\bar{c}) &= \lim_{\rx \nearrow 1/\bar{c}} \frac{\rx \|\sigma_{W_1,X}\| \sqrt{1 - \max\{\min\{\bar{c}, \rx\},\uc\}^2}}{1 -\rx \max\{\min\{\rx,\bar{c}\},\uc\}} = +\infty. 
\end{align*}
To see the second equality, note that $1/\bar{c} \geq 1$ since $\bar{c} \leq 1$. Hence $\max\{\min \{ \rx, \bar{c} \},\uc\} \rightarrow \max\{\bar{c},\uc\} = \bar{c}$ as $\rx \nearrow 1/\bar{c}$. So $1 - \rx \bar{c} \rightarrow 0$ along this sequence and thus the denominator converges to zero. The numerator converges the constant $(1/\bar{c}) \| \sigma_{W_1,X} \| \sqrt{ 1 - \bar{c}^2 }$. For $\bar{c} < 1$, putting these results together shows that the desired sequence $\rx^{(n)}$ exists. For $\bar{c} = 1$, any sequence $\rx^{(n)} \nearrow 1$ has $\max\{\min \{ \rx^{(n)}, \bar{c} \},\uc\} = \rx^{(n)}$ and hence
$\bar{z}_X(\rx^{(n)},\uc,\bar{c}) = \frac{\rx^{(n)}}{\sqrt{1 - (\rx^{(n)})^2}} \| \sigma_{W_1,X} \|$ for $n$ sufficiently large, which converges to $+\infty$ as $\rx^{(n)} \nearrow 1$. The proof of $\underline{B}(\rx, \uc,\bar{c}) = -\infty$ is analogous.
\end{enumerate}
\end{proof}

\begin{proof}[Proof of Theorem \ref{cor:IdsetRyANDcFree}]
This result follows as a corollary of Theorem \ref{thm:IdsetRyFree} by setting $[\uc,\bar{c}] = [0,1]$, and by showing that the scalar $W_2$ assumption is a normalization, which we do next. Our approach here is similar to how \cite{AltonjiElderTaber2005} and \cite{Oster2019} allow for vector unobservables.

To distinguish the vector from scalar case, let $\widetilde{W}_2$ denote the vector of unobservables. Then equation \eqref{eq:XprojectionW1W2} is written as
\[
	X = \pi_1' W_1 + \widetilde{\pi}_2' \widetilde{W}_2 + X^{\perp W_1, \widetilde{W}_2}.
\]
Write
\begin{align*}
	\widetilde{\pi}_2' \widetilde{W}_2
	&= \sqrt{\var(\widetilde{\pi}_2' \widetilde{W}_2)} \cdot \frac{\widetilde{\pi}_2' \widetilde{W}_2}{\sqrt{\var(\widetilde{\pi}_2' \widetilde{W}_2)} } =: \pi_2 \cdot W_2.
\end{align*}
Thus we define $\pi_2$ to be the standard deviation of the index $\widetilde{\pi}_2' \widetilde{W}_2$ and define $W_2$ to be the standardized version of this index. Using this definition of $W_2$, we can now apply Theorem \ref{thm:IdsetRyFree}. Note that this approach does not require researchers to specify whether they think $W_2$ is a vector or not a priori. Finally, observe the following:
\begin{enumerate}
\item By construction, $\var(W_2) = 1$.

\item Next we'll show that this generalization is consistent with our analysis of the scalar $W_2$ case. Specifically, suppose $\widetilde{W}_2$ is scalar. Then
\begin{align*}
	\pi_2
	&= \sqrt{ \var(\widetilde{\pi}_2 \widetilde{W}_2) } = | \widetilde{\pi}_2 | \sqrt{\var( \widetilde{W}_2 )}
\end{align*}
and
\[
	W_2 = \frac{\widetilde{W}_2}{\sqrt{ \var(\widetilde{W}_2)}}.
\]
Thus $W_2$ is the standardized version of the original unobserved covariate and $\pi_2$ is the standardized version of its coefficient.

\item Consider \ref{assump:rx}. Using the above definition of $W_2$, it is equivalent to
\[
	\sqrt{ \var(\widetilde{\pi}_2' \widetilde{W}_2) }
	\leq
	\bar{r}_X \cdot \sqrt{\var(\pi_1' W_1)}.
\]
Thus the interpretation of this assumption does not change when the unobservables are a vector.

\item Consider the following two regressions:
\begin{enumerate}
\item OLS of $Y$ on $(1,X,W_1,\widetilde{W}_2)$. Denote the coefficient on $X$ by $\beta_{\text{long},\text{vec}}$. Let $(\gamma_{1,\text{vec}}, \gamma_{2,\text{vec}})$ denote the coefficients on $(W_1, \widetilde{W}_2)$.

\item OLS of $Y$ on $(1,X,W_1,\widetilde{\pi}_2' \widetilde{W}_2)$. Denote the coefficient on $X$ by $\beta_{\text{long},\text{index}}$. Let $(\gamma_{1,\text{index}}, \gamma_{2,\text{index}})$ denote the coefficients on $(W_1, \widetilde{\pi}_2' \widetilde{W}_2)$.
\end{enumerate}
Then we have $\beta_{\text{long},\text{vec}} = \beta_{\text{long},\text{index}}$. To see this, by definition of $(\pi_1, \widetilde{\pi}_2)$ as linear projection coefficients, we have
\[
	X = \pi_1' W_1 + \widetilde{\pi}_2' \widetilde{W}_2 + X^{\perp W_1, \widetilde{W}_2}. 
\]
Consequently, the coefficients on $(W_1, \widetilde{\pi}_2' \widetilde{W}_2)$ from a linear projection of $X$ on $(1,W_1, \widetilde{\pi}_2' \widetilde{W}_2)$ are $(\pi_1, 1)$. That is, we can write
\[
	X = \pi_1' W_1 + 1 \cdot (\widetilde{\pi}_2' \widetilde{W}_2) + X^{\perp W_1, \widetilde{\pi}_2' \widetilde{W}_2}. 
\]
This implies that $X^{\perp W_1, \widetilde{W}_2} = X^{\perp W_1, \widetilde{\pi}_2' \widetilde{W}_2}$. Hence, by the FWL theorem,
\[
	\beta_{\text{long},\text{vec}} = \frac{\cov(Y, X^{\perp W_1, \widetilde{W}_2})}{\var(X^{\perp W_1, \widetilde{W}_2})}
	= \frac{\cov(Y, X^{\perp W_1, \widetilde{\pi}_2' \widetilde{W}_2})}{\var(X^{\perp W_1, \widetilde{\pi}_2' \widetilde{W}_2})}
	= \beta_{\text{long},\text{index}}.	
\]
This shows that the definition of the parameter of interest, $\beta_\text{long}$, does not depend on whether $W_2$ is a vector or a scalar.
\end{enumerate}
\end{proof}

\begin{proof}[Proof of Corollary \ref{corr:breakdownPointRXonly}]
Note that we can equivalently write
\[
	\rx^\text{bp} = \inf \{ \rx \geq 0 : b \in \mathcal{B}_I(\rx) \text{ for some $b \leq 0$} \}.
\]
If $\beta_\text{med} \geq 0$, then 
\begin{align*}
	\rx^\text{bp}
	&= \inf \{\rx \geq 0: b \in [\underline{B}(\rx),\overline{B}(\rx)] \text{ for some } b \leq 0\}\\
	&= \inf \left\{ \rx \geq 0: \frac{\var(Y^{\perp X,W_1})}{\var(X^{\perp W_1})} \cdot \frac{\rx^2 R^2_{X \sim W_1} }{1 - R^2_{X \sim W_1} - \rx^2 } = \beta_\text{med}^2 \right\}.
\end{align*}
The second line follows by monotonicity and continuity of $\underline{B}(\bar{r}_X)$. The same equality holds when $\beta_\text{med} \leq 0$. Thus $\rx^\text{bp}$ is the unique non-negative solution to  $\beta_\text{med}^2 = \text{dev}(\rx)^2$. To complete the proof, we solve for this solution. First note that
\[
	\beta_\text{med}^2\frac{\var(X^{\perp W_1})}{\var(Y^{\perp W_1})} =: R^2_{Y\sim X \sbullet W_1}
 \qquad \text{and} \qquad
	\frac{\var(Y^{\perp X,W_1})}{\var(Y^{\perp W_1})} = \frac{\var(Y)(1 - R^2_{Y \sim X,W_1})}{\var(Y)(1 - R^2_{Y \sim W_1})} = 1 - R^2_{Y\sim X \sbullet W_1}.
\]
Thus, multiplying by $\var(X^{\perp W_1})/\var(Y^{\perp W_1})$,  $\rx^\text{bp}$ solves
\begin{align*}
	(1 - R^2_{Y\sim X \sbullet W_1})\frac{\big( \rx^\text{bp} \big)^2 R^2_{X \sim W_1} }{1 - R^2_{X \sim W_1} - \big( \rx^\text{bp} \big)^2 } = R^2_{Y\sim X \sbullet W_1}
\end{align*}
or
\begin{align*}
	\big( \rx^\text{bp} \big)^2 \left(R^2_{X \sim W_1}(1-R^2_{Y\sim X \sbullet W_1}) + R^2_{Y\sim X \sbullet W_1}\right) &= R^2_{Y\sim X \sbullet W_1}(1-R^2_{X \sim W_1}).
\end{align*}
Therefore
$
	\rx^\text{bp} = \left(R^2_{Y\sim X \sbullet W_1}/\left(\frac{R^2_{X \sim W_1}}{1 - R^2_{X \sim W_1}} + R^2_{Y\sim X \sbullet W_1}\right)\right)^{1/2}.
$
\end{proof}

\begin{proof}[Proof of Equation \eqref{eq:ourOVBformula}]
By the FWL theorem, $\beta_{\text{med}} = \cov(Y,X^{\perp W_1})/\var(X^{\perp W_1})$. Next,
\begin{align*}
	\cov(Y,X^{\perp W_1})
	&= \cov(X\beta_\text{long} + \gamma_1'W_1 + \gamma_2' W_2 + Y^{\perp X,W},X^{\perp W_1})\\
	&= \beta_\text{long} \var(X^{\perp W_1}) + \gamma_2'\cov(W_2,(\pi_1'W_1 + \pi_2'W_2 + X^{\perp W})^{\perp W_1})\\
	&= \beta_\text{long} \var(X^{\perp W_1}) + \gamma_2'\cov(W_2, \pi_2'W_2^{\perp W_1} + X^{\perp W})\\
	&= \beta_\text{long} \var(X^{\perp W_1}) + \gamma_2'\var(W_2^{\perp W_1}) \pi_2.
\end{align*}
When $W_2$ is scalar and $\var(W_2) = 1$, $\var(W_2^{\perp W_1}) = 1 - R^2_{W_2 \sim W_1}$. This gives the desired result. 
\end{proof}

\begin{proof}[Proof of Theorem \ref{cor:BFCalculation3D}]
Note that we can equivalently write
\[
	\bar{r}_Y^\text{bf}(\rx, \bar{c}, \underline{b})
	= \inf \{ \bar{r}_Y \geq 0 : b \in \mathcal{B}_I(\rx,\ry,\uc, \bar{c}) \text{ for some $b \leq \underline{b}$} \}.
\]
We use this version in our proof. We will consider each case stated in the theorem separately.

\bigskip

\noindent\textbf{Case 1} If $\underline{b} \ge \beta_\text{med}$, then $\ry^\text{bf}(\rx, \uc,\bar{c}; \underline{b}) = 0$.

\bigskip

By Theorem \ref{thm:IdsetRyFree}, $\beta_\text{med} \in\mathcal{B}_I(\rx,0,\uc,\bar{c})$. We then have
$
	\inf\{\ry \ge 0: b \in \mathcal{B}_I(\rx,\ry,\uc,\bar{c}) \text{ for some } b \le \underline{b}\} = 0
$
by letting $b = \beta_\text{med} \leq \underline{b}$.

\bigskip

\noindent\textbf{Case 2} If $\underline{B}(\rx,\uc,\bar{c}) > \underline{b}$, then $\ry^\text{bf}(\rx,\uc, \bar{c}; \underline{b}) = +\infty$.

\bigskip

Note that $\mathcal{B}_I(\rx, \ry,\uc, \bar{c}) \subseteq \mathcal{B}_I(\rx, +\infty,\uc, \bar{c}) \subseteq [\underline{B}(\rx,\uc,\bar{c}),\overline{B}(\rx,\uc,\bar{c})]$. Therefore, 
\begin{align*}
	&\{\ry \geq 0: b \in \mathcal{B}_I(\rx,\ry,\uc,\bar{c}) \text{ for some } b \le \underline{b}\} \\
	&\qquad \subseteq \left\{ \ry \geq 0: b \in [\underline{B}(\rx,\uc,\bar{c}),\overline{B}(\rx,\uc,\bar{c})] \text{ for some } b \le \underline{b} \right\} = \emptyset
\end{align*} since $\underline{B}(\rx,\uc,\bar{c}) > \underline{b}$.
Hence
$
	\inf\{\ry \geq 0: b \in \mathcal{B}_I(\rx,\ry,\uc,\bar{c}) \text{ for some } b \le \underline{b}\} = +\infty.
$

\bigskip

\noindent\textbf{Case 3} If $\underline{B}(\rx,\uc,\bar{c}) \leq \underline{b} < \beta_\text{med}$.

\bigskip

\noindent First, note that
\begin{align*}
	&\bar{r}_Y^\text{bf}(\rx,\uc, \bar{c}, \underline{b}) \\
	&= \inf \{\ry \geq 0: b \in \mathcal{B}_I(\rx,\ry,\uc,\bar{c}) \text{ for some } b \le \underline{b}\} \\
	&= \inf\{ \ry \geq 0: \text{There is a $(b,r_X,r_Y,c)$ s.t. } b \in \mathcal{B}(r_X,r_Y,c), \|r_X\| \le \rx, \|r_Y\| \le \ry, \|c\| \in[\uc, \bar{c}], b \le \underline{b} \} \\
	&= \inf\{ \|r_Y\| : \text{There is a $(b,r_X,r_Y,c)$ s.t. } b \in \mathcal{B}(r_X,r_Y,c), \|r_X\| \le \rx, r_Y \in \mathbb{R}^{d_1}, \|c\| \in [\uc, \bar{c}], b \le \underline{b} \} \\
	&= \inf \left\{\|r_Y\|: r_Y \in \bigcup_{\|r_X\| \leq \rx, \|c\| \in [\uc, \bar{c}], b \le \underline{b}}\{ r_Y \in \mathbb{R}^{d_1}: b \in \mathcal{B}(r_X,r_Y,c)\}\right\}.
\end{align*}
The second equality follows from the definition of $\mathcal{B}_I(\rx,\ry,\uc,\bar{c})$. The third holds because $\ry \ge \|r_Y\|$, so it is minimized by setting $\ry = \|r_Y\|$. 

Next we can write
\begin{align*}
	&\bigcup_{\|r_X\| \leq \rx, \|c\| \in [\uc, \bar{c}], b \le \underline{b}}\{ r_Y \in \mathbb{R}^{d_1}: b \in \mathcal{B}(r_X,r_Y,c)\}\\
	&= \bigcup_{\|r_X\| \leq \rx, \|c\| \in [\uc, \bar{c}], b \le \underline{b}}\{ r_Y \in \mathbb{R}^{d_1}: \text{\eqref{eq:IDsetEqAltA}--\eqref{eq:IDsetIneq_c} hold for $(r_X,r_Y,c,b,z)$ and for some } z \in \R\setminus \{0\}\}\\
	&= \bigcup_{\|r_X\| \leq \rx, \|c\| \in [\uc, \bar{c}], b \le \underline{b}}\{ r_Y \in \mathbb{R}^{d_1}: \eqref{eq:IDsetEqAltA} \text{ holds for $(r_X, r_Y, c, b, z)$ where } (r_X,c,b,z) \in A(\rx,\uc,\bar{c}), z \in \R\setminus \{0\}\}\\
	&= \bigcup_{(r_X,c,b,z) \in A(\rx,\uc,\bar{c}), b \le \underline{b}}\{ r_Y \in \mathbb{R}^{d_1}: \eqref{eq:IDsetEqAltA} \text{ holds for } (r_X, r_Y, c, b, z)\}.
\end{align*}
The first equality follows by Lemma \ref{lemma:IDsetAlt}, which we can apply since we consider $b$ satisfying $b \leq \underline{b} < \beta_\text{med}$. The second equality follows by definition of the set $A(\rx,\uc, \bar{c})$. The third equality follows since the inequalities $\|r_X\| \leq \rx$ and $\|c\| \in [\uc, \bar{c}]$ are part of the definition of $A(\rx, \uc, \bar{c})$.
Thus
\begin{align*}
	&\bar{r}_Y^\text{bf}(\rx, \uc, \bar{c}, \underline{b}) \\
	&= \inf \left\{\|r_Y\|: r_Y \in \bigcup_{(r_X,c,b,z) \in A(\rx,\uc,\bar{c}), b \le \underline{b}}\{ r_Y \in \mathbb{R}^{d_1}: \eqref{eq:IDsetEqAltA} \text{ holds for } (r_X, r_Y, c, b, z)\} \right\} \\
	&= \inf\Big\{ \inf\big\{ \|r_Y\| : r_Y \in \mathbb{R}^{d_1}, \eqref{eq:IDsetEqAltA} \text{ holds for } (r_X, r_Y, c, b, z) \big\} : (r_X,c,b,z) \in A(\rx,\uc, \bar{c}), b \le \underline{b}\Big\}  \\
	&= \inf \left\{\underline{r}_Y(z,c,b): (r_X,c,b,z) \in A(\rx,\uc,\bar{c}), b \le \underline{b} \right\}.
\end{align*}
The last line follows by Lemma \ref{lemma:IDsetRy_claim1}. We have just shown that $\bar{r}_Y^\text{bf}(\rx, \uc,\bar{c}, \underline{b})$ can be computed as the solution to a constrained minimization problem with objective function $\underline{r}_Y(z,c,b)$. We complete the proof by showing that the constraints $(r_X,c,b,z) \in A(\rx,\uc,\bar{c}), b \le \underline{b}$ here are equivalent to the constraints given in the statement of the theorem.

The constraint set is
\begin{align*}
	&\{(r_X,c,b,z) \in A(\rx,\uc,\bar{c}) : b \le \underline{b}\} \\[0.5em]
	&= \{ (r_X,c,b,z) : \text{\eqref{eq:IDsetEqAltB}--\eqref{eq:IDsetIneqBeta} hold}, \|r_X\| \leq \rx, \|c\| \in [\uc, \bar{c}], b \le \underline{b}, z \in \R\setminus\{0\}\} \\[0.5em]
	&= \{ (r_X,c,b,z) : (b - \beta_\text{med})^2 < \text{devsq}(z), z^2 < k_0, \\
	&\hspace{40mm} z = r_X'(\sqrt{1 - \|c\|^2}\sigma_{W_1,X} - cz), \|r_X\| \le \rx, \|c\| \in [\uc, \bar{c}]\cap[0,1), b \le \underline{b}, z \in \R\setminus\{0\}\}\\[0.5em]
	&= \{ (r_X,c,b,z) : (b - \beta_\text{med})^2 < \text{devsq}(z), 0<z^2 < k_0, p(z,c;\rx) \ge 0, \|c\| \in [\uc, \bar{c}]\cap[0,1), b \le \underline{b}, r_X \in \R^{d_1} \}.
\end{align*}
The first equality follows by definition of $A(\rx,\uc,\bar{c})$. The second equality follows by definition of equations \eqref{eq:IDsetEqAltB}--\eqref{eq:IDsetIneqBeta}. The third equality follows from Lemma \ref{lem:rxbound}.

Thus we have shown that
\begin{equation}\label{eq:Theorem4simpleCase}
	\bar{r}_Y^\text{bf}(\rx, \uc, \bar{c}, \underline{b})
	= \inf \{\underline{r}_Y(z,c,b): (b - \beta_\text{med})^2 < \text{devsq}(z), 0<z^2 < k_0, p(z,c;\rx) \ge 0, \|c\| \in [\uc, \bar{c}]\cap[0,1), b \le \underline{b}\}.
\end{equation}
We can further write this as
\begin{align*}
	&\bar{r}_Y^\text{bf}(\rx, \uc,\bar{c}, \underline{b}) \\
	&= \inf \{ \inf \{\underline{r}_Y(z,c,b): p(z,c;\rx) \ge 0, c \in \R^{d_1}, \|c\| \in [\uc, \bar{c}]\cap[0,1) \} : (b - \beta_\text{med})^2 < \text{devsq}(z), 0<z^2 < k_0, b \le \underline{b} \} \\
	&= \inf \{ \inf \{\underline{r}_Y(z,c,b): p(z,c;\rx) \ge 0, c \in \vspan \{ \sigma_{W_1,X}, \sigma_{W_1,Y} \}, \|c\| \in [\uc, \bar{c}]\cap[0,1) \} : \\
	&\hspace{80mm} (b - \beta_\text{med})^2 < \text{devsq}(z), 0<z^2 < k_0, b \le \underline{b} \} \\
  &= \inf\{\underline{r}_Y(z ,c_1\sigma_{W_1,Y} + c_2\sigma_{W_1,X}, b):  (b - \beta_\text{med})^2 < \text{devsq}(z), z^2 < k_0,\\
  &\hspace{15mm} p(z,c_1\sigma_{W_1,Y} + c_2\sigma_{W_1,X};\rx) \ge 0, \|c_1\sigma_{W_1,Y} + c_2\sigma_{W_1,X}\| \in [\uc, \bar{c}]\cap[0,1), (c_1, c_2) \in \mathbb{R}^2, b \le \underline{b} \}.
\end{align*}
The second equality follows by Lemma \ref{lem:c2dim}, which uses the assumptions $d_1 \geq 2$ and $(\sigma_{W_1,X},\sigma_{W_1,Y})$ are linearly independent. Moreover, note that the outer infimum constraint set implies that $z \neq 0$ and $b \neq \beta_\text{med}$ (since $\underline{b} < \beta_\text{med}$), which are also needed to apply Lemma \ref{lem:c2dim}. The third equality follows by combining the two infimums and using the definition of the span.
\end{proof}

\begin{proof}[Proof of Theorem \ref{cor:BFCalculation_rx_and_ry}]
This is obtained directly by setting $[\uc,\bar{c}] = [0,1]$ in the proof of Theorem \ref{cor:BFCalculation3D}.
\end{proof}

\subsection{Proofs for Section \ref{sec:causalModels}}

\begin{proof}[Proof of Proposition \ref{prop:unconfoundednessBaseline}]
We have
\begin{align*}
	\cov(Y, X^{\perp W_1,W_2})
	&= \cov(\beta_c X + \gamma_1' W_1 + \gamma_2' W_2 + U, X^{\perp W_1,W_2}) \\
	&= \beta_c \cov(X, X^{\perp W_1,W_2}) + \cov(U, X^{\perp W_1,W_2}) \\
	&= \beta_c \cov(X^{\perp W_1,W_2}, X^{\perp W_1,W_2}) + \cov(U^{\perp W_1,W_2}, X^{\perp W_1,W_2}) \\
	&= \beta_c \var(X^{\perp W_1,W_2}).
\end{align*}
The first line follows by the linear potential outcomes assumption. The last line follows by linear latent unconfoundedness. Thus
\[
	\beta_c = \frac{\cov(Y, X^{\perp W_1,W_2})}{ \var(X^{\perp W_1,W_2})}.
\]
The term on the right equals $\beta_\text{long}$ by the FWL theorem. Note that the denominator here is nonzero by \ref{assump:posdefVar}.
\end{proof}

\begin{proof}[Proof of Corollary \ref{corr:SelectionOnObsCausal}]
Proposition \ref{thm:baselineEndogControl1} implies that $\beta_\text{long} = \beta_\text{med}$. Proposition \ref{prop:unconfoundednessBaseline} implies that $\beta_c = \beta_\text{long}$.
\end{proof}

\begin{proof}[Proof of Proposition \ref{prop:DiD}]
We have
\begin{align*}
	\cov(\Delta Y, (\Delta X)^{\perp \Delta W_1, \Delta W_2})
	&= \beta_c \cov(\Delta X, (\Delta X)^{\perp \Delta W_1, \Delta W_2}))
	+ \cov(\Delta V, (\Delta X)^{\perp \Delta W_1, \Delta W_2}) \\
	&= \beta_c \var\big( (\Delta X)^{\perp \Delta W_1, \Delta W_2} \big).
\end{align*}
The first line follows by linear potential outcomes. The second line follows by exogeneity. The result then follows by the FWL theorem.
\end{proof}

\begin{proof}[Proof of Corollary \ref{corr:DiD}]
Proposition \ref{thm:baselineEndogControl1} implies that $\beta_\text{long} = \beta_\text{med}$. Proposition \ref{prop:DiD} implies that $\beta_c = \beta_\text{long}$.
\end{proof}

\subsection{Proofs for Appendix \ref{sec:generalIdentSet}}

\textbf{Note}: In this section we drop tildes from $\tilde{r}_X$ and $\tilde{r}_Y$ for brevity. 

\smallskip

We use the following lemma to prove Theorem \ref{thm:mainMultivariateResult}.

\begin{lemma}\label{lemma:relativeBoundEquivalences}
Let $W_2$ be scalar with $\var(W_2) = 1$.
\begin{enumerate}
\item \ref{assump:rx} holds if and only if there exists $r_X \in \R^{d_1}$ such that $\pi_2 = r_X' \pi_1$ and $\| r_X \|_{\Sigma_{\text{obs}}^{-1}} \leq \rx$.

\item \ref{assump:ry} holds if and only if there exists $r_Y \in \R^{d_1}$ such that $\gamma_2 = r_Y' \gamma_1$ and $\| r_Y \|_{\Sigma_{\text{obs}}^{-1}} \leq \ry$.
\end{enumerate}
\end{lemma}

\begin{proof}[Proof]
We prove part 1; the proof of part 2 is analogous. First, let \ref{assump:rx} hold. If $\pi_1 = 0$, then the conclusion trivially holds by setting $r_X = 0$. If $\pi_1 \neq 0$, let
\[
	r_X = \frac{\pi_2}{\|\pi_1\|_{\Sigma_{\text{obs}}}^2} \Sigma_\text{obs}\pi_1.
\]
We can see that $r_X'\pi_1  = \pi_2$ and $\| r_X \|_{\Sigma_{\text{obs}}^{-1}} = \frac{|\pi_2|}{\|\pi_1\|_{\Sigma_{\text{obs}}}^2} \cdot \| \Sigma_{\text{obs}} \pi_1 \|_{\Sigma_\text{obs}^{-1}} = \frac{|\pi_2|}{\|\pi_1\|_{\Sigma_{\text{obs}}}}\leq \rx$. The inequality follows by \ref{assump:rx}. Thus the $\Rightarrow$ direction holds.

Next assume there is an $r_X \in \R^{d_1}$ such that $\pi_2 = r_X' \pi_1$ and $\| r_X \|_{\Sigma_{\text{obs}}^{-1}} \leq \rx$. Let $\Sigma_\text{obs}^{1/2}$ denote the unique square root of the positive-definite matrix $\Sigma_\text{obs}$. Then
$$|\pi_2| = |r_X'\pi_1| = |r_X'\Sigma_\text{obs}^{-1/2}\Sigma_\text{obs}^{1/2}\pi_1| \leq \|r_X'\Sigma_\text{obs}^{-1/2}\| \cdot \|\Sigma_\text{obs}^{1/2}\pi_1\| = \|r_X\|_{\Sigma_\text{obs}^{-1}} \|\pi_1\|_{\Sigma_\text{obs}} \leq \rx \|\pi_1\|_{\Sigma_\text{obs}}.$$ The first inequality follows from the Cauchy-Schwarz inequality. The second inequality was assumed. Therefore \ref{assump:rx} holds.
\end{proof}

\begin{proof}[Proof of Theorem \ref{thm:mainMultivariateResult}]
We prove this result in two parts. First we show that the true value of $\beta_\text{long}$ is in $\mathcal{B}_I(\rx,\ry,\uc,\bar{c})$. Second we show sharpness: Any $b$ in $\mathcal{B}_I(\rx,\ry,\uc,\bar{c})$ is consistent with the distribution of the observables and the model assumptions.

\bigskip

\emph{Part 1 (Outer set).} By \ref{assump:rx} and \ref{assump:ry}, Lemma \ref{lemma:relativeBoundEquivalences} implies that there exists $r_X, r_Y \in \R^{d_1}$ such that $\pi_2 = r_X' \pi_1$ and $\gamma_2 = r_Y' \gamma_1$ with $\| r_X \|_{\Sigma_{\text{obs}}^{-1}} \leq \rx$ and $\| r_Y \|_{\Sigma_{\text{obs}}^{-1}} \leq \ry$. Let $c \coloneqq \cov(W_1,W_2)$ and note that $\| c \|_{\Sigma_{\text{obs}}^{-1}} \in [\uc,\bar{c}]$ holds by \ref{assump:corr}. By the definition of $\mathcal{B}_I(\rx, \bar{r}_Y,\uc, \bar{c})$, it therefore suffices to show that $\beta_\text{long} \in \mathcal{B}(r_X, r_Y, c)$ for these values of $(r_X, r_Y, c)$.

We have
\begin{align*}
	\cov(X,W_1) &= \cov(\pi_1' W_1 + \pi_2 W_2 + X^{\perp W}, W_1) = \pi_1' \var(W_1) + \pi_2 c' 
= \pi_1' (\var(W_1) + r_X c').
\end{align*}
Similarly,
\begin{align*}
	\cov(X,W_2) &= \cov(\pi_1' W_1 + \pi_2 W_2 + X^{\perp W}, W_2) = \pi_1' c + (\pi_1'r_X) \var(W_2) = \pi_1' (c + r_X).
\end{align*}
We used $\var(W_2) = 1$ in the third line. Next consider
\begin{align*}
	\cov(Y,X) &= \cov(\beta_\text{long} X + \gamma_1'W_1 + \gamma_2 W_2 + Y^{\perp X,W}, X) \\
	&= \beta_\text{long} \var(X) + \gamma_1' \cov(W_1,X) + \gamma_2 \cov(W_2, X) \\
	&= \beta_\text{long} \var(X)  + \gamma_1'(\var(W_1) + c r_X' + r_Y c' + r_Yr_X')\pi_1
\end{align*}
and
\begin{align*}
	\cov(Y,W_1)	&= \cov(\beta_\text{long} X + \gamma_1' W_1 + \gamma_2 W_2 + Y^{\perp X,W}, W_1) 
= \beta_\text{long} \cov(X,W_1) + \gamma_1' (\var(W_1) + r_Y c').
\end{align*}
These results imply that equalities \eqref{eq:IDseteq1}--\eqref{eq:IDseteq3} hold when $b = \beta_\text{long}$ by setting $(p_1,g_1) = (\pi_1,\gamma_1)$. 

Next we verify that inequalities \eqref{eq:IDseteq4}--\eqref{eq:IDseteq6} hold when $(b,p_1,g_1) = (\beta_\text{long},\pi_1, \gamma_1)$.
\begin{enumerate}[wide, labelwidth=!, labelindent=0pt]
\item Consider inequality \eqref{eq:IDseteq4}. By \ref{assump:posdefVar}, $\var(Y^{\perp X,W}) > 0$. Therefore
\begin{align*}
	\var(Y)
	&= \var(X\beta_\text{long} + \gamma_1' W_1 + \gamma_2 W_2) + \var(Y^{\perp X,W}) \\[0.5em]
	&> \beta_\text{long}^2 \var(X) + \gamma_1'\var(W_1)\gamma_1 + \gamma_2^2  + 2 \beta_\text{long} \gamma_1'\cov(W_1,X) + 2 \beta_\text{long} \gamma_2 \cov(X,W_2) + 2 \gamma_1' c\gamma_2\\[0.5em]
	&= \beta_\text{long}^2\var(X) + \gamma_1'\left(\var(W_1) + r_Yr_Y' + 2 c r_Y'\right)\gamma_1 + 2\beta_\text{long}\pi_1'( \var(W_1) +  r_Xc' + (c+r_X)r_Y')\gamma_1.
\end{align*}
Hence inequality \eqref{eq:IDseteq4} holds.

\item Consider inequality \eqref{eq:IDseteq5}. By \ref{assump:posdefVar}, $\var(X^{\perp W}) > 0$. Therefore
\begin{align*}
	\var(X) &= \var(\pi_1' W_1 + \pi_2 W_2) + \var(X^{\perp W})\\
	&> \pi_1'\var(W_1) + 2\pi_1'c \pi_2 + \pi_2^2\\
	&= \pi_1'(\var(W_1) + 2 c r_X' + r_Xr_X')\pi_1.
\end{align*}
Hence inequality \eqref{eq:IDseteq5} holds.

\item Consider inequality \eqref{eq:IDseteq6}. By \ref{assump:posdefVar}, $R^2_{W_2\sim W_1} <1$. Therefore
\begin{align*}
	1 > R^2_{W_2\sim W_1} = \cov(W_2,W_1) \var(W_1)^{-1} \cov(W_1,W_2) = c'\var(W_1)^{-1}c.
\end{align*}
Hence inequality \eqref{eq:IDseteq6} holds. Thus $\beta_\text{long} \in \mathcal{B}(r_X, r_Y, c)$, which implies $\beta_\text{long} \in \mathcal{B}_I(\rx,\ry,\uc,\bar{c})$.
\end{enumerate}

\bigskip

\emph{Part 2 (Sharpness)}. Let $b \in \mathcal{B}_I(\rx,\ry,\uc,\bar{c})$. We will construct a distribution for $(Y,X,W_1,W_2)$ consistent with $\beta_\text{long} = b$ and with all of our assumptions and the known distribution of $(Y,X,W_1)$.

\bigskip

\textbf{Part 1: Constructing the variance matrix}. Since $b \in \mathcal{B}_I(\rx,\ry,\uc,\bar{c})$, let $(r_X,r_Y,c)$ be such that $b \in \mathcal{B}(r_X,r_Y,c)$, $\| r_X \|_{\Sigma_{\text{obs}}^{-1}} \leq \rx$, $\| r_Y \|_{\Sigma_{\text{obs}}^{-1}} \leq \ry$, and $\| c \|_{\Sigma_{\text{obs}}^{-1}} \in [\uc,\bar{c}]$. Using the definition of $\mathcal{B}(r_X,r_Y,c)$, let $(p_1,g_1) \in \R^{d_1} \times \R^{d_1}$ be such that equations \eqref{eq:IDseteq1}--\eqref{eq:IDseteq6} hold.

Define the matrix
\begin{align*}
	\Sigma
	&\coloneqq
	\begin{pmatrix}
		1 & c' & p_1'(c+r_X) & b p_1'(c+r_X) + g_1'(c+r_Y)\\
		c & \var(W_1) & \cov(W_1,X) & \cov(W_1,Y)\\
		p_1'(c+r_X) & \cov(X,W_1) & \var(X) & \cov(X,Y)\\
		b p_1'(c+r_X) + g_1'(c+r_Y) & \cov(Y,W_1) & \cov(Y,X) & \var(Y)
	\end{pmatrix}.
\end{align*}
We will use this matrix to construct $W_2$. Note that the bottom right $(d_1+2) \times (d_1+2)$ block is $\var(W_1,X,Y)$. We next show that $\Sigma$ is positive definite. To do this, we apply Sylvester's criterion as in the proof of Proposition \ref{thm:baselineEndogControl1}. First, consider $\Sigma_2$, which we define as the $(1+d_1)\times (1+d_1)$ leading principal minor of $\Sigma$. Repeating the arguments from part 2(i) of the proof of Proposition \ref{thm:baselineEndogControl1}, $\Sigma_2$ is positive definite when $c'\var(W_1)^{-1} c < 1$, which holds by equation \eqref{eq:IDseteq6}.

Next consider the $(1+d_1+1)\times (1+d_1+1)$ leading principal minor of $\Sigma$,  denoted by $\Sigma_3$. Its determinant equals
\begin{align*}
	\det(\Sigma_3)
	&=\det \begin{pmatrix}
		1 & c' & (c+r_X)'p_1 \\
		c & \var(W_1) & (\var(W_1) + c r_X') p_1 \\
		p_1'(c+r_X) & p_1'(\var(W_1) + r_X c') & \var(X)
	\end{pmatrix} \\
	&= \det(\Sigma_2) \cdot \left(\var(X) - \begin{pmatrix}
	p_1'(c+r_X) & p_1'(\var(W_1) + r_X c')
	\end{pmatrix}\Sigma_2^{-1} \begin{pmatrix}
	(c+r_X)'p_1 \\ (\var(W_1) + c r_X') p_1
	\end{pmatrix}\right)\\
	&= \det (\Sigma_2) \cdot (\var(X) - p_1'\left(r_X r_X' + 2 r_X c' + \var(W_1) \right)p_1)\\
	&>0.
\end{align*}
In the first line we expanded $\cov(X,W_1)$. In the third line we used
\[
	\Sigma_2^{-1}
	=
	\frac{1}{1 - \| c \|^2_{\Sigma_{\text{obs}}^{-1}}}
	\begin{pmatrix}
		1 & -c' \var(W_1)^{-1} \\
		- \var(W_1)^{-1} c & \var(W_1)^{-1} (1- c' \var(W_1)^{-1} c) + \var(W_1)^{-1} c c' \var(W_1)^{-1}
	\end{pmatrix}.
\]
The inequality follows from $\det(\Sigma_2) > 0$, and from inequality \eqref{eq:IDseteq5} in the definition of $b \in \mathcal{B}(r_X,r_Y,c)$.

Finally, the determinant of the entire matrix is
\begin{align*}
	\det(\Sigma)
	&= \det(\Sigma_3)
	\cdot \left( \var(Y) 
	- \left\|\begin{pmatrix}
	b p_1'(c+r_X) + g_1'(c+r_Y) \\
	b \cov(X,W_1) + g_1' (\var(W_1) + r_Y c') \\
	b \var(X)  + g_1'(\var(W_1) + c r_X' + r_Y c' + r_Yr_X')p_1
	\end{pmatrix}\right\|_{\Sigma_3^{-1}}\right)\\
	&= \det (\Sigma_3) \cdot \Big( \var(Y) - (b^2\var(X) + g_1'\var(W_1)g_1 \\
	&\qquad \qquad \qquad + 2bp_1'(\var(W_1)+r_X c')g_1 + (g_1'r_Y)^2 + 2bg_1' r_Y p_1'(c+r_X) + 2g_1'r_Yg_1'c) \Big) >0.
\end{align*}
The first line follows since $\Sigma_3$ is invertible and by the determinant formula for partitioned matrices.  The second line follows by similar calculations as earlier. The last line follows since we have already shown that $\det(\Sigma_3) > 0$, and from inequality \eqref{eq:IDseteq4} in the definition of $\mathcal{B}(r_X,r_Y,c)$. Thus the matrix $\Sigma$ is positive definite by Sylvester's Criterion.

\bigskip

\textbf{Part 2: Constructing the joint distribution of $(Y,X,W_1,W_2)$.} 
We next use $\Sigma$ to construct the unobservable $\Wu$. Since $\Sigma$ and $\var(W_1,X,Y)$ are positive definite, the partitioned inverse matrix formula yields that
\begin{align*}
	v \coloneqq 1 - \begin{pmatrix}
	c' & p_1'(c + r_X) & bp_1'(c + r_X) + g_1'(c+r_Y)
	\end{pmatrix}\var(W_1,X,Y)^{-1}\begin{pmatrix}
	c \\ p_1'(c + r_X) \\ bp_1'(c + r_X) + g_1'(c+r_Y)
	\end{pmatrix} > 0.
\end{align*}
Let $\widetilde{W_2}$ be a unit-variance random variable that is uncorrelated with $(W_1,X,Y)$ and let
\begin{align*}
	W_2 &= \begin{pmatrix}
	c' & p_1'(c + r_X) & bp_1'(c + r_X) + g_1'(c+r_Y)
	\end{pmatrix}\var(W_1,X,Y)^{-1}\begin{pmatrix}
		W_1\\
		X\\
		Y
	\end{pmatrix} + v^{1/2} \widetilde{W}_2.
\end{align*}
We can verify that $\var(W_2,W_1,X,Y) = \Sigma$. Thus we have constructed a joint distribution of $(W_2,W_1,X,Y)$. By construction, this distribution is consistent with the distribution of $(Y,X,\Wob)$.

\bigskip

\textbf{Part 3: Verifying consistency with the assumptions and equations \eqref{eq:outcome} and \eqref{eq:XprojectionW1W2}}. 

We can immediately see that \ref{assump:posdefVar} holds since $\Sigma$ is positive definite. $W_2$ is a scalar with $\var(W_2)=1$ by definition of $W_2$. \ref{assump:corr} holds by $\cov(W_1,W_2) = c$ and $\|c\|_{\Sigma_{\text{obs}}^{-1}} \in [\uc,\bar{c}]$. We show that \ref{assump:rx} and \ref{assump:ry} hold below.

Next we show that $Y^{\perp X,W} = Y - b X - g_1'W_1 - (r_Y'g_1) W_2$ by verifying that it is uncorrelated with $(X,W_1,W_2)$. First,
\begin{align*}
	&\cov(Y - b X - g_1' W_1 - (r_Y'g_1)W_2,X)
	= (\cov(Y,X) - b \var(X)) - g_1'\cov(W_1,X) -(r_Y'g_1) p_1'(r_X + c)\\
	&= g_1'(\var(W_1) + c r_X' + r_Y c' + r_Yr_X')p_1  - g_1'(\var(W_1) + c r_X')p_1 - g_1' (r_Y r_X' + r_Yc')p_1\\
	&= 0
\end{align*}
where we used equations \eqref{eq:IDseteq1}--\eqref{eq:IDseteq3}. Using these equations again,
\begin{align*}
	&\cov(Y - b X - g_1'W_1 - (r_Y'g_1) W_2,W_1) = \cov(Y,W_1) - b \cov(X,W_1) -g_1'\var(W_1) - (r_Y'g_1) c'\\
	&= \left(b \cov(X,W_1) + g_1' (\var(W_1) + r_Y c')\right) - b \cov(X,W_1) -g_1'\var(W_1) - (g_1' r_Y) c'\\
	&= 0.
\end{align*}
Finally,
\begin{align*}
	\cov(Y - b X - g_1'W_1 - (r_Y'g_1) W_2,W_2) &= \cov(Y,W_2) - b \cov(X,W_2) -g_1'c -r_Y'g_1 \\
	&= \left(bp_1'(c+r_X) + g_1'(c+r_Y)\right) - bp_1'(c+r_X) - g_1'c - g_1' r_Y\\
	&= 0.
\end{align*}
Thus equation \eqref{eq:outcome} holds with $(\beta_\text{long},\gamma_1,\gamma_2) = (b,g_1,r_Y'g_1)$. By Lemma \ref{lemma:relativeBoundEquivalences}, this implies \ref{assump:ry} holds. Finally, we show that $X^{\perp W} = X - p_1'W_1 - (r_X'p_1) W_2$ by verifying it is uncorrelated with $(W_1,W_2)$:
\begin{align*}
	\cov(X - p_1'W_1 - (r_X'p_1) W_2,W_1) &= p_1'(\var(W_1)+r_X c') - p_1'\var(W_1) - r_X' p_1 c' = 0\\
	\cov(X - p_1'W_1 - (r_X'p_1) W_2,W_2) &= p_1' (r_X + c) - p_1'c - r_X' p_1 =0.
\end{align*}
Here we used $\cov(X,W_1) = p_1'(\var(W_1)+r_X c')$. Therefore equation \eqref{eq:XprojectionW1W2} holds with $(\pi_1,\pi_2) = (p_1,p_1'r_X)$. By Lemma \ref{lemma:relativeBoundEquivalences}, \ref{assump:rx} holds.

Overall, we have constructed a joint distribution of $(W_2,W_1,X,Y)$ that is consistent with assumptions \ref{assump:posdefVar} and \ref{assump:rx}--\ref{assump:corr}, the known distribution of $(Y,X,W_1)$, and projection equations \eqref{eq:outcome} and \eqref{eq:XprojectionW1W2} with $\beta_\text{long} = b$. It follows that the set $\mathcal{B}_I(\rx, \bar{r}_Y, \bar{c})$ is sharp.
\end{proof}

\bibliographystyle{econometrica}
\bibliography{BadControls_paper}
\makeatletter\@input{DMP1aux.tex}\makeatother